\documentclass[twocolumn,showkeys,showpacs,preprintnumbers,superscriptaddress,nofootinbib,floatfix]
{revtex4}

\usepackage[dvips]{graphicx} 
\usepackage{color}
\usepackage{amsmath}
\usepackage{amssymb}
\newcommand{\beq}{\begin{equation}}
\newcommand{\eeq}{\end{equation}}
\newcommand{\bdm}{\begin{displaymath}}
\newcommand{\edm}{\end{displaymath}}
\newcommand{\bea}{\begin{eqnarray}}
\newcommand{\eea}{\end{eqnarray}}
\newcommand{\bt}{\begin{tabular}}
\newcommand{\et}{\end{tabular}}
\newcommand{\xv}{{\bf x}}
\newcommand{\rv}{{\bf r}}
\newcommand{\yv}{{\bf y}}
\newcommand{\kv}{{\bf k}}

\renewcommand{\v}[1]{\mathbf{#1}}
\newcommand{\xvp}{{\bf x}_\perp}
\newcommand{\kvp}{{\bf k}_\perp}
\newcommand{\xpe}{x_\perp}
\newcommand{\rpe}{r_\perp}
\newcommand{\kpe}{k_\perp}
\newcommand{\rpa}{r_\parallel}

\newcommand{\xit}{\widetilde{\xi}}
\newcommand{\zetat}{\widetilde{\zeta}}

\newcommand{\refeq}[1]{equation~(\ref{eq:#1})}          
         
\newcommand{\evfrac}[2]{\left\langle\frac{#1}{#2}\right\rangle}

\def\<{\langle}
\def\>{\rangle}
\def\d{\delta}
\def\k{\kappa}
\def\g{\gamma}

\def\De{\Delta}

\def\Mpc{\, h^{-1} \, {\rm Mpc}}
\def\cMpc{\, h^{-3} \, {\rm Mpc}^3}

\def\cGpc{\, h^{-3} \, {\rm Gpc}^3}

\def\icMpc{\, h^3 \, {\rm Mpc}^{-3}}

\begin{document}
\preprint{FERMILAB-PUB-08-076-A}

\title{Weak Lensing Effects on the Galaxy Three-Point Correlation Function}

\author{Fabian Schmidt}
\affiliation{Department of Astronomy \& Astrophysics, The University of
Chicago, Chicago, IL 60637-1433}
\affiliation{Kavli Institute for Cosmological Physics, Chicago, IL 
60637-1433}
\author{Alberto Vallinotto}
\affiliation{Institute d'Astrophysique de Paris, CNRS-UMR 7095, 
Universit\' e Paris VI Pierre et Marie Curie, Paris, France}
\author{Emiliano Sefusatti}
\affiliation{Particle Astrophysics Center, Fermi National Accelerator 
Laboratory, Batavia, IL 60510-0500}
\author{Scott Dodelson}
\affiliation{Department of Astronomy \& Astrophysics, The University of
Chicago, Chicago, IL 60637-1433}
\affiliation{Kavli Institute for Cosmological Physics, Chicago, IL 
60637-1433}
\affiliation{Particle Astrophysics Center, Fermi National Accelerator 
Laboratory, Batavia, IL 60510-0500}

\begin{abstract}
We study the corrections to the galaxy three-point correlation function (3PCF)
induced by weak lensing magnification due to the matter distribution along the 
line of sight. We consistently derive all the correction terms arising up to 
second order in perturbation theory and provide analytic expressions as well
as order of magnitude estimates for their relative importance. The magnification
contributions depend on the geometry of the projected triangle on the sky
plane, and scale with different powers of the number count slope and redshift
of the galaxy sample considered.
We evaluate all terms numerically and show that,  
depending on the triangle configuration as well as the galaxy sample considered, 
weak lensing can in general significantly contribute to and alter the 
three-point correlation function observed through galaxy and quasar catalogs. 
\end{abstract}


\date{\today}

\maketitle

\section{Introduction}

The detection and measurement of the three-point and higher order correlation
functions in position space and of their Fourier counterparts in the 
large-scale galaxy distribution date back to the late seventies and early 
eighthies, see, e.g. \cite{SoneiraPeebles1977,BeanEtal1983,JingMoBorner1991,
BaumgartFry1991}. At that time such observations served in the first place as 
a confirmation of the hierarchical model of structure formation. Cosmological
perturbation theory, for instance, makes specific predictions for matter 
higher order correlators which are generated by non-linearities in the 
equations describing gravitational evolution, even when Gaussian initial 
conditions are assumed \cite{Fry1984,BernardeauEtal2002}. An extra source of 
non-Gaussianity is given by non-linearities in the galaxy bias, i.e., in the 
relation between the observed galaxy distribution and the underlying dark matter 
fluctuations \cite{FryGaztanaga1993}.

The measurement of the galaxy bispectrum or 
of the three-point function in redshift surveys can provide a determination of 
linear galaxy bias independently of the normalization of the dark matter 
perturbations \cite{Fry1994,MatarreseVerdeHeavens1997,ScoccimarroEtal2001B}, 
and, in the framework of the halo model, on the parameters controlling the halo 
occupation distribution \cite{SefusattiScoccimarro2005,MarinEtal2008}. This 
kind of analysis has been performed for current large-scale galaxy surveys such 
as the 2dF Galaxy Redshift Survey (2dFGRS) \cite{VerdeEtal2002,
GaztanagaEtal2005,PanSzapudi2005}, and the Sloan Digital Sky Survey (SDSS), 
\cite{KulkarniEtal2007,NishimichiEtal2006}. More recently, it has been shown
that the bispectrum can also provide insights into the non-linear nature of 
galaxy bias which can be used to constrain models of the power spectrum in the 
range of the acoustic oscillations \cite{SmithScoccimarroSheth2006,
SmithScoccimarroSheth2007,McDonald2006}. In general, it has been shown that 
in surveys such as the SDSS, adding bispectrum information allows for an 
improvement in the determination of cosmological parameters as compared
to using the power spectrum alone \cite{SefusattiEtal2006}.

An additional contribution to the galaxy three-point 
function is expected from primordial non-Gaussianity, i.e. if perturbations 
at early times are not perfectly Gaussian \cite{FryScherrer1994,ChodorowskiBouchet1996,Scoccimarro2000A,VerdeEtal2000,DurrerEtal2000,
ScoccimarroSefusattiZaldarriaga2004,HikageKomatsuMatsubara2006}. Current 
constraints on primordial non-Gaussianities derived from CMB observations, 
\cite{KomatsuEtal2003,SpergelEtal2006,CreminelliEtal2007,KomatsuEtal2008,
YadavWandelt2007} still allow for a component potentially 
observable in future large-scale galaxy surveys, 
\cite{ScoccimarroSefusattiZaldarriaga2004,SefusattiKomatsu2007}.

Recent interest in the observation of acoustic features in the galaxy correlation
as a means to accurately map the expansion history of the Universe and thereby constrain 
dark energy parameters \cite{ColeEtal2005,EisensteinEtal2005,Hutsi2006,
PercivalEtal2007}, has motivated several proposals for future large surveys
at high redshift, as, for instance, the Hobby-Eberly Dark Energy Experiment 
(HETDEX) \cite{HillEtal2004}, the Wide-Field Multi-Object Spectrograph (WFMOS) 
\cite{GlazebrookEtal2005}, the Advanced Dark Energy Physics Telescope (ADEPT) 
mission, the ESA Euclid mission \cite{Euclid}, or the Cosmic Inflation Probe 
(CIP) \cite{MelnickEtal2004}.

One of the technical and theoretical challenges that these projects will have to 
face is that the galaxy power 
spectrum or correlation function of high-redshift objects is affected by weak lensing magnification due to the intervening
matter along the line-of-sight \cite{Matsubara2000,VallinottoEtal2007,
HuiGaztanagaLoVerde2007,HuiGaztanagaLoVerde2007B,LoVerdeHuiGaztanaga2008}. 
When a magnitude limited survey is 
carried out, weak gravitational lensing acts on what is observed in several 
ways. The first effect arises because a congruence of photons 
traveling from a distant galaxy to the observer is focused or defocused
by gravitational potentials along the way,
thus resulting in magnification or demagnification of the object observed. 
Weak lensing therefore pushes objects over the threshold magnitude of the survey,
in either direction. The resulting catalog does not contain
all the objects that are intrinsically brighter than the limiting magnitude, 
and contains some objects that are intrinsically fainter than this magnitude.
The size of this effect depends on the slope of the magnitude distribution
of galaxies: the steeper the slope, the stronger the effect of weak lensing
magnification on the observed number density will be.
The second effect is due to the fact that weak lensing also 
stretches or contracts patches of the sky, thus varying the density of 
objects observed. Through these two first-order effects, lensing affects the 
observation of 
galaxy overdensities. Since these are used as a way of quantifying matter 
overdensities, weak lensing enters the measurement of any correlation function 
that is based on a magnitude limited survey. We do not consider the effects of 
the lensing-induced displacement of sources here (i.e. the 
``smoothing effect''), as we expect it to 
be smaller than the magnification contribution.

This work attempts a first step
towards quantifying and possibly exploiting the weak lensing 
contribution to the three point correlation function of pointlike objects like 
galaxies or quasars. 

The intrinsic galaxy three-point function depends on 
three parameters (e.g., the three sides of a triangle). The lensing effects
induce an orientation dependence in the observed 3PCF, specified e.g. by
two angles. Hence, the lensing effects depend on five parameters, in addition
to the number count slope and the redshift considered. In this paper, we
do not attempt to fully explore this vast parameter space. Rather, after 
presenting
the full set of equations, we focus on specific parameter sets
chosen to allow for an understanding of the basic features of the lensing
effects.

For this work, we use the framework of second order cosmological 
perturbation theory \cite{Fry1984,BernardeauEtal2002}. While the validity of 
this approach is limited to large scales, it allows for a straightforward and 
consistent analytical treatment of the magnification correction. Moreover, the 
results are physically intuitive, and, as we will show, all relevant expressions 
reduce to manageable integrals over a small number of auxiliary functions.

The paper is organized as follows. In Sec.~\ref{sec:2} we present the un-lensed 
3PCF, we then derive all the terms arising because of weak lensing and  
provide order of magnitude estimates for the different correction terms. 
In Sec.~\ref{sec:3} we present results for four specific triangular 
configurations and for the reduced three-point function. We also
obtain an estimate of the observability of the lensing effect for future 
surveys. We conclude in Sec.~\ref{sec:4}. The details of the calculation as
well as discussions of approximations used are relegated to the Appendices.

\section{Magnification of the 3PCF}
\label{sec:2}

\subsection{The un-lensed galaxy 3-point correlation function}

We want to study the effects of lensing magnification on the observable
galaxy 3-point correlation function (3PCF) on large scales. A non-vanishing
three-point correlation function, even for Gaussian initial conditions, is 
expected due to non-linearities in the dark matter evolution, as well as due to 
non-linearities in the galaxy bias, i.e. the relation between the observed 
galaxy distribution and the underlying dark matter field.

Here we consider the expression for the dark matter 3PCF predicted by 
tree-level perturbation theory (PT) \cite{Fry1984}, for a review see 
\cite{BernardeauEtal2002}. The dark matter bispectrum, with Gaussian initial 
conditions, is then given by
\beq\label{Btreelevel}
B(k_1,k_2,k_3) = F_2(\kv_1,\kv_2) P_L(k_1)P_L(k_2)+{\rm 2~ perm.},
\eeq
where $P_L(k)$ is the dark matter {\it linear} power spectrum and the PT 
kernel $F_2$ is given by
\beq
F_2(\kv_1,\kv_2)=\frac{5}{7}
+\frac{1}{2}\frac{\kv_1\cdot\kv_2}{k_1 k_2}
+\frac{2}{7}\left(\frac{\kv_1\cdot\kv_2}{k_1 k_2}\right)^2.
\eeq
The three-point matter correlation 
function in position space is defined as
\bea
\zeta(\xv_1,\xv_2,\xv_3) &\!\! \equiv\!\! & \<\d(\xv_1)\d(\xv_2)\d(\xv_3)\>
\nonumber\\
 &\!\! =\!\! &\!\!
\int\!\! d^3 k_1 ~e^{-i\kv_1\cdot\xv_1}\!\!\!\!
\int d^3 k_2 ~e^{-i\kv_2\cdot\xv_2}
\nonumber\\
&\!\!\!\! & \!\!
\times
\int d^3 k_3 ~e^{-i\kv_3\cdot\xv_3}
\d_D(\kv_{123})~B(k_1,k_2,k_3)
\nonumber\\
 &\!\! =\!\! & \!\!
\int\!\! d^3 k_1 e^{-i\kv_1\cdot\xv_{13}}\!\!\!\!
\int\!\! d^3 k_2 e^{-i\kv_2\cdot\xv_{23}}\!
B(\kv_1,\kv_2).\quad\;\;
\eea
Here, $\xv_1$, $\xv_2$, $\xv_3$ denote the vertices of the triangle;
\beq
\xv_{ij} \equiv \xv_i - \xv_j, \quad x_{ij} \equiv |\xv_i-\xv_j|;
\eeq
and $\d_D(\kv_{123})$
is the Dirac delta function for the sum of the three wavevectors $\kv_{123}=\kv_1+\kv_2+\kv_3$.

A simple expression for $\zeta(\xv_1,\xv_2,\xv_3)$ has been obtained by Jing and 
B\"orner \cite{JingBoerner1997} assuming the tree-level matter bispectrum given 
in equation~(\ref{Btreelevel}),
\bea
\zeta & = & \frac{10}{7}\xi(x_{12})\xi(x_{13})
\nonumber\\
& & 
-\left[\eta_2(x_{12})\eta_0(x_{13})+\eta_0(x_{12})\eta_2(x_{13})\right]
 \xv_{12}\cdot\xv_{13}
\nonumber\\
& & +\frac{4}{7}\left[\epsilon(x_{12})\epsilon(x_{13})(\xv_{12}\cdot\xv_{13})^2
\right.
\nonumber\\
& & 
+\eta_2(x_{13})\epsilon(x_{12})x_{12}^2+\eta_2(x_{12})\epsilon(x_{13})x_{13}^2
\nonumber\\
& & 
\left.+3\eta_2(x_{12})\eta_2(x_{13})\right]+ {\rm 2~perm.},
\eea
where the auxiliary functions $\eta_l(x)$ and $\epsilon(x)$ are defined in 
Appendix~\ref{app:3PCFaux}. They are related to derivatives of the two-point
correlation function.

The matter 3PCF is usually measured using survey data, under the assumption 
that galaxy overdensities trace matter overdensities. A simple, local model for
the galaxy bias, consistent with the perturbative treatment
of dark matter evolution, is given by the expansion in terms of the matter
overdensity $\d(\xv)$ \cite{FryGaztanaga1993}, relating
(up to second order) the galaxy overdensity $\d_g(\xv)$ to $\d(\xv)$ as
\beq\label{eq:local_bias}
\d_g(\xv)\simeq b_1\d(\xv)+\frac{b_2}{2}\d^2(\xv),
\eeq
where the two constants $b_1$ and $b_2$ are the linear and quadratic bias 
parameters, respectively. In order to be consistent with the PT expression for 
the dark matter bispectrum it is necessary to retain contributions proportional to 
$b_2$. This is because the galaxy 3PCF receives two contributions of the same 
order: the first one from the dark matter, tree-level, 3PCF and the second one 
from the non-linear bias parameter $b_2$. From equation~(\ref{eq:local_bias}) one obtains
\bea
\<\d_g(\xv_1)\d_g(\xv_2)\d_g(\xv_3)\> & = & b_1^3~\zeta(\xv_1,\xv_2,\xv_3)
\nonumber\\ 
&+& b_1^2 b_2~[\xi_{12}\; \xi_{13} + {\rm 2~perm.}],
\eea
where $\zeta(\xv_1,\xv_2,\xv_3)$ is the matter 3PCF and $\xi_{ij}$ is a 
shorthand notation for the matter two-point function for a given separation 
\beq
\xi_{ij}\equiv\xi(x_{ij})=\<\d(\xv_i)\d(\xv_j)\>.
\eeq

\subsection{Magnification Corrections}\label{sec:2b}

Due to the effect of weak lensing magnification, 
the observed galaxy number density $n(<m,\hat{n})$ below a certain limiting 
magnitude $m$ in a given direction $\hat{n}$ in the sky 
is not equal to the actual density $n_0$, but is related to it via
\cite{Broadhurst1994}
\beq\label{eq:numberdensity}
n(<m,\hat{n})=n_0(<m,\hat{n})\mu(\hat{n})^{2.5 s-1},
\eeq
where $n_0=\bar{n}[1+\d_g(\hat{n})]$, $\mu(\hat{n})$ is the magnification, and 
$s$ is the slope of the number density as a function of the magnitude $m$
\beq
s\equiv\frac{d\ln[n_0(m)]}{dm}.
\eeq 
The magnification $\mu$ is given by
\beq\label{eq:mu}
\mu(\hat{n})=\frac{1}{(1-\k)^2-\g^2},
\eeq
where $\k(\xv)$ and $\g(\xv)$ represent, respectively, the convergence and 
shear (see equations~(\ref{eq:convergence}) and (\ref{eq:shear}) and 
appendix~\ref{app:shearconv} for the definitions). Following the 
perturbative approach adopted for the estimation of the large-scale 3-point 
correlation function, and using the fact that $\k$ and $\g$ are linear in the 
matter density $\d$, we expand the magnification factor in equation~(\ref{eq:mu}) to second 
order as
\beq\label{eq:mu_approx}
\mu\simeq 1+2\k+3\k^2+\g^2.
\eeq
From equations~(\ref{eq:numberdensity}, \ref{eq:mu_approx}), the observed galaxy 
number density can then be expressed as
\bea
n(<m,\hat{n}) & \simeq & \bar{n}(<m)\left(1+b_1~\d+\frac{b_2}{2}\d^2\right)
\nonumber\\
& &\!\! \times 
\left(\! 1+c_1\;\k+\frac{1}{2}c_2\;\k^2+\frac{1}{2}c_1\;\g^2\!\right),
\eea
where we have defined the constants $c_1\equiv(5s-2)$ and 
$c_2\equiv c_1(c_1+1)=(5s-2)(5s-1)$. Up to second order in $\d$, the observed 
galaxy overdensity is then
\bea
\d_{obs}(\xv) & \simeq & b_1~\d +c_1~\k+\frac{1}{2}b_2~\d^2+b_1~c_1~\d~\k
\label{eq:expansion} \nonumber\\
& & +\frac{1}{2}c_2~\k^2 +\frac{1}{2}c_1~\g^2.
\eea

From this expression one derives sixteen different contributions to the 
observed 3PCF, all corresponding to terms of the same order in perturbation
theory. Making use of the
short-hand notation $\d_i\equiv\d(\xv_i)$, $\k_i\equiv\k(\xv_i)$ and 
$\g_i\equiv\g(\xv_i)$, we have
\begin{widetext}
\bea\label{eq:contributions}
\<\d_{obs}(\xv_1)\d_{obs}(\xv_2)\d_{obs}(\xv_3)\> 
& = & 
~b_1^3~\<\d_1\d_2\d_3\> 
+ b_1^2~b_2~[~\frac{1}{2}~\<\d_1\d_2\d^2_3\> + 2~{\rm perm.}]
\nonumber\\
& & 
+ b_1^2~c_1 ~[\<\d_1\d_2\k_3\> + 2~{\rm perm.}]
+ b_1~b_2~c_1 ~[\frac{1}{2}\<\d_1\d^2_2\k_3\> + 5~{\rm perm.}]
\nonumber\\
& & 
+ b_1^3 ~c_1 ~[\<\d_1\d_2\d_3\k_3\> + 2~{\rm perm.}]
+ b_1^2~ c_2 ~[\frac{1}{2}\<\d_1\d_2\k^2_3\> + 2~{\rm perm.}]
\nonumber\\
& & 
+ b_1^2 ~c_1 ~[\frac{1}{2}\<\d_1\d_2\g^2_3\> + 2~{\rm perm.}]
+ b_1 ~c_1^2 ~[\<\d_1\k_2\k_3\>+2~{\rm perm.}]
\nonumber\\
& & 
+ b_2 ~c_1^2 ~[\frac{1}{2}\<\d^2_1\k_2\k_3\>+2~{\rm perm.}]
+ b_1^2 ~c_1^2 ~[\<\d_1\d_2\k_2\k_3\> + 5~{\rm perm.}]
\nonumber\\
& &
+ b_1~c_1~c_2~[\frac{1}{2}\<\d_1\k_2\k^2_3\>+5~{\rm perm.}]
+ b_1~c_1^2~[\frac{1}{2}\<\d_1\k_2\g^2_3\>+5~{\rm perm.}]
\nonumber\\
& & 
+ c_1^3~\<\k_1\k_2\k_3\>
+ c_1^2~c_2~[\frac{1}{2}\<\k_1\k_2\k^2_3\>+2~{\rm perm.}]
\nonumber\\
& & 
+ c_1^3~[\frac{1}{2}\<\k_1\k_2\g^2_3\>+2~{\rm perm.}]
+ b_1~c_1^3~[\<\d_1\k_1\k_2\k_3\>+2~{\rm perm.}].
\eea
\end{widetext}

Each of the contributions may involve several permutations of vertices.
Note that, as one takes into account second order terms in the convergence, 
one cannot distinguish unambiguously between ``galaxy-galaxy-lens'',
``galaxy-lens-lens'', and ``lens-lens-lens'' contributions. For practical 
purposes \textit{only}, we will nevertheless identify 
four groups of terms, the first, referred to as the ``galaxy-galaxy-galaxy'' 
group (or GGG), being given by the intrinsic matter 3PCF plus
contributions due to non-linear bias:
\begin{widetext}
\bea\label{eq:GGG-A}
b_1^3~\<\d_1\d_2\d_3\> 
& = & b_1^3~\zeta(\xv_1,\xv_2,\xv_3) 
\qquad[{\rm GGG-A}],\\
b_1^2~b_2~\left[~\frac{1}{2}~\<\d_1\d_2\d^2_3\> + 2~{\rm perm.}\right]
& = & b_1^2~b_2~(\xi_{12}~\xi_{13}+\xi_{12}~\xi_{23}+\xi_{13}~\xi_{23})
\qquad[{\rm GGG-B}].\label{eq:GGG-B}
\eea
\end{widetext}

For the rest of the paper, we assume without loss of generality that 
$\chi_1\le\chi_2\le\chi_3$, where 
$\chi_i$ represents the comoving distance to the point $\xv_i$.
In the context of the Limber approximation, we can then neglect
all terms where lenses in the direction of closer vertices are correlated
with overdensities further away, such as $\<\k_1 \d_2\>$, $\<\k_1\k_2\d_3\>$,
and so on. The applicability and limitations of the Limber approximation in this
context are discussed in Appendix~\ref{app:Limber}.

We can now
express the non-vanishing contributions due to lensing magnification as 
follows, distinguishing a ``galaxy-galaxy-lens'' (GGL) group,
\begin{widetext}
\bea
\label{eq:GGL-A}
b_1^2~c_1 ~[\<\d_1\d_2\k_3\> + 2~{\rm perm.}]
& = & b_1^2~c_1 ~\<\d_1\d_2\k_3\>
\qquad[{\rm GGL-A}],\\
b_1~b_2~c_1 ~\left[\frac{1}{2}\<\d_1\d^2_2\k_3\> + 5~{\rm perm.}\right]
& = & b_1~b_2~c_1 ~[\xi_{12}(\<\d_2\k_3\>+\<\d_1\k_3\>)+\xi_{13}\<\d_1\k_2\>]
\qquad[{\rm GGL-B}],\label{eq:GGL-B}\\
b_1^3 ~c_1 ~[\<\d_1\d_2\d_3\k_3\> + 2~{\rm perm.}]
& = & b_1^3~c_1~(\xi_{23}\<\d_1\k_3\>+\xi_{13}\<\d_2\k_3\>+\xi_{23}\<\d_1\k_2\>)
\qquad[{\rm GGL-C}],\label{eq:GGL-C}\\
b_1^2~ c_2 ~\left[\frac{1}{2}\<\d_1\d_2\k^2_3\> + 2~{\rm perm.}\right]
& = & b_1^2~c_2~\<\d_1\k_3\>\<\d_2\k_3\>
\qquad[{\rm GGL-D}],\\
b_1^2 ~c_1 ~\left[\frac{1}{2}\<\d_1\d_2\g^2_3\> + 2~{\rm perm.}\right]
& = & b_1^2~c_1~\<\d_1\g_3\>\<\d_2\g_3\>
\qquad[{\rm GGL-E}],
\eea
a ``galaxy-lens-lens'' (GLL) group,
\bea
\label{eq:GLL-A}
b_1 ~c_1^2 ~[\<\d_1\k_2\k_3\>+2~{\rm perm.}]
& = & b_1~c_1^2~\<\d_1\k_2\k_3\>
\qquad[{\rm GLL-A}],\\
b_2 ~c_1^2 ~\left[\frac{1}{2}\<\d^2_1\k_2\k_3\>+2~{\rm perm.}\right]
& = & b_2~c_1^2~\<\d_1\k_2\>\<\d_1\k_3\>
\qquad[{\rm GLL-B}],\\
b_1^2 ~c_1^2 ~\left[\<\d_1\d_2\k_2\k_3\> + 5~{\rm perm.}\right]
& = & b_1^2~c_1^2~[\<\d_1\k_2\>\<\d_2\k_3\>+(\xi_{12}+\xi_{13})\<\k_2\k_3\>
\nonumber\\
& & +(\xi_{12}+\xi_{23})\<\k_1\k_3\>+(\xi_{13}+\xi_{23})\<\k_1\k_2\>]
\qquad[{\rm GLL-C}],\\
b_1~c_1~c_2~\left[\frac{1}{2}\<\d_1\k_2\k^2_3\>+5~{\rm perm.}\right]
& = & b_1~c_1~c_2~[(\<\d_1\k_3\>+\<\d_1\k_2\>)\<\k_2\k_3\>
+\<\d_2\k_3\>\<\k_1\k_3\>]
\qquad[{\rm GLL-D}],\\
b_1~c_1^2~\left[\frac{1}{2}\<\d_1\k_2\g^2_3\>+5~{\rm perm.}\right]
& = & b_1~c_1^2~[\<\d_1\g_3\>\<\k_2\g_3\>+\<\d_1\g_2\>\<\g_2\k_3\>
+\<\d_2\g_3\>\<\k_1\g_3\>]
\qquad[{\rm GLL-E}],
\eea
and a ``lens-lens-lens'' (LLL) group,
\bea
c_1^3~\<\k_1\k_2\k_3\>
& = & c_1^3~\<\k_1\k_2\k_3\>
\qquad[{\rm LLL-A}],\\
c_1^2~c_2~\left[\frac{1}{2}\<\k_1\k_2\k^2_3\>+2~{\rm perm.}\right]
& = & c_1^2~c_2~(\<\k_1\k_2\>\<\k_1\k_3\>+\<\k_1\k_2\>\<\k_2\k_3\>
+\<\k_1\k_3\>\<\k_2\k_3\>)
\qquad[{\rm LLL-B}],\\
c_1^3~\left[\frac{1}{2}\<\k_1\k_2\g^2_3\>+2~{\rm perm.}\right]
& = & c_1^3~(\<\g_1\k_2\>\<\g_1\k_3\>+\<\k_1\g_2\>\<\g_2\k_3\>
+\<\k_1\g_3\>\<\k_2\g_3\>)
\qquad[{\rm LLL-C}],\\
b_1~c_1^3~[\<\d_1\k_1\k_2\k_3\>+2~{\rm perm.}]
& = & b_1~c_1^3~(\<\d_1\k_2\>\<\k_1\k_3\>+\<\d_1\k_3\>\<\k_1\k_2\>
+\<\d_2\k_3\>\<\k_1\k_2\>)
\quad[{\rm LLL-D}].
\label{eq:LLL-D}
\eea
\vspace{0.05cm}
\end{widetext}

The right hand sides in the equations above include explicitly all 
non-vanishing permutations. Weak lensing and non-linear bias each
couple terms which, as ``disconnected'' terms, would not
contribute to the observed 3PCF otherwise \cite{BernardeauEtal2002}.
The non-linear bias couples the matter overdensity at a fixed location with 
itself, see e.g. the GGG-B contribution [equation~(\ref{eq:GGG-B})].
The magnification effect couples a lens along the line of sight to
a galaxy with that galaxy, with a ``coupling strength'' given by
$c_1=5s-2$. For example, for the first term of the GGL-C contribution
[equation~(\ref{eq:GGL-C})], a galaxy at $\xv_1$ correlating with a lens
along the line of sight to the galaxy at $\xv_3$, and a galaxy at $\xv_2$
correlating with the one at $\xv_3$, result in a contribution to the observed
galaxy three-point function. In addition, there are mixed terms where
one coupling is due to non-linear bias while the other is due to lensing
[e.g., GGL-B, equation~(\ref{eq:GGL-B})].

\subsection{Evaluation of Corrections}

We now proceed to evaluate the terms contributing to the observed 3PCF, 
equations~(\ref{eq:GGG-A}--\ref{eq:LLL-D}). As shown in Appendix~\ref{app:shearconv}, 
the magnification and shear perturbations are given as the following 
line-of-sight integrals:
\bea
\label{eq:convergence}
\k(\xv_i) &=& \frac{3}{2}\Omega_m H_0^2
\int_0^{\chi_i}\!\!\!d\chi \frac{W_L(\chi_i,\chi)}{a(\chi)}
\d\left(\yv_i(\chi); \chi\right), \\
\g(\xv_i) &=& \frac{3}{2}\Omega_m H_0^2 \int_0^{\chi_i}d\chi
\frac{W_L(\chi_i,\chi)}{a(\chi)}
\nonumber\\
& & \times \int d^3 k\frac{k_1^2-k_2^2+2ik_1k_2}{k^2}\d_\kv(\chi)
e^{-i\kv\cdot\yv_i(\chi)}.\quad\;\;
\label{eq:shear}
\eea
Here, $W_L(\chi_i,\chi) = \chi/\chi_i(\chi_i-\chi)$, and $\chi_i$ is the 
distance to $\xv_i$, while $\chi$ is the lens distance. 
$\yv_i(\chi) = (\chi/\chi_i)\xv_i$ denotes a point 
moving along the line of sight connecting the observer and the source located at 
$\xv_i$ at a distance $\chi_i$. 
In the following, we also use the shorthand notation $C\equiv 3\Omega_m H_0^2/2$.

In addition to the matter two-point and three-point functions $\xi$ and $\zeta$, 
we need to calculate permutations of 3 different connected three-point terms, 
$\<\d\d\k\>$, $\<\d\k\k\>$, $\<\k\k\k\>$, and of four different two-point terms, 
$\<\d\k\>$, $\<\d\g\>$, $\<\k\k\>$, $\<\k\g\>$. It is quite remarkable that 
they can all be calculated with the help of only four auxiliary functions, as 
detailed in Appendix~\ref{app:3PCFaux}.

We will make use of the Limber approximation \cite{BartelmannSchneider2001} in 
the evaluation of the magnification terms. Each application of the approximation 
sets the lens distance $\chi$ in equations~(\ref{eq:convergence}) and (\ref{eq:shear}) 
to the value of the observable (e.g., $\d$) correlated with, thereby reducing 
the number of line-of-sight integrals by one, and fixing the separation $r$ to 
$\rpe$, the projected (or transverse) separation. See Appendix~\ref{app:Limber} 
for a discussion.

By defining the ``transverse'' two- and three-point correlation functions, 
$\tilde{\xi}$ and $\tilde{\zeta}$, respectively, and the auxiliary function
$\tilde{\epsilon}$  (see Appendix \ref{app:3PCFaux} for more details), one
obtains very similar expressions for almost all magnification terms. The
projected functions are given by
\bea
\tilde{\xi}(\xv_1,\xv_2;\chi)&\equiv&(2\pi)^2\!\!\int d^2k_{\perp}
e^{-i\kv_{\perp}\cdot(\xv_{1\perp}-\xv_{2\perp})} \nonumber\\
& & \times P_L(k_{\perp};\chi),
\label{eq:xi_tilde}
\\
\zetat(\xv_1,\xv_2,\xv_3;\chi)&\equiv&
(2\pi)^2\!\!
\int d^2k_{1\perp} e^{-i\kv_{1\perp}\cdot(\xv_{1\perp}-\xv_{3\perp})}
\nonumber\\
& & \times
\int d^2 k_{2\perp} e^{-i\kv_{2\perp}\cdot(\xv_{2\perp} -\xv_{3\perp})} 
\nonumber\\
& & \times
B(\kv_{1\perp},\kv_{2\perp};\chi),\label{eq:zeta_tilde}
\eea
and
\beq
\widetilde{\epsilon}(x; \chi) \equiv
\frac{(2\pi)^2}{x^3}\int dk P_L(k;\chi) 
[k xJ_0(kx) - 2J_1(kr)].
\eeq
We give all relevant quantities in terms of these functions below, 
first as the exact integral, then using the Limber approximation (L.A.).
In all instances, $x_{ij\perp}$ denotes the transverse component of the 
separation, i.e. the projection onto the sky plane.
\begin{widetext}
\bea
\<\d(\xv_1)\d(\xv_2)\k(\xv_3)\> & = & 
C\int_0^{\chi_3}\!\!\!\!\! d\chi\frac{W_L(\chi_3,\chi)}{a(\chi)}
\zeta(\xv_1,\xv_2,\yv_3(\chi)) \nonumber\\
& \stackrel{\mbox{\tiny L.A.}}{\simeq}& 
C \frac{W_L(\chi_3,\chi_1)}{a(\chi_1)}
\zetat(\xv_{1\perp},\xv_{2\perp},\xv_{3\perp};\chi_1)\:\delta_D(\chi_1-\chi_2),
\label{eq:ggl}\\
\<\d(\xv_1)\k(\xv_2)\k(\xv_3)\> & = &
C^2 
\int_0^{\chi_2} d\chi\frac{W_L(\chi_2,\chi)}{a(\chi)}
\int_0^{\chi_3} d\chi'\frac{W_L(\chi_3,\chi')}{a(\chi')}
\zeta(\xv_1,\yv_2(\chi),\yv_3(\chi'))\nonumber\\
& \stackrel{\mbox{\tiny L.A.}}{\simeq}&
C^2 
\frac{W_L(\chi_2,\chi_1)W_L(\chi_3,\chi_1)}{a^2(\chi_1)} 
\zetat(\xv_{1\perp},\xv_{2\perp},\xv_{3\perp};\chi_1),\label{eq:dkk}
\eea

\bea
\<\k(\xv_1)\k(\xv_2)\k(\xv_3)\>  
& = & 
C^3
\int_0^{\chi_1} d\chi\frac{W_L(\chi_1,\chi)}{a(\chi)}
\int_0^{\chi_2} d\chi'\frac{W_L(\chi_2,\chi')}{a(\chi')}
\int_0^{\chi_3} d\chi''\frac{W_L(\chi_3,\chi'')}{a(\chi'')}
\zeta(\yv_1(\chi),\yv_2(\chi'),\yv_3(\chi''))\nonumber\\
& \stackrel{\mbox{\tiny L.A.}}{\simeq} &
C^3
\int_0^{\chi_1} d\chi
\frac{W_L(\chi_1,\chi)W_L(\chi_2,\chi)W_L(\chi_3,\chi)}{a^3(\chi)}
\zetat(\yv_{1\perp}(\chi),\yv_{2\perp}(\chi),\yv_{3\perp}(\chi);\chi),\label{eq:kkk}
\\
\<\d(\xv_1)\k(\xv_2)\>  
& = & 
C
\int_0^{\chi_2} d\chi\frac{W_L(\chi_2,\chi)}{a(\chi)}
~\xi(\xv_1,\yv_2(\chi))\nonumber\\
& \stackrel{\mbox{\tiny L.A.}}{\simeq} &
C~
\frac{W_L(\chi_2,\chi_1)}{a(\chi_1)}
~\widetilde{\xi}(\xv_{1\perp},\xv_{2\perp};\chi_1),\label{eq:dk}\\
\<\k(\xv_1)\k(\xv_2)\>  
& = & 
C^2
\int_0^{\chi_1} d\chi\frac{W_L(\chi_1,\chi)}{a(\chi)}
\int_0^{\chi_2} d\chi'\frac{W_L(\chi_2,\chi')}{a(\chi')}
\xi(\yv_1(\chi),\yv_2(\chi'))\nonumber\\
& \stackrel{\mbox{\tiny L.A.}}{\simeq} &
C^2
\int_0^{\chi_1} d\chi
\frac{W_L(\chi_1,\chi)W_L(\chi_2,\chi)}{a^2(\chi)}
\widetilde{\xi}(\yv_{1\perp}(\chi),\yv_{2\perp}(\chi);\chi),\label{eq:kk}\\
\<\d(\xv_1)\g(\xv_2)\> 
& = & C
\int_0^{\chi_2}d\chi\frac{W_L(\chi_2,\chi)}{a(\chi)}
\int d^3 k P_L(k,\chi)\frac{k_1^2-k_2^2+2ik_1k_2}{k^2}
e^{-i\kv\cdot(\yv_2(\chi)-\xv_1)}\nonumber\\
&\stackrel{\mbox{\tiny L.A.}}{\simeq}&
C
\frac{W_L(\chi_2,\chi_1)}{a(\chi_1)}
x_{12,\perp}^2 \widetilde{\epsilon}(x_{12,\perp};\chi_1),\\
\<\k(\xv_1)\g(\xv_2)\> & = & C^2
\int_0^{\chi_1}d\chi\frac{W_L(\chi_1,\chi)}{a(\chi)}
\int_0^{\chi_2}d\chi'\frac{W_L(\chi_2,\chi')}{a(\chi')} \nonumber\\
& & \times
\int d^3 k P_L(k;\chi;\chi')\frac{k_1^2-k_2^2+2ik_1k_2}{k^2}
e^{-i\kv\cdot(\yv_2(\chi')-\yv_1(\chi))}\nonumber\\
&\stackrel{\mbox{\tiny L.A.}}{\simeq}&
C^2
\int_0^{\min(\chi_1,\chi_2)}d\chi
\frac{W_L(\chi_1,\chi)W_L(\chi_2,\chi)}{a^2(\chi)}
y_{12,\perp}^2 \widetilde{\epsilon}(y_{12,\perp}(\chi);\chi).\label{eq:kg}
\eea
\end{widetext}

A few comments are in order. To gain a better understanding of the effect of
applying the Limber approximation, consider equation~(\ref{eq:dkk}). The 
first, exact, line requires the integration of the matter three-point function 
-- weighted by the appropriate geometrical lensing factors -- over all triangles 
that have one vertex fixed at $\xv_1$ while the other two (located at $\yv_2$ and 
$\yv_3$) are moving along the lines of sight connecting the observer to 
the source locations at $\xv_2$ and $\xv_3$, respectively. Similarly, in 
equation~(\ref{eq:kkk}) we integrate the unlensed 3PCF over all triangles with 
vertices ($\yv_1$,$\yv_2$,$\yv_3$) lying on the three lines of sight connecting 
the observer to the sources. For all these terms, the Limber approximation
counts the largest contributions to the integrals, which arise from 
configurations that are perpendicular to the line of sight, 
i.e. where the separation $x_{ij\parallel}$ along the line of sight is of 
order $x_{ij\perp}$ or less. This is because the two and three point correlation 
functions increase as the separations get smaller. The largest contributions to 
the integrals will therefore arise when the distance between the points $\yv_i$
moving along the lines of sight are smallest, which corresponds to transverse
separation vectors, i.e. perpendicular to the line of sight. Hence, 
after the Limber approximation all terms are proportional to the 
``transverse'' correlation functions evaluated at the transverse displacements 
$x_{i\perp}$, $y_{i\perp}$. Clearly, the magnification terms depend on the size 
and configuration of the \textit{projected} triangle, rather than the 
three-dimensional triangle itself.  

However, it is important to note that the Limber approximation has limitations.
Most severely, for the GGL-A term [equation~(\ref{eq:ggl})], applying the Limber
approximation yields an
unphysical $\d_D(\chi_1-\chi_2)$, as the lens is forced to be at the same 
distance $\chi$ as \textit{both} $\d_1$ and $\d_2$. To avoid this artifact of
the approximation, we evaluate the exact integral for this term. For all other
terms, we discuss the
accuracy of the
Limber approximation in Appendix~\ref{app:Limber}. Briefly, we found that,
whenever an integral over the line of sight remains in the final expression, the 
Limber approximation is accurate. This is the case for the
$\<\k\k\k\>$, $\<\k\k\>$, and $\<\k\g\>$ terms. For the other terms, the 
approximation can become invalid in certain limits.

After specifying the vertices of the triangle, 
equations~(\ref{eq:ggl}--\ref{eq:kg}) above then allow a calculation of all the 
terms contributing to the observed galaxy 3PCF. The results of this calculation 
are reported in Sec.~\ref{sec:3}. Before plunging into the full calculation of 
all the terms in equations~(\ref{eq:GGG-A}--\ref{eq:LLL-D}), however, it is useful to 
assess the relative magnitude of the different terms, in order to have a better 
understanding of which of them provide the most relevant contribution to the 
observed 3PCF.

\subsection{Order of magnitude estimates}\label{subsec:estimates}

We will give an estimate of the order of magnitude of the various contributions 
in terms of the matter correlation function $\xi(r)$, or, equivalently, the 
dimensionless matter power spectrum $\Delta^2(k)$, since $\xi(r)\sim 
\Delta^2(1/r)$ (see also section 3 of \cite{HuiGaztanagaLoVerde2007}). 

In general, we consider a single scale $r$ for the triangular configuration, 
so that we might expect these estimates to be more accurate for equilateral
configurations. We let $r_\perp$ denote the typical extent of the triangle
in the direction transverse to the line-of-sight, while $r_\parallel$ 
represents the longitudinal extent. The relative magnitude of these is related
to the orientation of the triangle, with $r_\perp=r,\:\rpa=0$ for a triangle
in the sky plane, and $\rpe\sim 0,\:\rpa\sim r$ for a triangle perpendicular to the
sky plane.

We approximate the matter 3PCF simply by the hierarchical relation  
$\zeta(r,r,r)\sim\xi^2(r)$ so that the $\<\d\d\d\>$ contribution is given by
\beq
\zeta_{ggg}\equiv b_1^3\<\d_1\d_2\d_3\> \sim b_1^3 \xi^2(r).
\eeq

The contributions due to magnification involve the projected functions 
$\widetilde{\xi}$ and $\widetilde{\zeta}$ which can be approximated as
\beq
\widetilde{\xi}(\rpe)\sim r_{\perp}\xi(r_{\perp})\simeq 
r_{\perp}\Delta^2(1/r_{\perp}),
\eeq
and
\beq
\widetilde{\zeta}(\rpe)\sim r^2_{\perp}\xi^2(r_{\perp})\simeq 
r^2_{\perp}[\Delta^2(1/r_{\perp})]^2.
\eeq
Note that $\widetilde{\xi}$ and $\widetilde{\zeta}$, as opposed to $\xi$ and 
$\zeta$, are {\it not} dimensionless, but rather have dimensions of length and 
length squared, respectively. We use these approximations in 
equations~(\ref{eq:ggl}--\ref{eq:kg}). Whenever there is an integral, we pull the
correlation functions out, setting $\chi\rightarrow \chi_1/2$, 
$y_{i\perp}\rightarrow \rpe/2$, and integrate over the lensing weight functions
(which can give appreciable numerical factors). When not integrated over, the 
lensing weight function is approximated as
\beq
W_L(\chi_2,\chi_1)\simeq \chi_2-\chi_1= r_\parallel.
\eeq

As an example of this approximation scheme, consider the GLL-A term in equation~(\ref{eq:GLL-A}),
\begin{eqnarray}
b_1 c_1^2 \langle \delta_1 \kappa_2 \kappa_3 \rangle
&=& b_1 c_1^2 C^2\frac{W_L(\chi_2,\chi_1)W_L(\chi_3,\chi_1)}{a^2(\chi_1)}
\widetilde\zeta 
\nonumber\\
&\sim& b_1 c_1^2 C^2 (1+z)^2 r_\parallel^2
r^2_{\perp} \xi^2(r_{\perp}),
\end{eqnarray}
where the first equality follows from the Limber equation in equation~(\ref{eq:dkk}) and the second line 
implements the order of magnitude estimates. Here, $z$ denotes the redshift at which the triangle is located, and the
correlation functions are to be evaluated at that redshift.
Recall that $C\propto H_0^2$ and the canonical GGG term is
of order $b_1^3\xi^2(r)$, so the ratio of this contribution to the canonical term is
\beq
\frac{\mbox{GLL-A}}{\mbox{GGG-A}}
\sim \left(  \frac{c_1}{b_1} (1+z) (H_0 r_\parallel) (H_0r_\perp)  \frac{\xi^2(r_{\perp})}{\xi^2(r)} \right)^2.
\eeq\vspace*{0.1cm}

The order-of-magnitude
expressions can be used to order the contributions according to
significance. Since $H_0\chi$ is of order 1, the main suppressing factors here 
are terms like $H_0 r$, while the ratios of correlation functions can enhance the 
expressions somewhat (since $\rpe < r$ and $\xi$ rises rapidly for small $r$). 
The GLL-A contribution is suppressed by a factor of order $(r/\chi)^4$, denoted $\mathit{O(4)}$ below,
and so is expected to be very small.

Applying this approximation to all contributions leads to
\begin{widetext}
\bea
\frac{b_1^2 c_1\langle \delta_1\delta_2\kappa_3\rangle}{\zeta_{ggg}} 
&\simeq & 
\frac{c_1}{b_1}(1+z) (H_0 \rpa) (H_0 \rpe) 
\left[\frac{\xi(\rpe)}{\xi(r)} \right]^2 \qquad[{\rm GGL-A}\;\;\mathit{O(2)} ],
\\
\frac{b_1 b_2 c_1\langle \delta_1 \delta_2^2 \kappa_3 \rangle}{\zeta_{ggg}} 
&\simeq & 
\frac{b_2c_1}{b_1^2}(1+z) (H_0 \rpa)
(H_0 \rpe) \frac{\xi(\rpe)}{\xi(r)} \qquad[{\rm GGL-B}\;\;\mathit{O(2)} ],
\\
\frac{b_1^3 c_1\langle \delta_1 \delta_2\d_3 \kappa_3 \rangle}{\zeta_{ggg}} 
&\simeq & 
c_1 (1+z) (H_0 \rpa)
(H_0 \rpe) \frac{\xi(\rpe)}{\xi(r)} \qquad[{\rm GGL-C}\;\;\mathit{O(2)} ],
\\
\frac{b_1^2 c_2\langle \delta_1 \delta_2 \kappa_3^2\rangle}{\zeta_{ggg}}  
&\simeq & 
\frac{c_2}{b_1}(1+z)^2 (H_0 \rpa)^2
(H_0 \rpe)^2 \left[ \frac{\xi(\rpe)}{\xi(r)} \right]^2 
\qquad[{\rm GGL-D}\;\;\mathit{O(4)} ],
\\
\frac{b_1^2 c_1\langle \delta_1 \delta_2 \gamma_3^2 \rangle}{\zeta_{ggg}} 
&\simeq & 
\frac{c_1}{b_1} (1+z)^2 (H_0 \rpa)^2
(H_0 \rpe)^2 \left[ \frac{\xi(\rpe)}{\xi(r)} \right]^2  
\qquad[{\rm GGL-E}\;\;\mathit{O(4)} ],
\\
\frac{b_1 c_1^2\langle \delta_1\kappa_2\kappa_3\rangle}{\zeta_{ggg}} 
&\simeq & 
\frac{c_1^2}{b_1^2}(1+z)^2 (H_0 \rpa)^2 (H_0 \rpe)^2 
\left[\frac{\xi(\rpe)}{\xi(r)} \right]^2 \qquad[{\rm GLL-A}\;\;\mathit{O(4)} ],
\\
\frac{b_2 c_1^2\langle \delta_1^2 \kappa_2 \kappa_3 \rangle}{\zeta_{ggg}} 
&\simeq & 
\frac{b_2 c_1^2}{b_1^3}(1+z)^2 (H_0 \rpa)^2
(H_0 \rpe)^2 \left[ \frac{\xi(\rpe)}{\xi(r)} \right]^2 
\qquad[{\rm GLL-B}\;\;\mathit{O(4)} ],
\eea

\bea
\frac{b_1^2 c_1^2\langle \d_1\delta_2 \kappa_2 \kappa_3 \rangle}{\zeta_{ggg}} 
&\simeq & 
\frac{c_1^2}{b_1}\frac{1}{30} (1+z)^2 (H_0\chi)^3 (H_0 \rpe)
~\frac{\xi(\rpe/2)}{\xi(r)}
\qquad[{\rm GLL-C}\;\;\mathit{O(1)} ],
\\
\frac{b_1 c_1 c_2\langle \delta_1 \kappa_2 \kappa_3^2\rangle}{\zeta_{ggg}} 
&\simeq & 
\frac{c_1c_2}{b_1^2}\frac{1}{30}(1+z)^3 (H_0 \chi)^3 
(H_0 \rpa) (H_0 \rpe)^2  \left[ \frac{\xi(\rpe/2)}{\xi(r)} \right]^2 
\qquad[{\rm GLL-D}\;\;\mathit{O(3)} ],
\\
\frac{b_1 c_1^2\langle \delta_1 \kappa_2 \gamma_3^2 \rangle}{\zeta_{ggg}} 
&\simeq & 
\frac{c_1^2}{b_1^2}\frac{1}{30}(1+z)^3 (H_0 \chi)^3 
(H_0 \rpa) (H_0 \rpe)^2  \left[\frac{\xi(\rpe/2)}{\xi(r)} \right]^2 
\qquad[{\rm GLL-E}\;\;\mathit{O(3)} ],
\\
\frac{c_1^3\langle \kappa_1\kappa_2\kappa_3\rangle}{\zeta_{ggg}} 
&\simeq & 
\frac{c_1^3}{b_1^3} \frac{1}{140} 
(1+z)^3 (H_0 \chi)^4 (H_0 \rpe)^2 
\left[\frac{\xi(\rpe/2)}{\xi(r)} \right]^2 
\qquad[{\rm LLL-A}\;\;\mathit{O(2)} ],
\\
\frac{c_1^2 c_2\langle \kappa_1 \kappa_2 \kappa_3^2\rangle}{\zeta_{ggg}} 
&\simeq & 
\frac{c_1^2c_2}{b_1^3}\frac{1}{30^2}(1+z)^4 (H_0 \chi)^6 
(H_0 \rpe)^2  \left[\frac{\xi(\rpe/2)}{\xi(r)} \right]^2 
\qquad[{\rm LLL-B}\;\;\mathit{O(2)} ],
\\
\frac{c_1^3\langle \kappa_1 \kappa_2 \gamma_3^2\rangle}{\zeta_{ggg}} 
&\simeq & 
\frac{c_1^3}{b_1^3}\frac{1}{30^2}(1+z)^4 (H_0 \chi)^6 
(H_0 \rpe)^2  \left[ \frac{\xi(\rpe/2)}{\xi(r)} \right]^2 
\qquad[{\rm LLL-C}\;\;\mathit{O(2)} ],
\\
\frac{b_1 c_1^3\langle \delta_1 \k_1\kappa_2\kappa_3\rangle}{\zeta_{ggg}} 
&\simeq & 
\frac{c_1^3}{b_1^2}\frac{1}{30}(1+z)^3 (H_0 \chi)^3 
(H_0 \rpa) (H_0 \rpe)^2  \left[ \frac{\xi(\rpe/2)}{\xi(r)} \right]^2 
\qquad[{\rm LLL-D}\;\;\mathit{O(3)} ].
\eea
\end{widetext}

The order in powers of $H_0 r$ is indicated by the symbols $\mathit{O(n)}$ in 
each line. Comparing with the exact expressions, we have verified that the 
above expressions correctly reproduce the relative ordering of the different 
terms. The largest contributions to the magnification corrections of the 3PCF 
arise from the $\mathit{O(1)}$ and $\mathit{O(2)}$ terms. Terms of order
$\mathit{O(3)}$ and $\mathit{O(4)}$ are significantly suppressed.

The significant GGL and GLL terms do not increase dramatically with 
redshift, as they scale with $(1+z)$ or $(1+z)^2$. The terms 
belonging to the LLL group on the other hand scale with third and fourth power 
of the redshift: one expects their weight relative to the other terms to 
increase with increasing redshift. This is consistent with a similar behavior 
of the GL and LL magnification terms in the two-point correlation function, as already 
noticed by \cite{Matsubara2000,VallinottoEtal2007,HuiGaztanagaLoVerde2007}. 

\section{Results}\label{sec:3}

The relative magnitude of the magnification corrections to the galaxy 3PCF 
derived in the previous section is highly dependent on the shape of the
specific triangular configuration considered, as well as on its orientation
with respect to the line-of-sight. Typically, as shown by the estimates 
presented above, we expect a large effect for triangles whose projection onto
the sky is much smaller than the three-dimensional extent, i.e. elongated 
triangles oriented along the line-of-sight. Another factor to be taken 
into account is the specific value for the linear and quadratic bias parameters 
as well as the number count slope that we can expect for the sample under 
consideration.

\subsection{Specific triangle configurations}

We will not present here a complete analysis of the detectability of this effect 
as this is beyond the scope of the present work. We will limit ourselves to 
consider some specific configurations at different redshifts and compute the 
relative magnification contributions. In section~\ref{sec:obs} we will comment
on the uncertainty of the 3PCF that one might expect in future
high-redshift surveys for these specific triangles.

We consider four classes of triangular configurations. The first
three correspond to triangles lying in a plane that contains the line-of-sight. 
In this case, the projections of the three galaxy positions on the sky plane
will be aligned. We fix the transverse 
separation between the points and vary in different ways their line-of-sight 
separation. Let us use $r_{ij,\perp}$ to denote the transverse separation 
between galaxies at positions $\xv_i$ and $\xv_j$, while $r_{ij,\parallel}=
\chi_j-\chi_i$ is the separation along the line of sight. 

We carefully chose triangle configurations and scales to ensure 
that the tree-level
perturbation theory approach remains valid. We can assume loop corrections 
in perturbation 
theory to be subdominant above scales of the order of $20\Mpc$. At the same 
time, while the lensing signal becomes relatively larger at large scales, it 
will still be difficult to measure the three-point correlation at scales of 
hundreds of Mpc, even in  forthcoming redshift surveys.
\begin{figure}[t]
\begin{center}
\includegraphics[width=.48\textwidth]{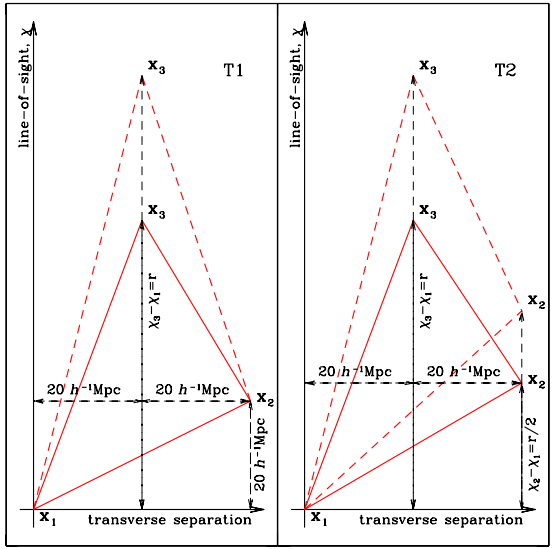}
\caption{\label{fig:triangles_t1t2} Triangular configurations T1 and T2. The 
dashed (red) line shows a second triangle of the same class.}
\end{center}
\end{figure}
\begin{figure}[t]
\begin{center}
\includegraphics[width=.48\textwidth]{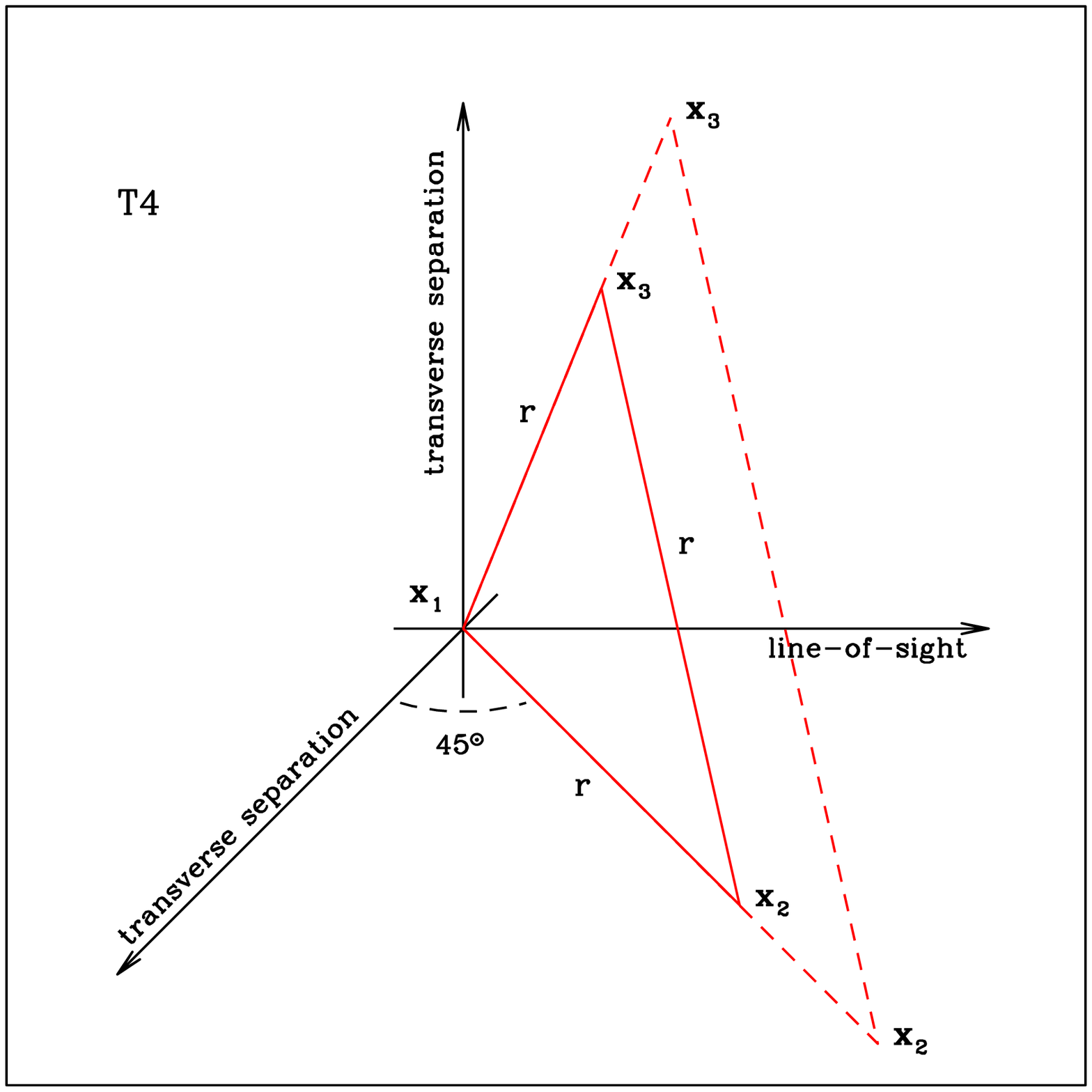}
\caption{\label{fig:triangles_t4} Triangular configurations T4. The dashed
(red) line shows a second triangle of the same class.}
\end{center}
\end{figure}

\begin{figure*}[t]
\begin{center}
\includegraphics[width=.32\textwidth]{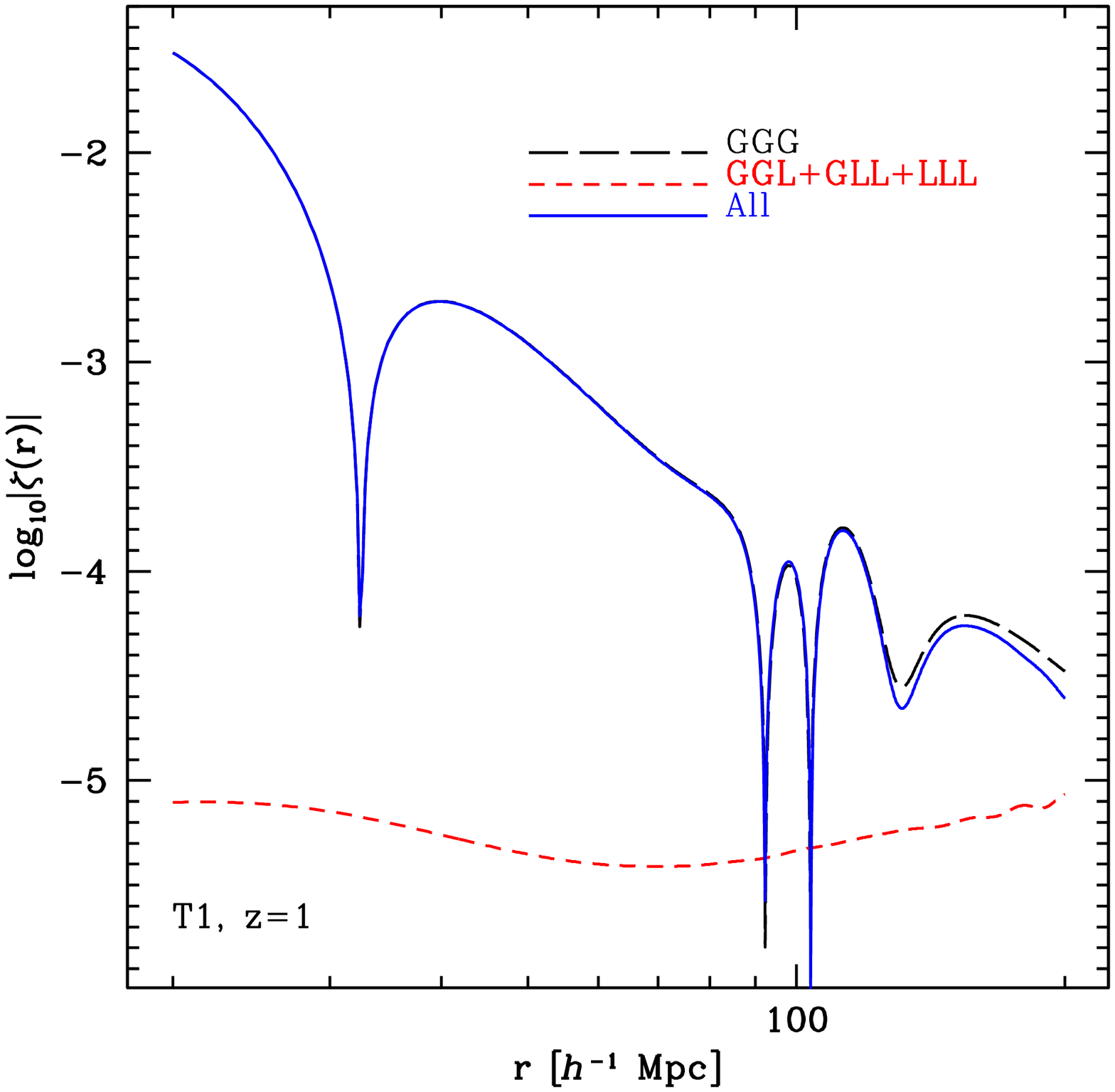}
\includegraphics[width=.32\textwidth]{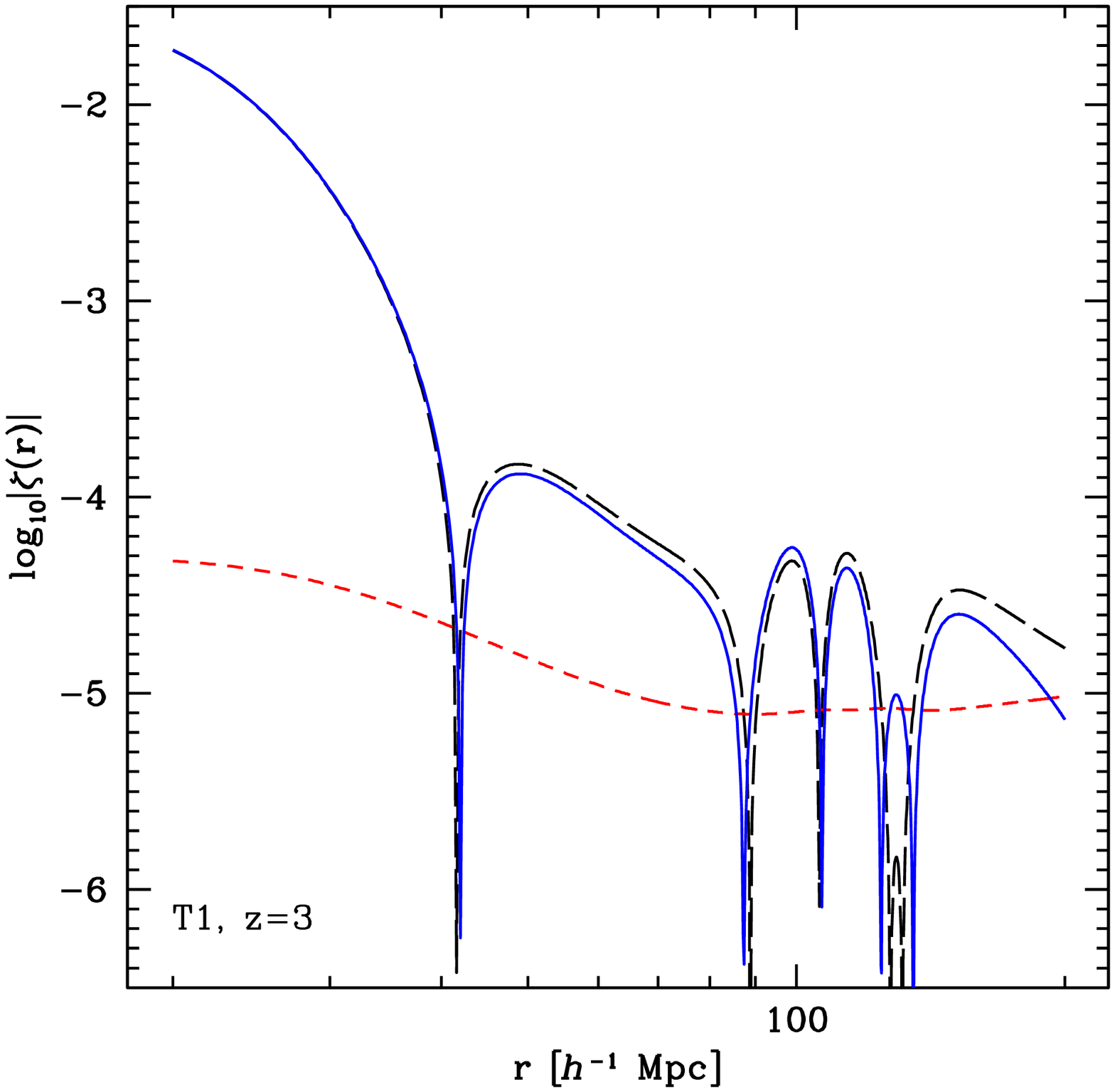}
\includegraphics[width=.32\textwidth]{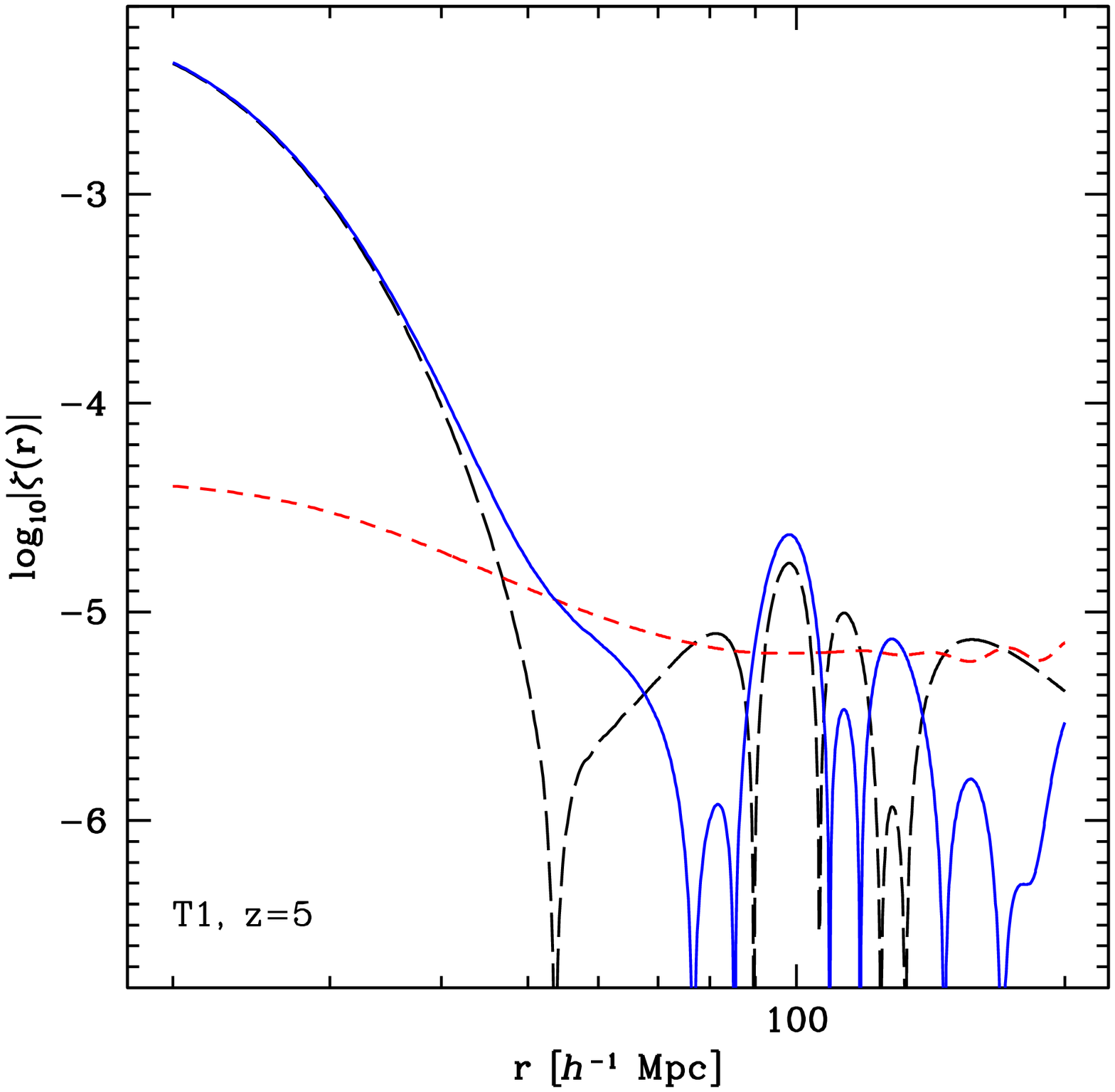}
\includegraphics[width=.32\textwidth]{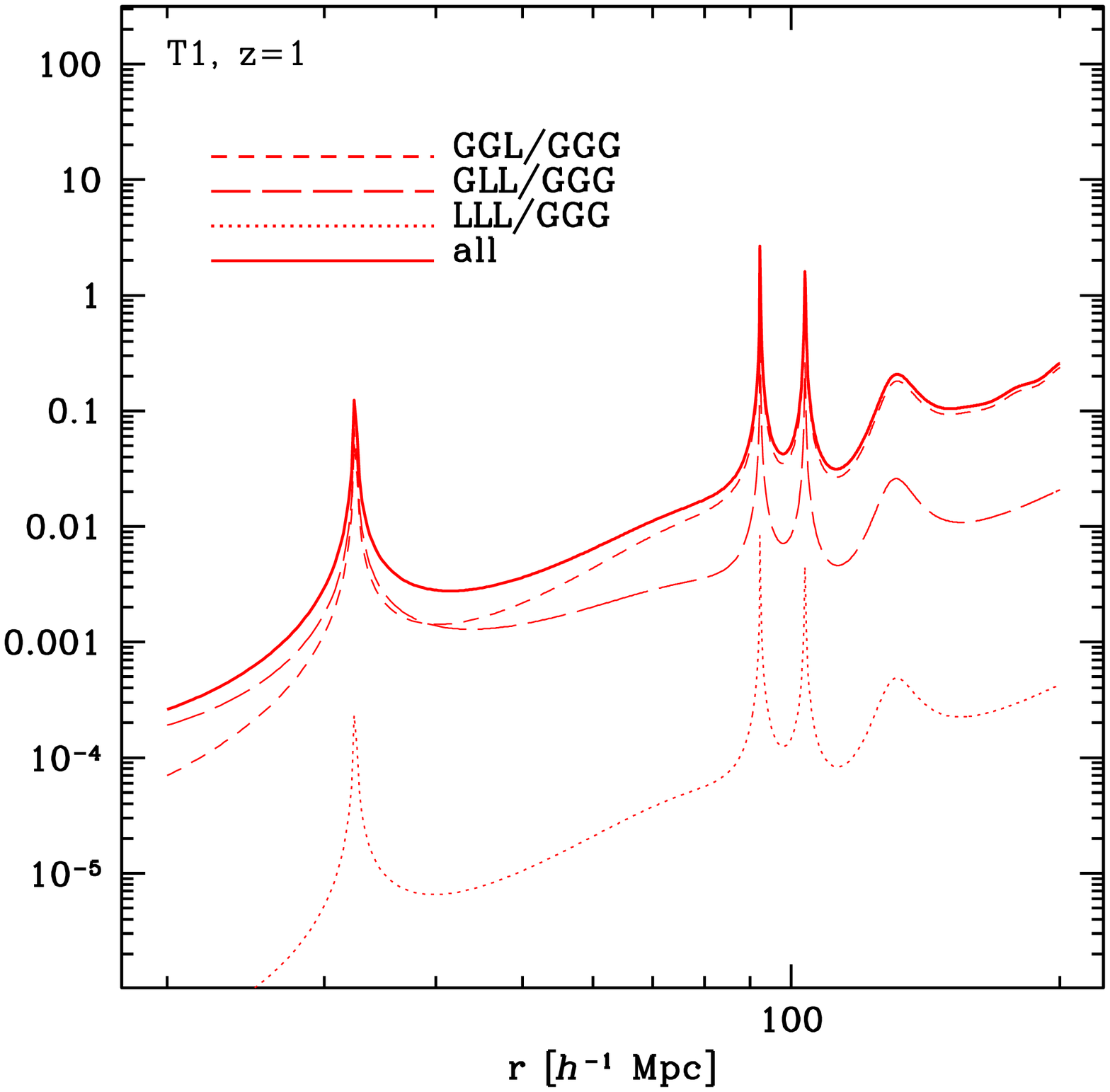}
\includegraphics[width=.32\textwidth]{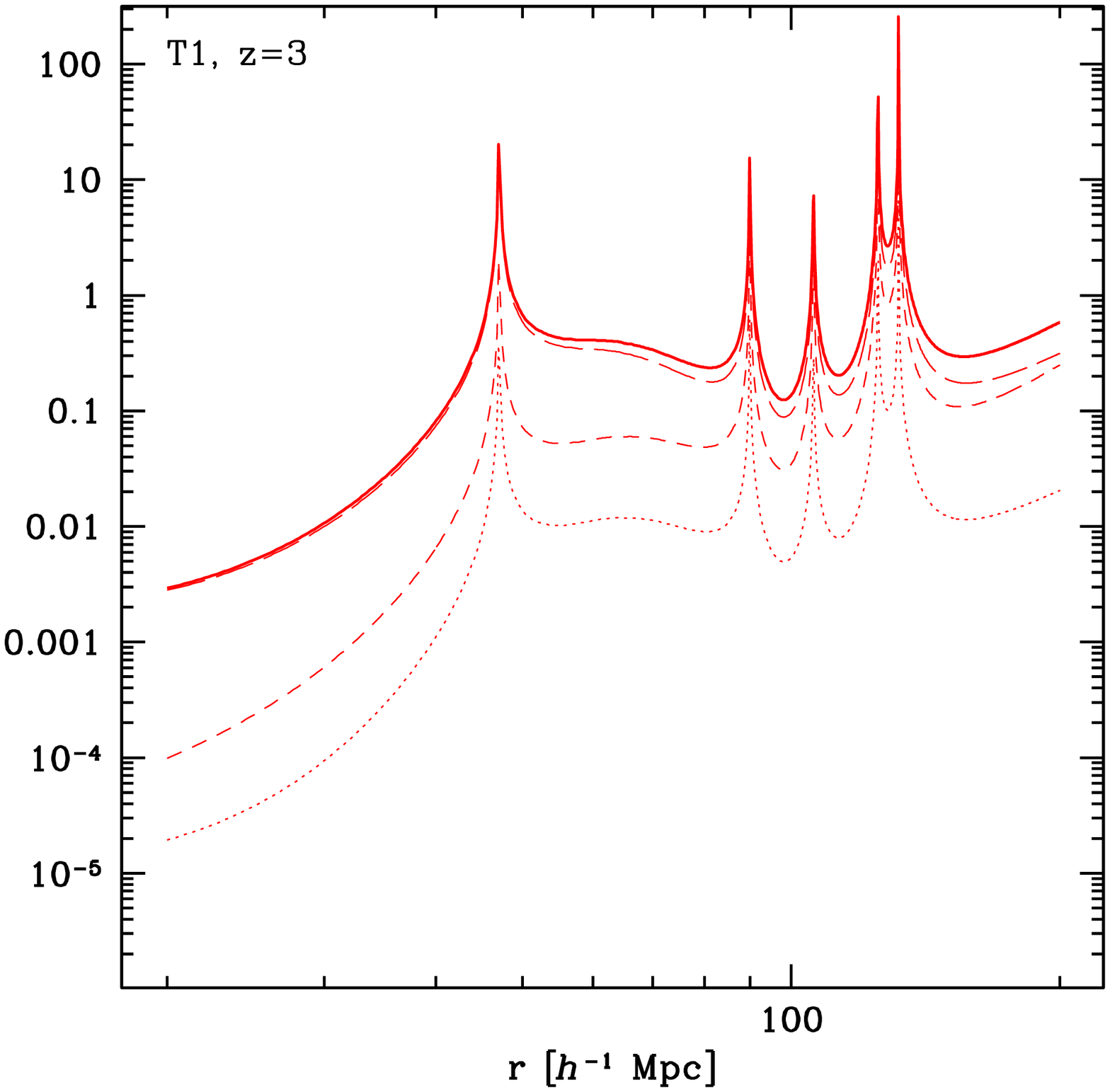}
\includegraphics[width=.32\textwidth]{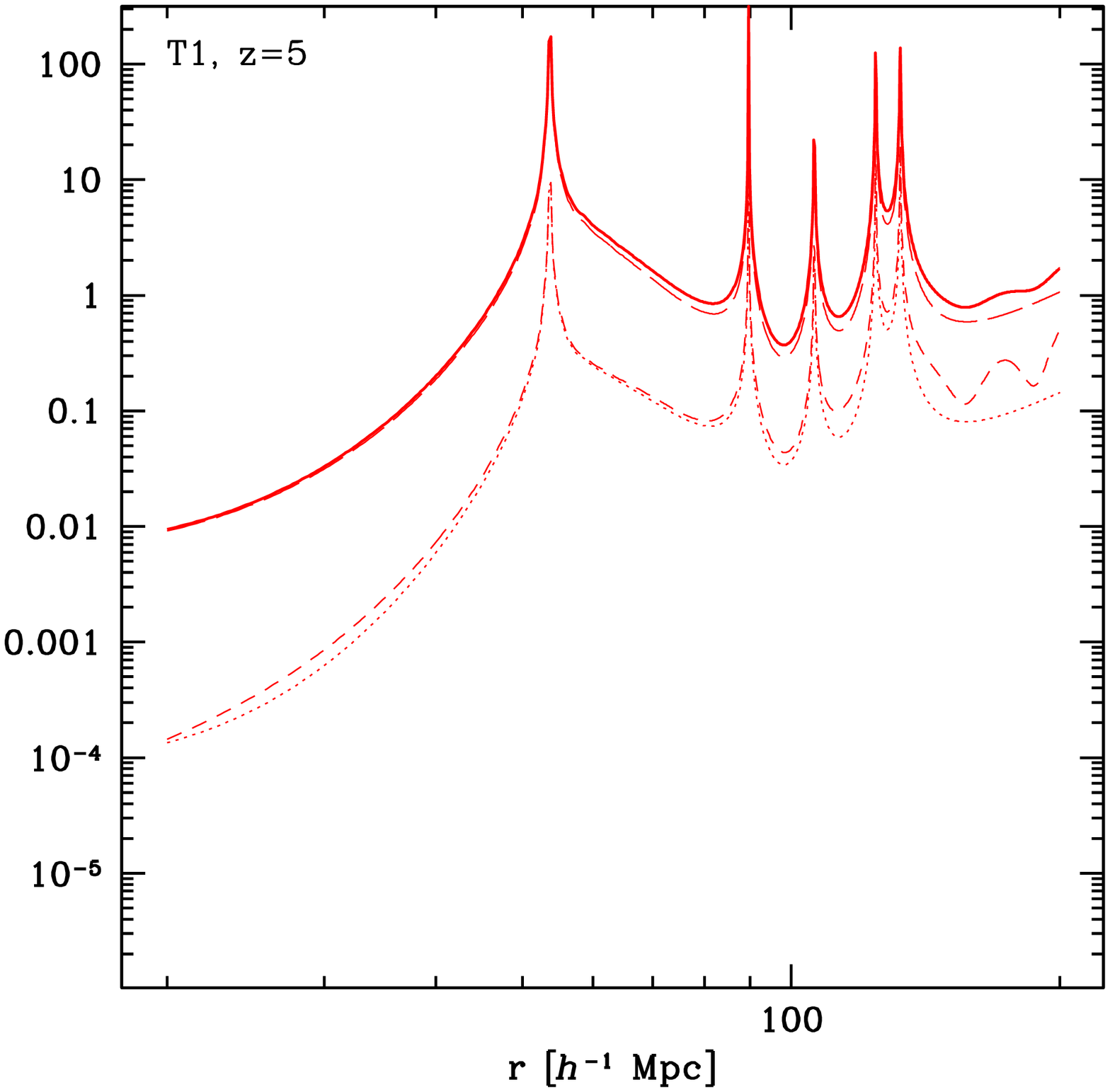}
\caption{\label{fig:t1} Triangular configuration T1. The upper panels show
the absolute value of the unlensed 3PCF, the GGG group (black, long-dashed
line), the contributions due to magnification, sum of the GGL, GLL and 
LLL terms (red, short-dashed line) and the total (blue, continuous line). 
The lower panels show the ratio between the magnification 
contributions and the unlensed 3PCF (continuous line), again as absolute value.
We also distinguish the 
GGL (short-dashed line), GLL (long-dashed line) and
LLL terms (dotted line). First, second and third columns correspond
respectively to low redshift ($z=1$), intermediate redshift ($z=3$), and high 
redshift ($z=5$). Specific values for the bias parameters are included as 
discussed in the text.}
\end{center}
\end{figure*}

\begin{figure*}[t]
\begin{center}
\includegraphics[width=.32\textwidth]{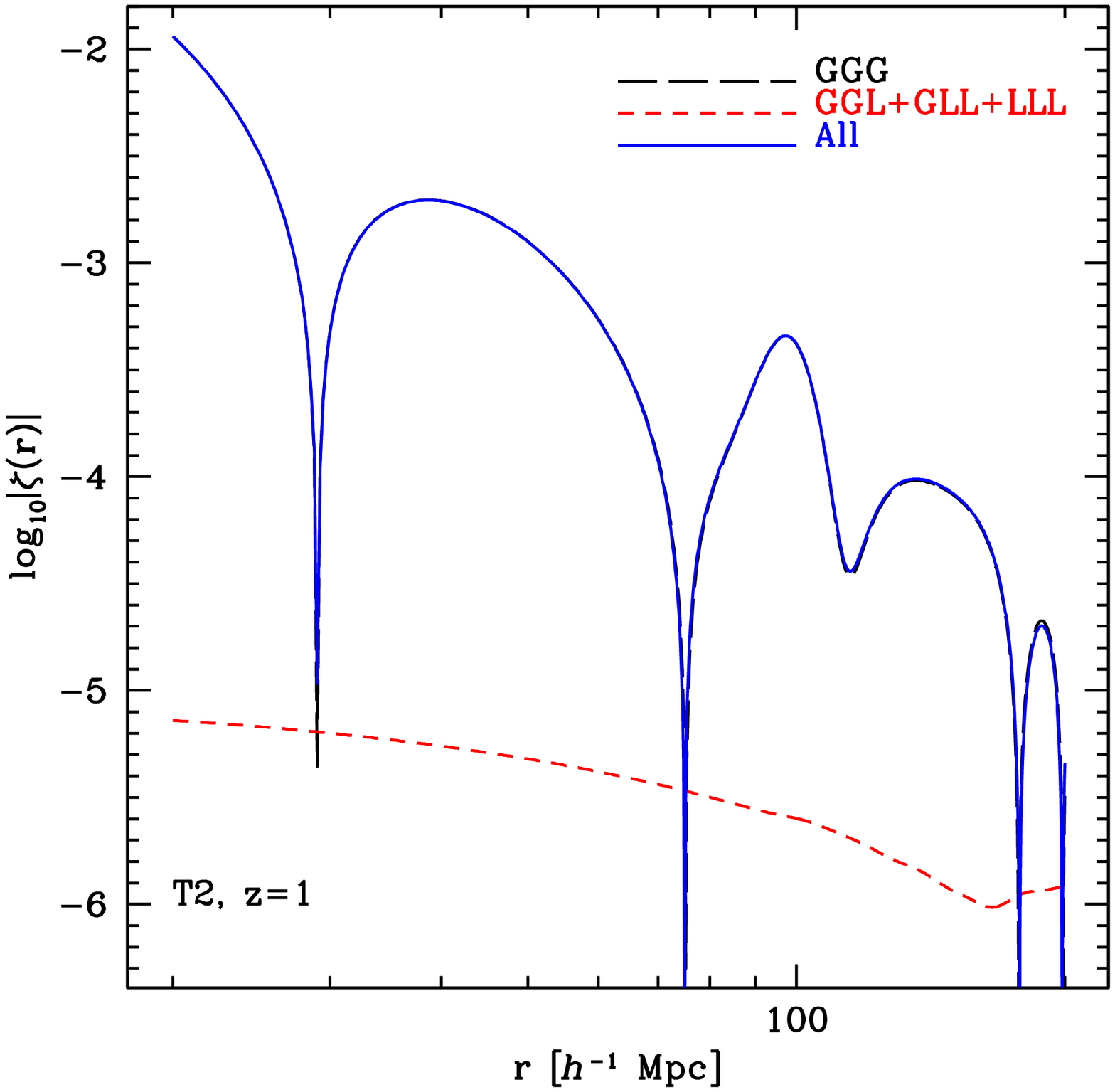}
\includegraphics[width=.32\textwidth]{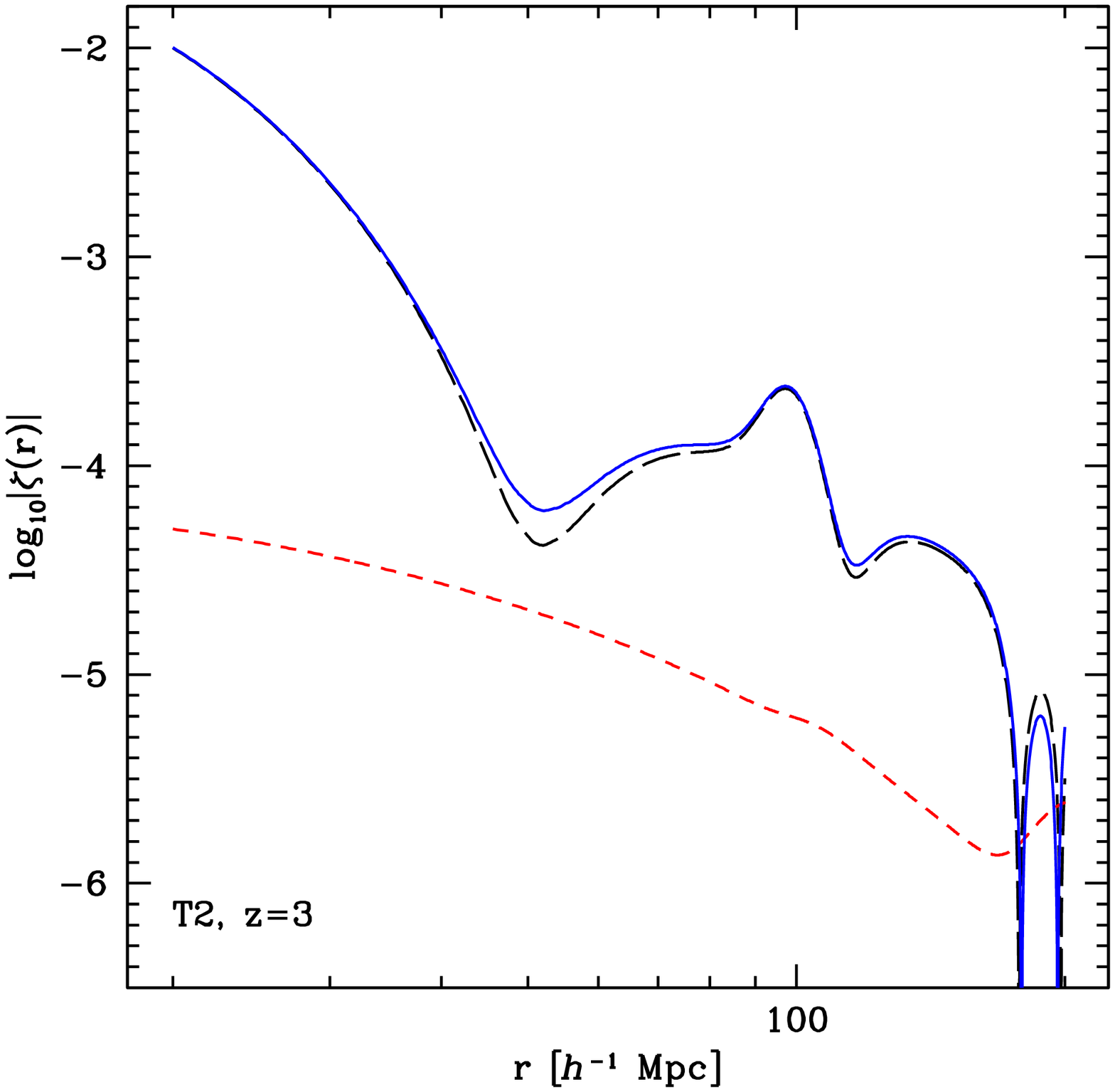}
\includegraphics[width=.32\textwidth]{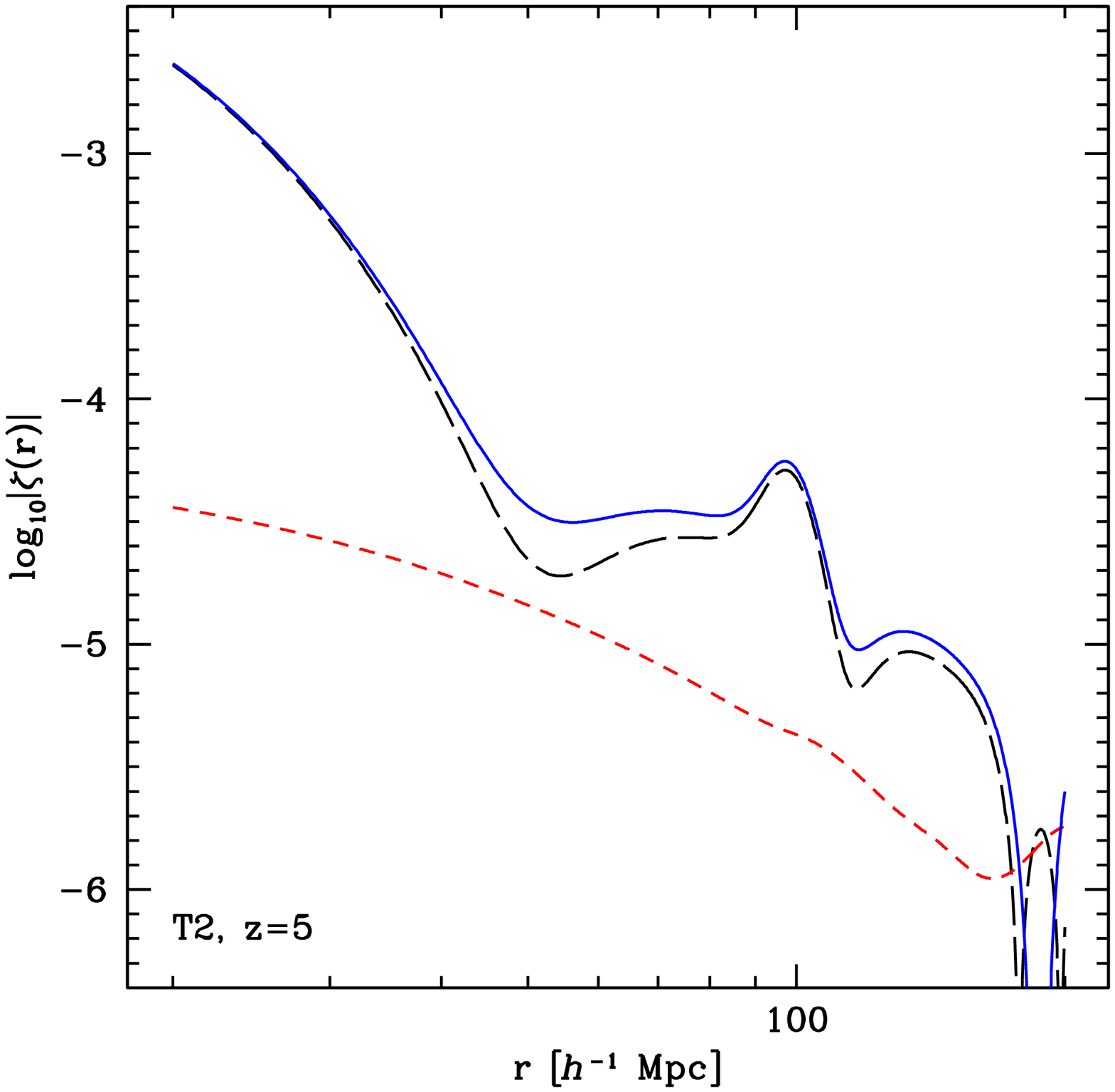}
\includegraphics[width=.32\textwidth]{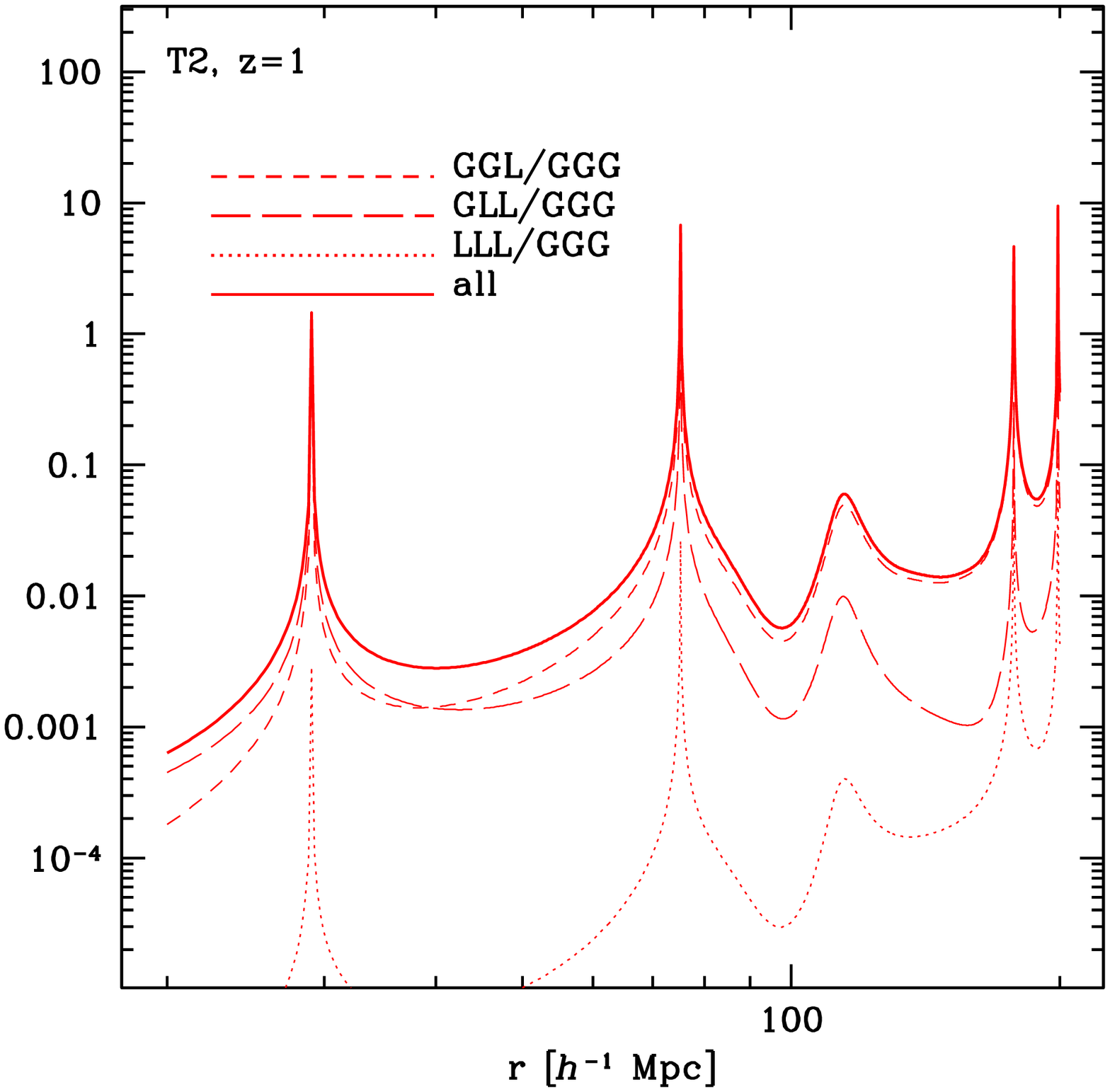}
\includegraphics[width=.32\textwidth]{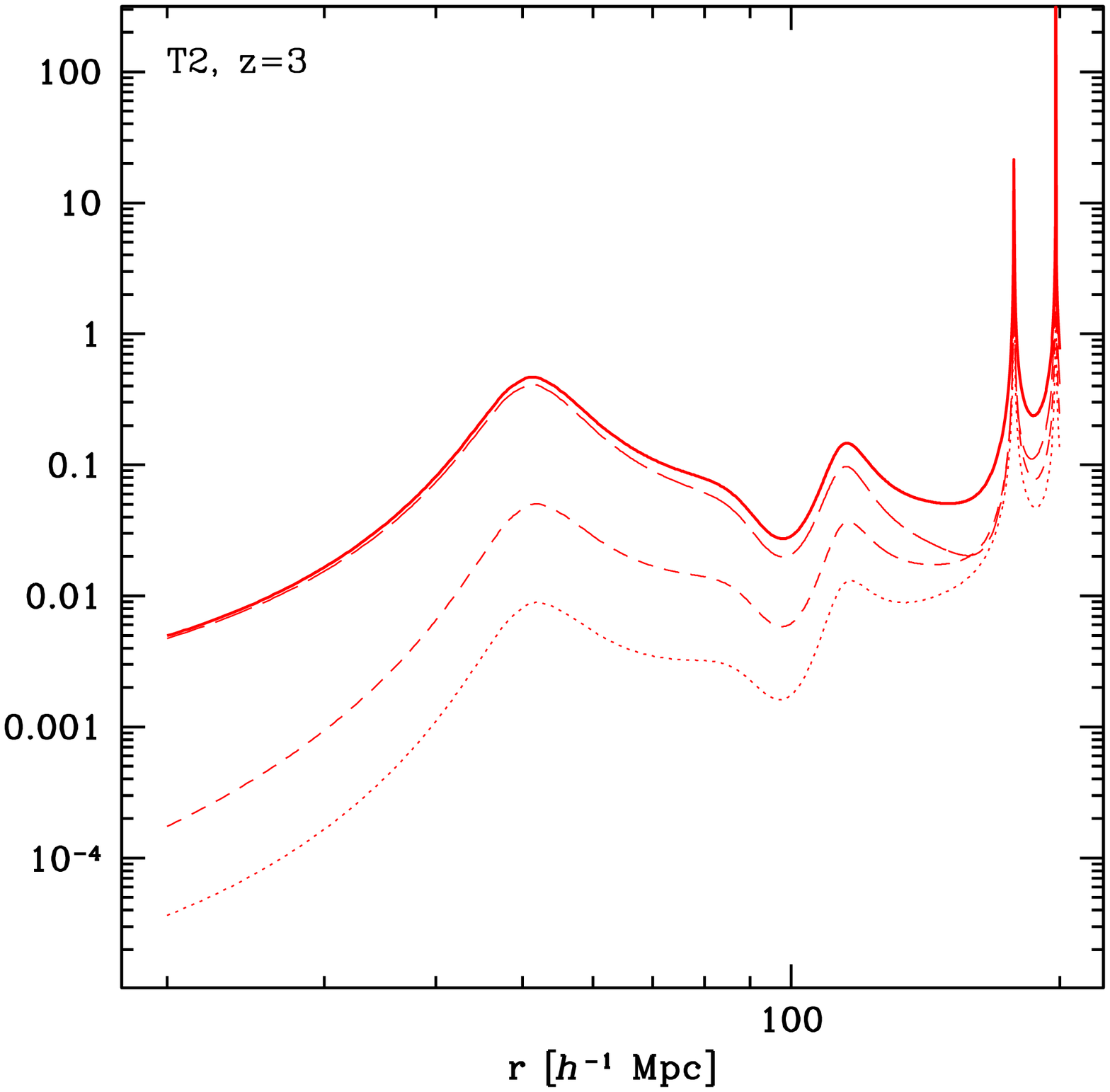}
\includegraphics[width=.32\textwidth]{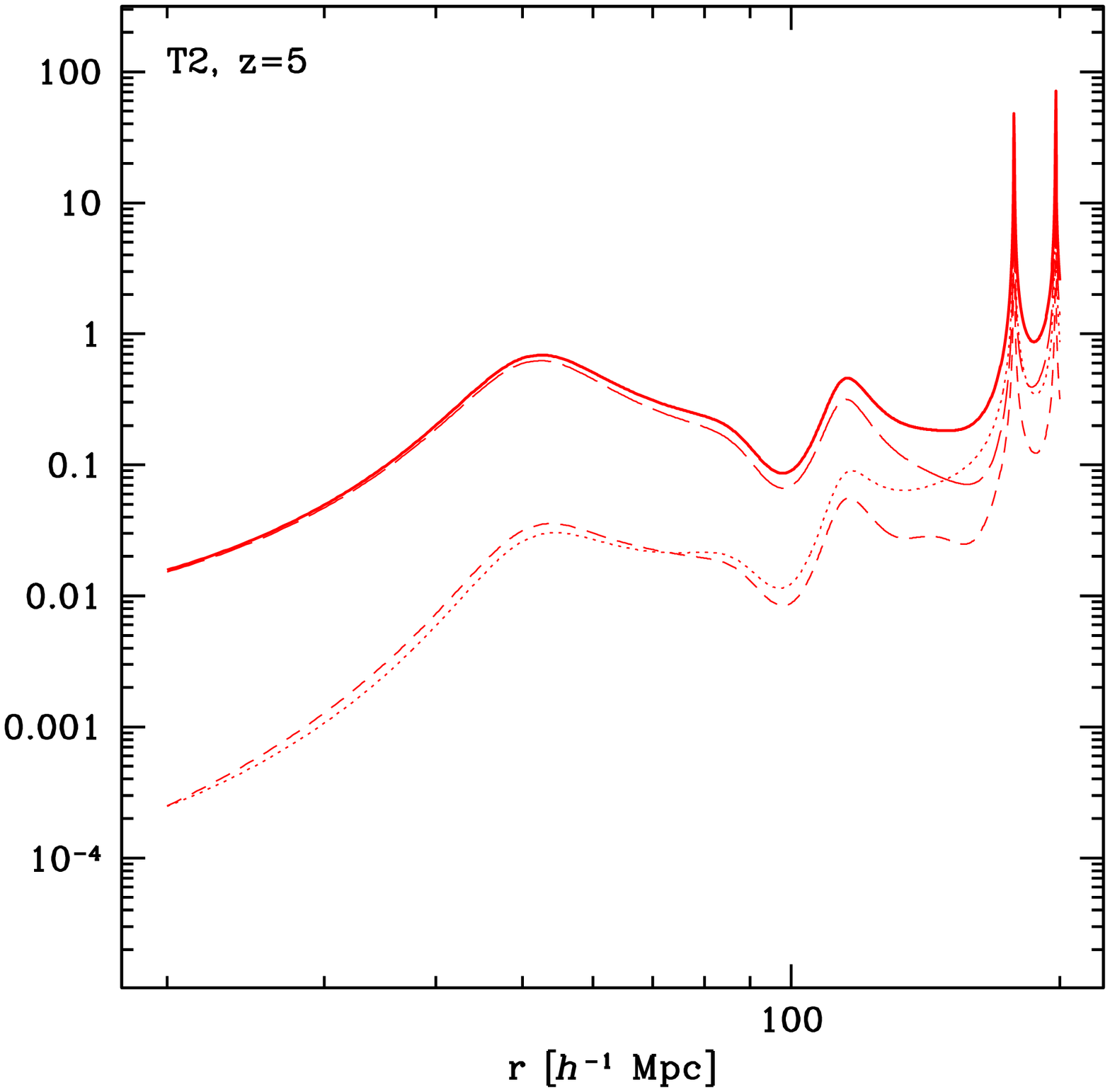}
\caption{\label{fig:t2} Triangular configuration T2, $z=1$, $3$, and $5$.}
\end{center}
\end{figure*}
\begin{figure*}[t]
\begin{center}
\includegraphics[width=.32\textwidth]{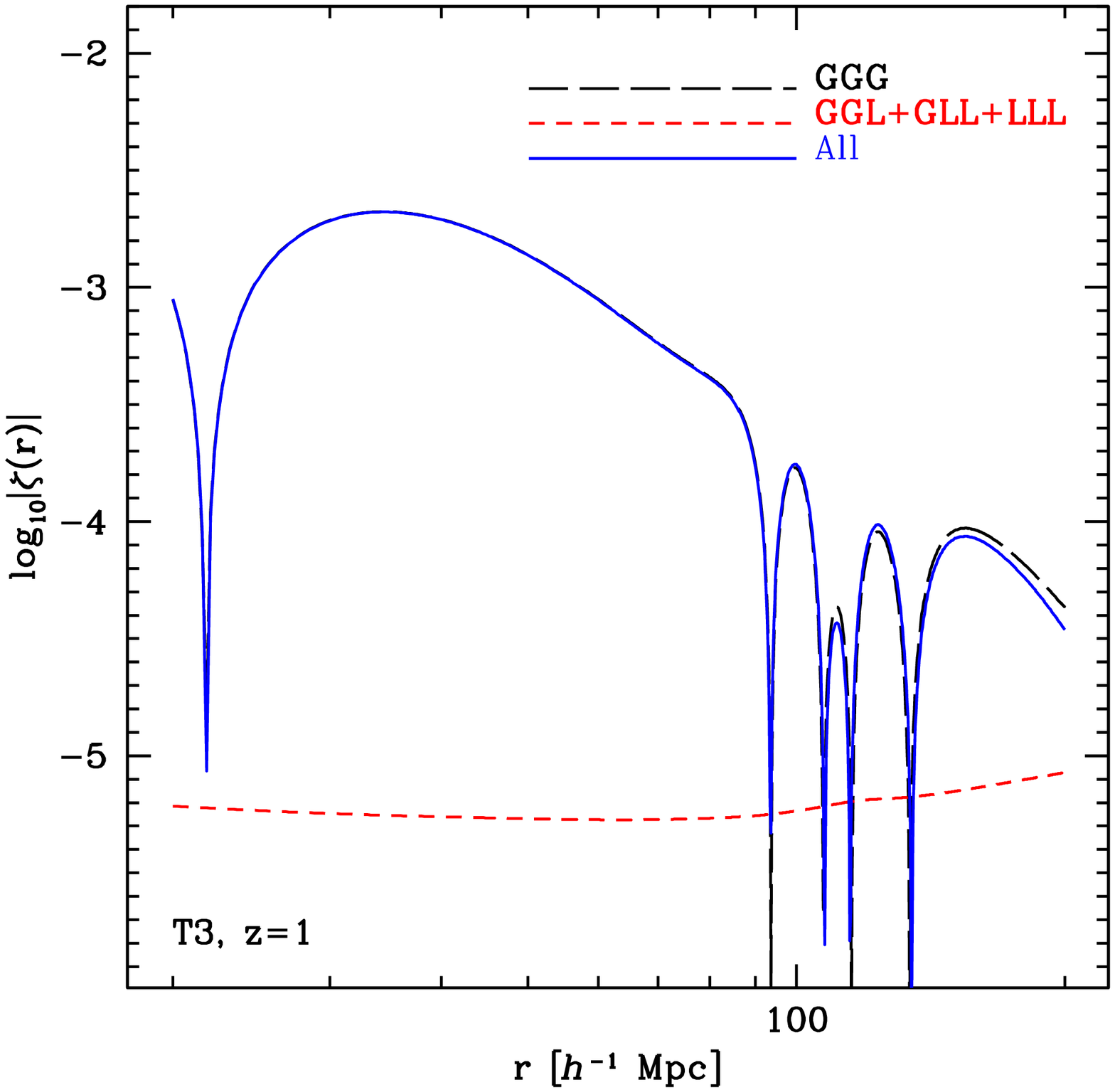}
\includegraphics[width=.32\textwidth]{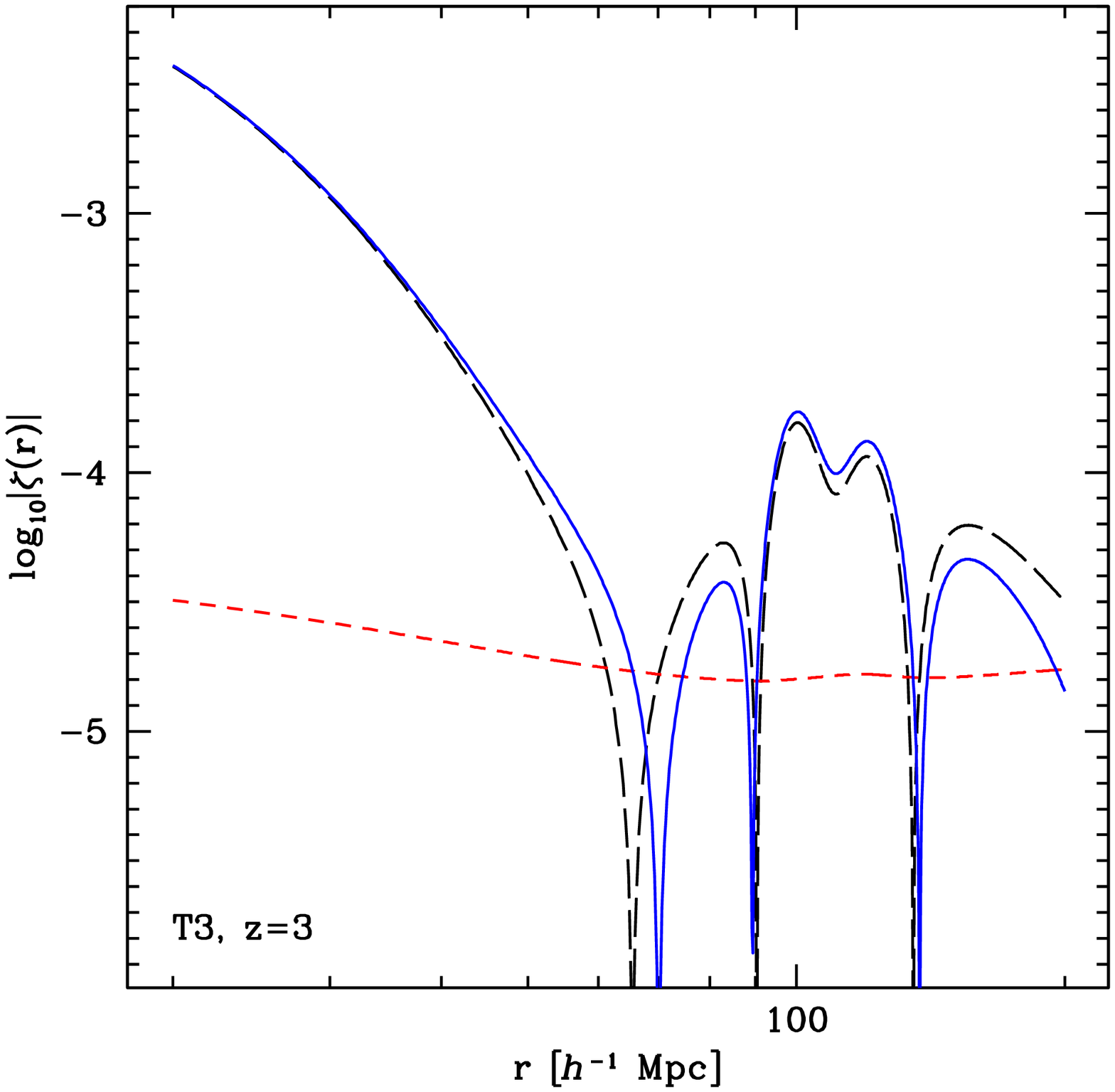}
\includegraphics[width=.32\textwidth]{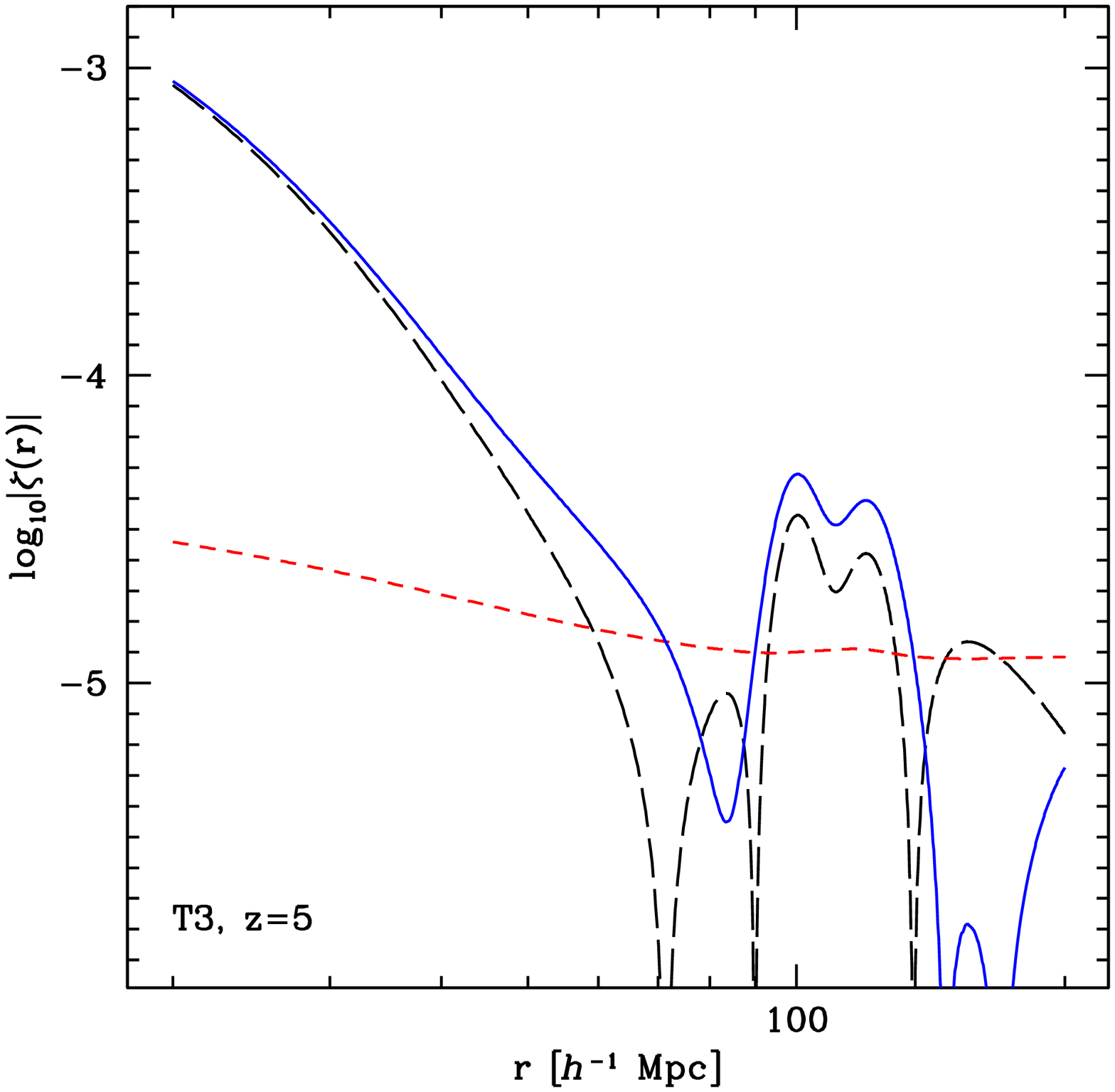}
\includegraphics[width=.32\textwidth]{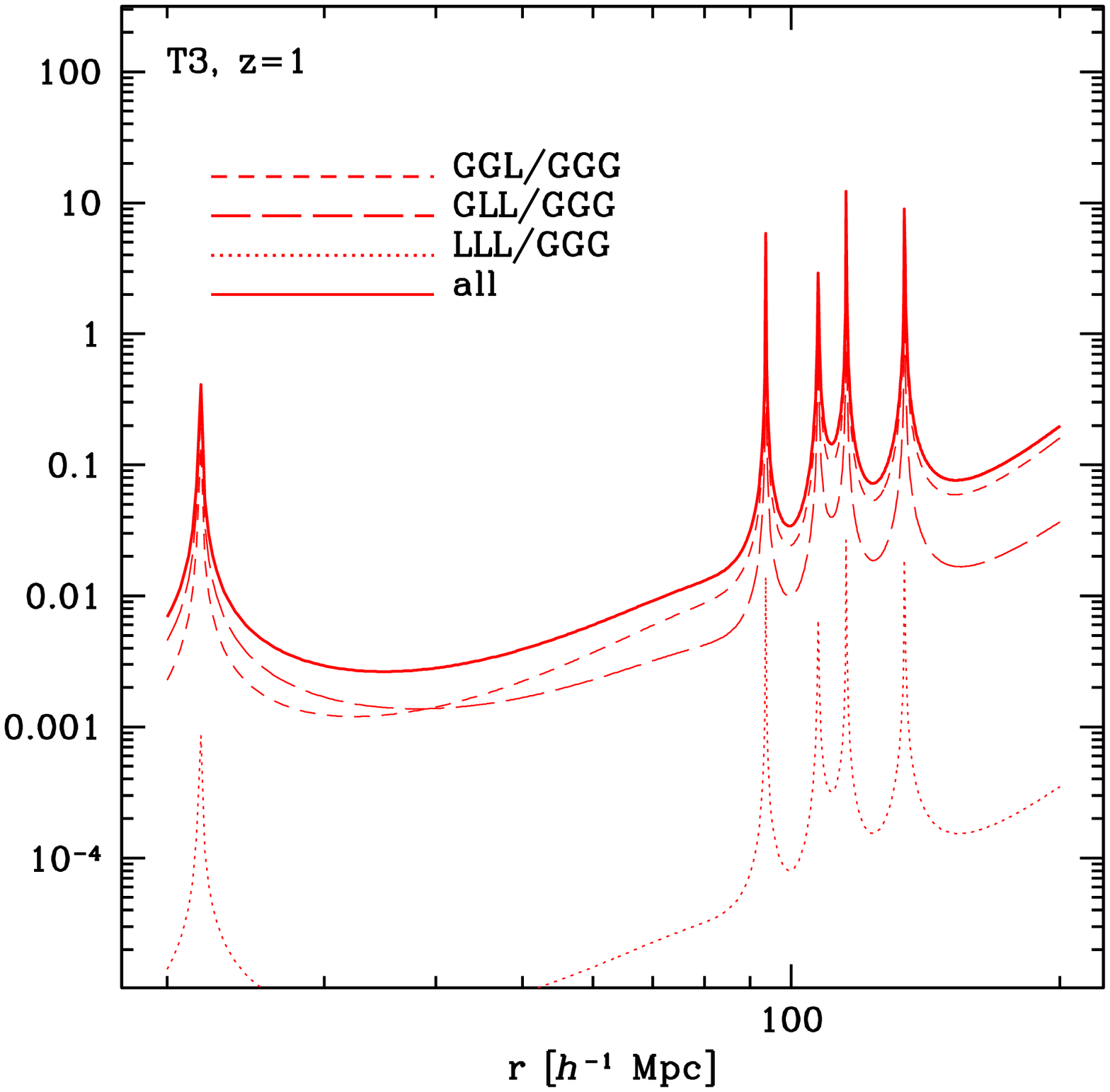}
\includegraphics[width=.32\textwidth]{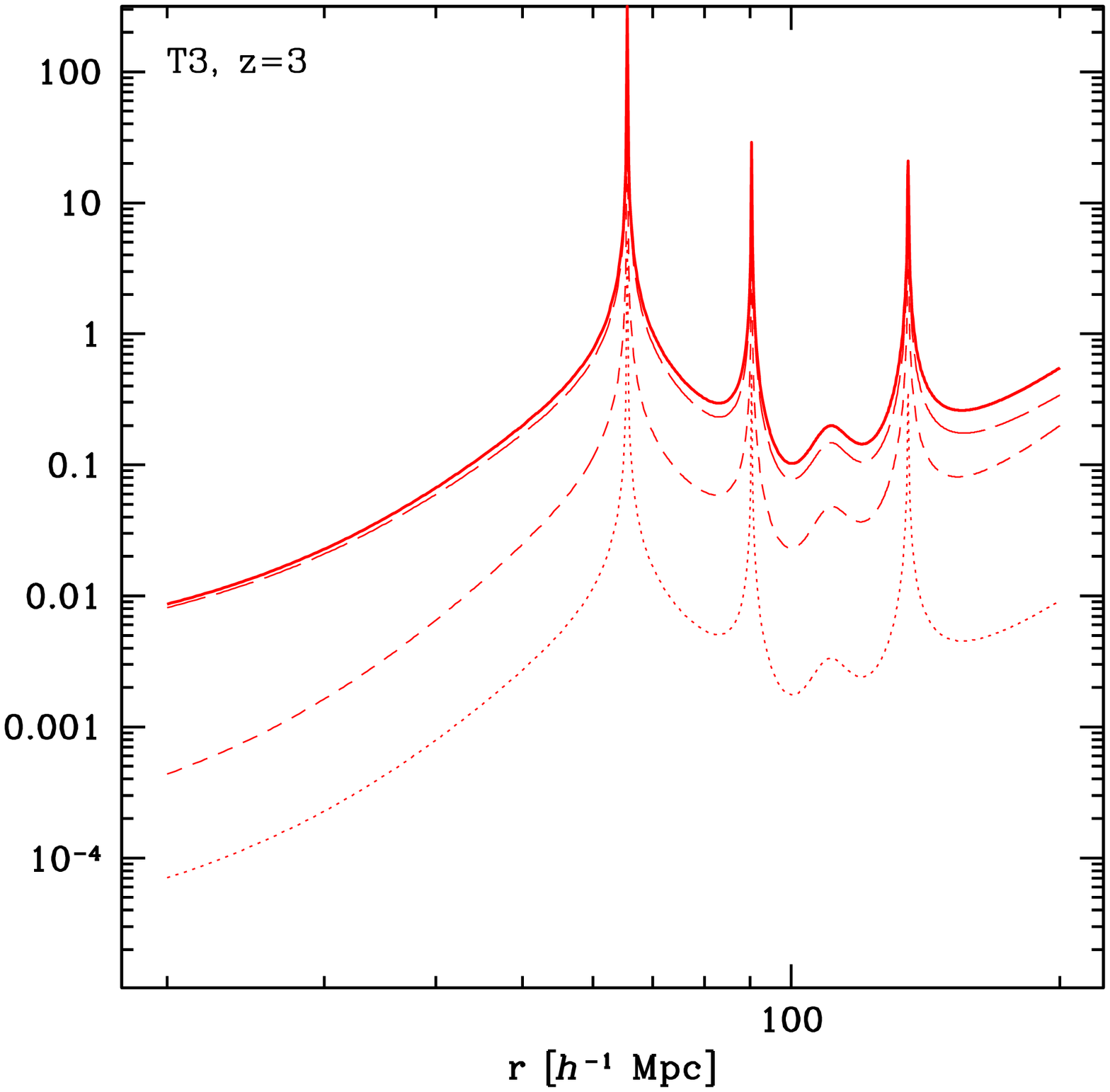}
\includegraphics[width=.32\textwidth]{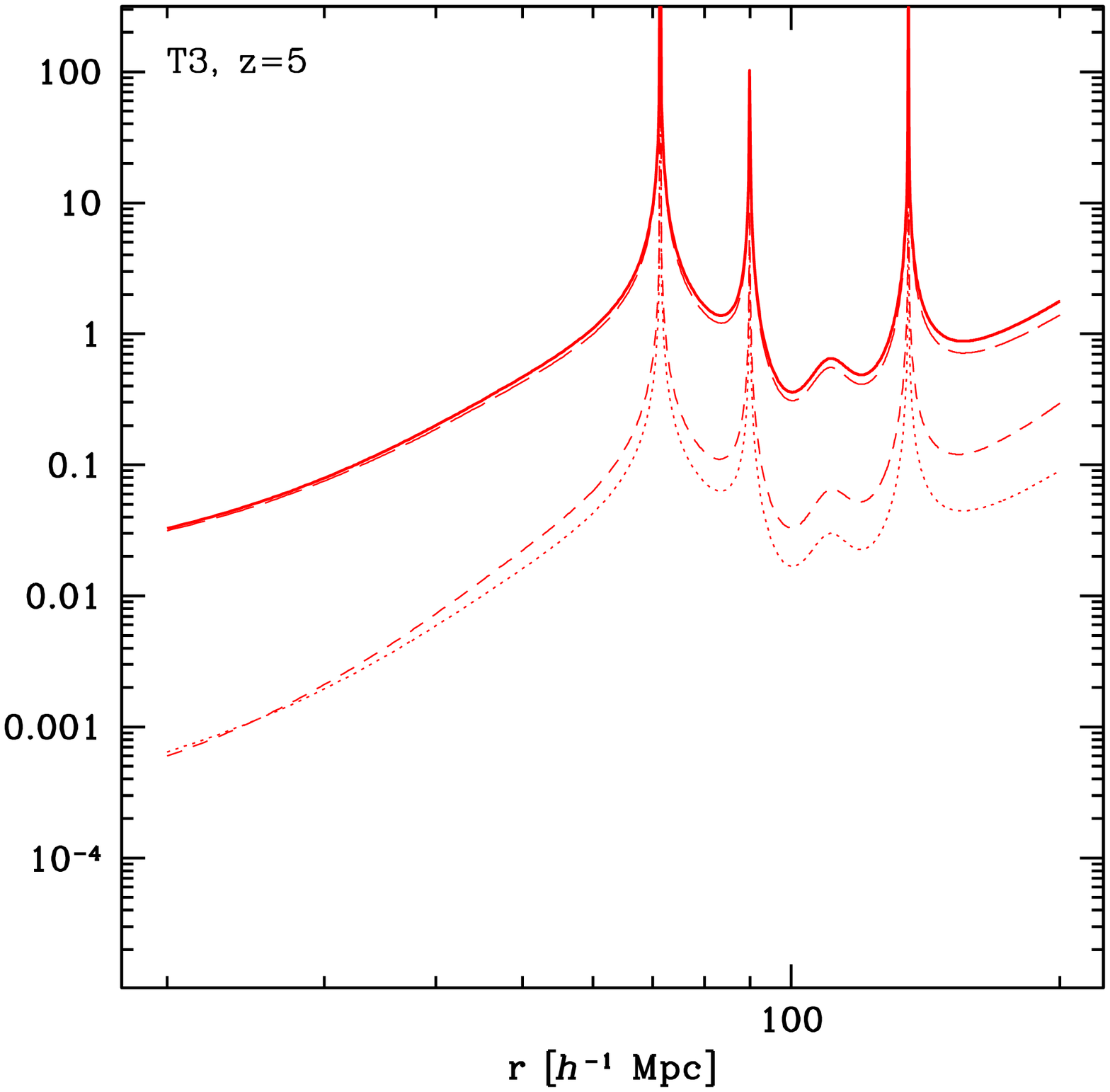}
\caption{\label{fig:t3} Triangular configuration T3, $z=1$, $3$, and $5$.}
\end{center}
\end{figure*}

For the first triangle configuration (``T1''), schematically represented in the 
left panel of
Fig.~\ref{fig:triangles_t1t2}, we keep $r_{13,\perp}=r_{23,\perp}=20\Mpc$, 
$r_{12,\perp}=40\Mpc$ and $\chi_2-\chi_1=20\Mpc$ and we evaluate the 3PCF and 
its magnification corrections as the line-of-sight separation $r\equiv 
\chi_3-\chi_1$ of the third galaxy is varied from $20$ to $200\Mpc$. 
 
In the second class of configurations (``T2'') we keep the same values for the 
transverse separations as for T1, but we consider the position $\xv_2$ to be
halfway between $\xv_1$ and $\xv_3$ (see Fig.~\ref{fig:triangles_t1t2}, right 
panel), with $r_{13,\parallel}$ again varying between $20$ and $200\Mpc$. The 
third one (``T3'') is analogous to T1 but with fixed $r_{23,\parallel}=20\Mpc$ 
and $r_{12,\parallel}$, $r_{13,\parallel}$ varying. All these classes represent a continuous set of 
triangles starting from an overall size of tens of megaparsec which become
progressively elongated and aligned with the line-of-sight as $\rpa$ increases. From 
equations~(\ref{eq:ggl}--\ref{eq:kg}) above, we expect an increasing contribution from 
magnification for such elongated configurations: while the intrinsic (unlensed) 
correlation decreases with the triangle size, the lensing terms will remain 
roughly constant, since they depend on the \textit{projected} triangle which we 
keep fixed.

In addition to these three peculiar classes of configurations, we evaluate 
the lensing contributions for a set of equilateral triangles of increasing
size, lying on a plane subtending a $45^{\circ}$ angle with the line-of-sight
(see Fig.~\ref{fig:triangles_t4}). One can take this to be an example of a 
triangle with generic orientation, where the parallel and perpendicular 
separations increase proportionally. The relative lensing contributions are 
in this case expected to be nearly independent of the side $r$ of the triangle.

Assuming a flat $\Lambda$CDM cosmology with parameters: $\Omega_m=0.27$,
$\Omega_b=0.046$, $h=0.72$, $n_s=1$, $\sigma_8=0.9$, we evaluate the galaxy 3PCF 
at three indicative redshift values: $z=1$, $3$ and $5$ (for definiteness, we 
refer to them as ``low'', ``intermediate'' and ``high'' redshift in what 
follows). Specifically, this is assumed to be 
the redshift of the galaxy at location $\xv_1$, while without loss of generality we always assume that the
distances of the three points from the observer obey the condition 
$\chi_1\le\chi_2\le\chi_3$. Small corrections due to the different values of
the growth factor at the three locations are easily included since they 
correspond to multiplicative factors at tree-level in PT. 
We compute the correlation function $\xi$ and the auxiliary functions $\eta$ and
$\epsilon$ from the power spectrum as described in the Appendix, assuming the 
matter transfer function provided by the CAMB code 
\cite{LewisChallinorLasenby2000}. 
Therefore, acoustic features in the power spectrum are included. However, we
do not include any smoothing effects due to non-linear gravitational clustering.
These are beyond the scope of this paper, and should in any case be quite 
small at the redshifts considered.

The last ingredient required to evaluate a realistic assessment of the  
lensing contribution to the measured galaxy 3PCF is a reasonable estimate of the bias
parameters $b_1$, $b_2$ and the number count slope $s$. We adopt the expressions 
for the bias parameters derived in the 
framework of the Halo Model and of the Halo Occupation Distribution (HOD), 
\cite{MoJingWhite1997,ScoccimarroEtal2001A}. We will refer in particular to 
the values computed in \cite{SefusattiKomatsu2007} for several forthcoming 
high-redshift spectroscopic surveys, where the HOD assumed is the one proposed
by \cite{TinkerEtal2005}. They are derived in terms of the survey mean 
redshift and expected galaxy density; we assume
$n_g\simeq 5\times 10^{-4}\icMpc$ for the $z=1$ and $z=3$ examples, and 
$n_g\simeq 50\times 10^{-4}\icMpc$ for the $z=5$ case. We will also consider
values for the survey volume of $V=4\cGpc$, $V=2.7\cGpc$, and $V=3.4\cGpc$ 
respectively for low, intermediate and high redshift cases, which will be used
later for an estimate of the expected error on the 3PCF measurement. These 
numbers roughly correspond to the characteristics of the proposed 
surveys WFMOS ($0.5<z<1.3$), HETDEX ($2<z<4$), and CIP ($3.5<z<6.5$). Under 
these assumptions we obtain bias parameter values of $b_1\simeq 2.0$ and 
$b_2\simeq 1.0$ at $z=1$, $b_1\simeq 3.6$ and $b_2\simeq 6.2$ at $z=3$ and 
$b_1\simeq 3.7$ and $b_2\simeq 6.8$ at $z=5$. 

%

The value of the number count slope $s$ strongly depends on the particular 
sample, i.e. type of object considered as well as the magnitude cut applied.
For galaxies, typical values are in the range $0.2 \lesssim s\lesssim 0.6$ for faint magnitude
cuts, and $0.4 \lesssim s \lesssim 1.5$ for bright magnitudes 
\cite{LoVerdeHuiGaztanaga2007,HuiGaztanagaLoVerde2007}. For QSO, number count 
slopes range from $s=-0.2$ to $s=0.8$ \cite{Gaztanaga2003,ScrantonEtal2005}.
In contrast to the bias parameters, the number count slope can be inferred from 
the observed distribution of galaxy apparent magnitudes. Furthermore, by 
changing the magnitude cut, it is possible to vary $s$ within a limited range
\cite{ScrantonEtal2005}. 
As a fiducial value, we assume $s=0.6$ here, so that $c_1=1$ and $c_2=2$. 

In Fig.~\ref{fig:t1} (upper panels) we plot the absolute value of the 
different contributions to the 3PCF for $z=1$, $3$ and $5$ (left to right
column) for the T1 class of triangular configurations. The 
long-dashed (black) curve represents the unlensed 3PCF, including the non-linear
bias contribution, as a function of $r$, i.e. the line-of-sight separation 
between the points $\xv_1$ and $\xv_3$ (see Fig.~\ref{fig:triangles_t1t2}, left 
panel), while the short-dashed (red) and the continuous (blue) curves represent 
the magnification contribution (all terms) and the total, 
{\it lensed} 3PCF, respectively. Note that, for this specific configuration, the 3PCF 
changes sign at about $r\sim 30\Mpc$ and several times in the range of scales 
corresponding to the acoustic features. This is due to the fact that the 
auxiliary functions $\eta$ and $\epsilon$ in the tree-level PT expression for the 
3PCF are related to derivatives of the matter correlation function which 
exhibits a 
local minimum and maximum in that range. In the lower panels we plot the ratio 
of the magnification contributions to the un-lensed 3PCF including the bias. 
We also distinguish the contributions of the different groups identified in the previous 
section, that is the GGL terms (short-dashed curve), the GLL terms (long-dashed),
and the LLL terms (dotted), while the solid line represents the sum of all 
of them.

For most of the configurations and redshifts considered, the magnification 
corrections to the 3PCF 
are above the percent level, especially at line-of-sight separations greater
than $\sim 50\Mpc$. 
Magnification corrections exceed $10\%$ for separations above $40~h^{-1}$ Mpc 
at 
redshift $z=3$, and above $30~h^{-1}$ Mpc at redshift $z=5$. 
As mentioned before, these trends are expected: while the 
magnification terms stay roughly constant, the unlensed 3PCF is oscillating and 
decreasing in amplitude. This fact implies that the magnification 
corrections become more important for increasing separations, and that in 
the regions where the unlensed 3PCF changes sign the observed 3PCF will be 
completely dominated by lensing effects. Furthermore, it is apparent that the 
magnification effect increases with redshift. This fact has a 
straightforward physical explanation: the further away the sources are located, the 
more cosmological structure responsible for lensing is present between the 
source and the observer, which affects several of the correction terms.
Finally, it is also clear from Fig.~\ref{fig:t1} that while the 
terms belonging to the GLL group, especially GLL-C, are dominant, the terms 
belonging to the LLL group have the strongest redshift evolution, and will
eventually overtake the other terms at sufficiently high redshift.

\begin{figure*}[t]
\begin{center}
\includegraphics[width=.32\textwidth]{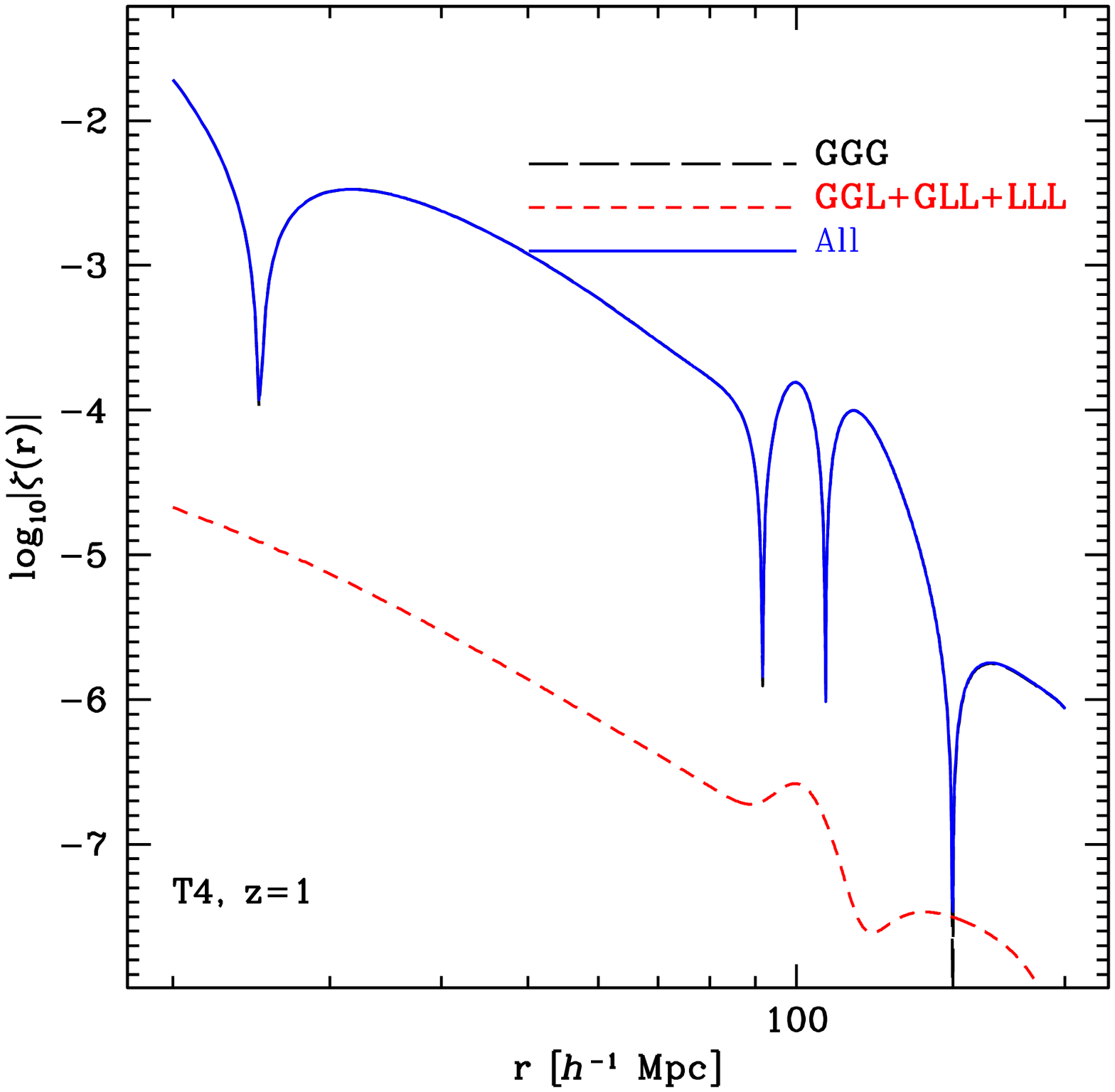}
\includegraphics[width=.32\textwidth]{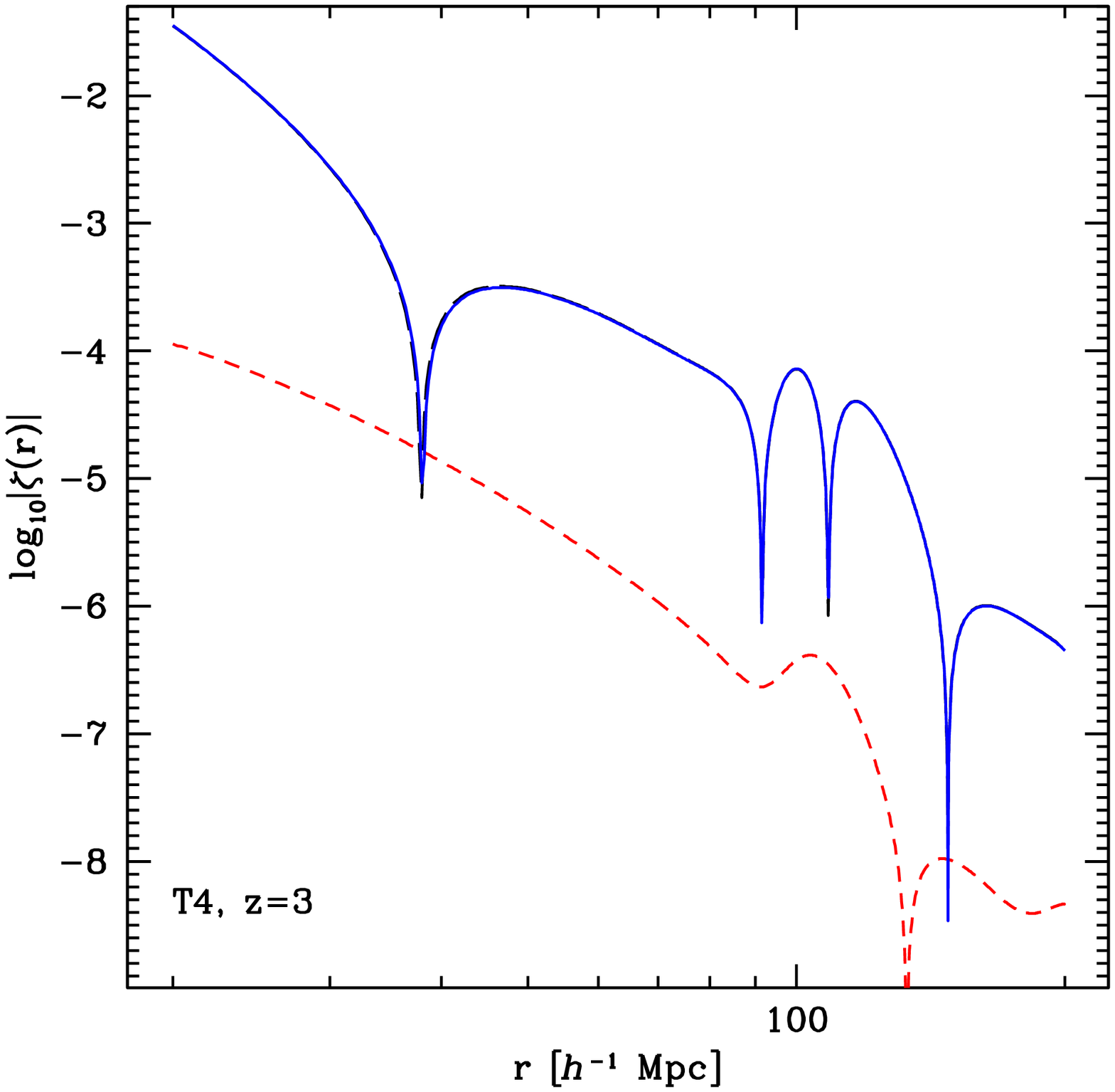}
\includegraphics[width=.32\textwidth]{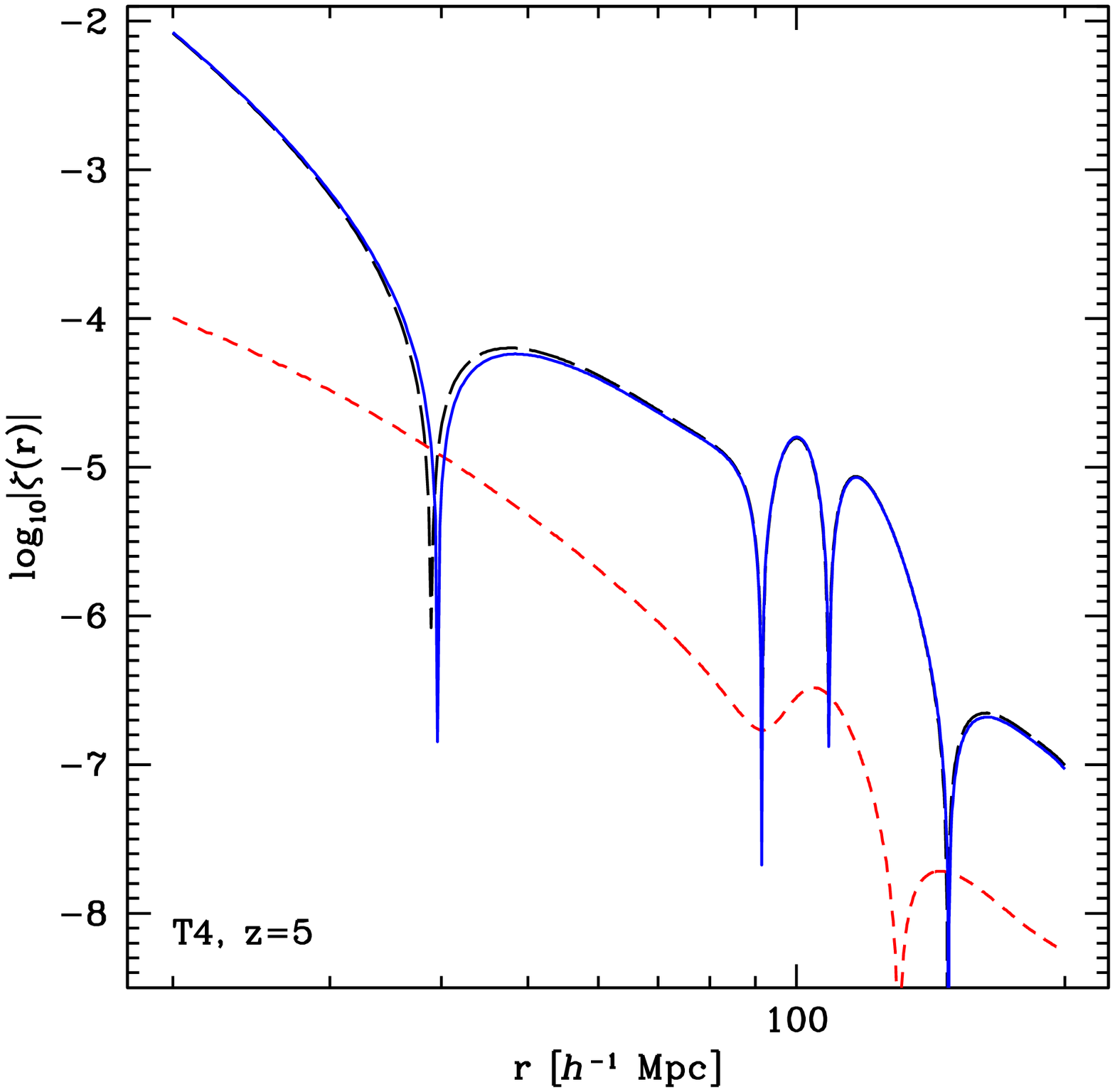}
\includegraphics[width=.32\textwidth]{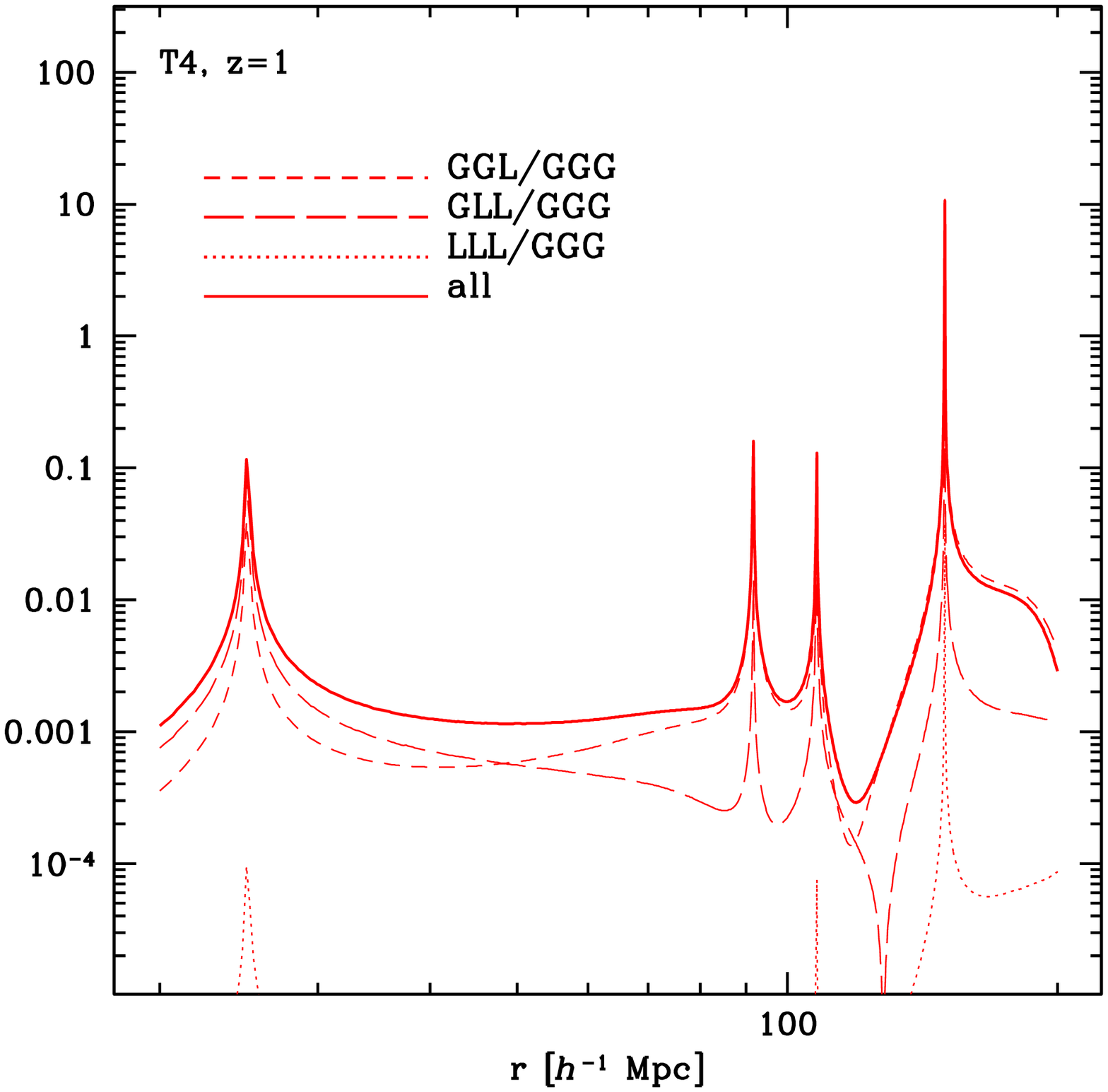}
\includegraphics[width=.32\textwidth]{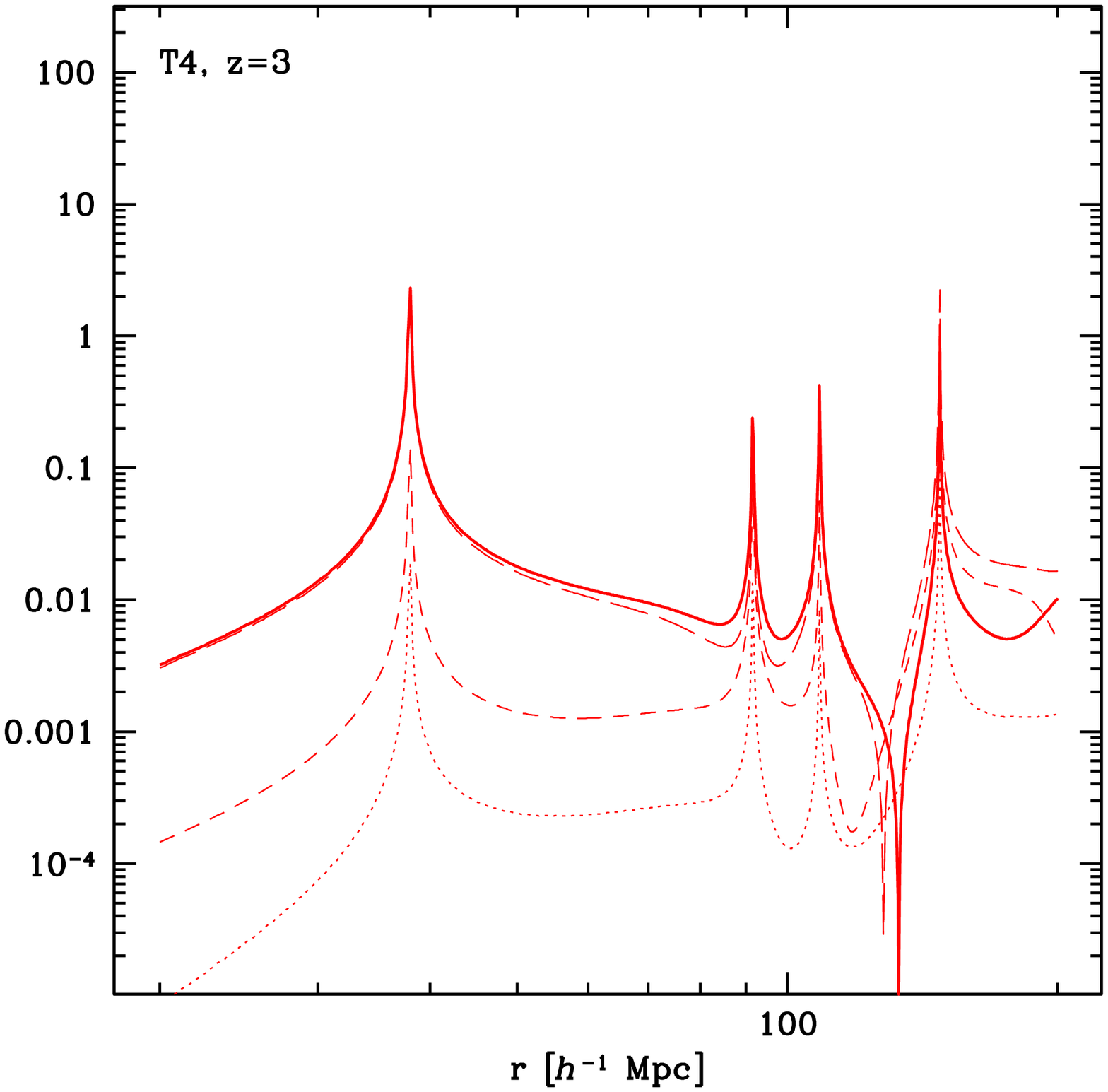}
\includegraphics[width=.32\textwidth]{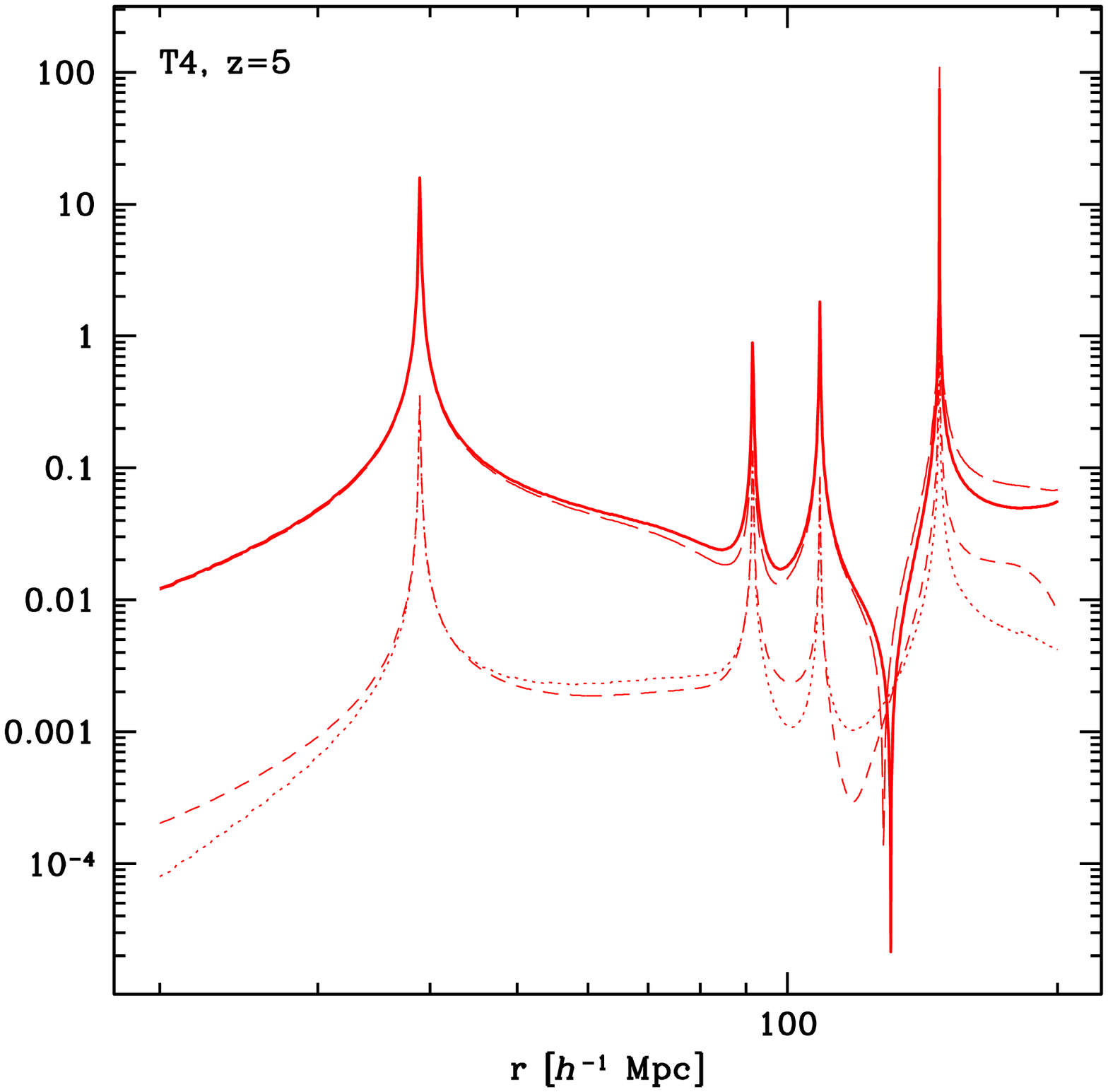}
\caption{\label{fig:t4} Triangular configuration T4, $z=1$, $3$ and $5$.}
\end{center}
\end{figure*}

In Fig.~\ref{fig:t2} and \ref{fig:t3} the same plots are presented for the 
triangular configurations T2 and T3, respectively. Most of the comments made 
about T1 are also valid for the triangular configurations T2 and T3. 
At intermediate and high redshifts, the magnification corrections exceed 
$10\%$ for separations of $40~h^{-1}$ Mpc and eventually dominate the 
observed 3PCF. For T2 and T3, the unlensed 3PCF is again oscillating and 
decreasing in amplitude while the magnification corrections stay roughly 
constant.

In Fig.~\ref{fig:t4} we finally present the results for the equilateral 
configuration T4. In this case, the magnification corrections behave somewhat
differently as function of the separation. This is 
because in this particular case the transverse separation between the sources is 
not held fixed. Hence, the magnification correction is decreasing with 
increasing separation roughly in the same way as the overall 3PCF. In this case, 
then, the lensing contributions to the 3PCF become overwhelming only when the 
unlensed 3PCF is changing sign and therefore going through zero. 
As before, the lensing contributions increase with redshift: at $\rpa\gtrsim 40\Mpc$,
they amount to less than $1\%$ for $z=1$, a few percent at $z=3$ and about 10\% 
at $z=5$.

\begin{figure}[t]
\begin{center}
\includegraphics[width=.48\textwidth]{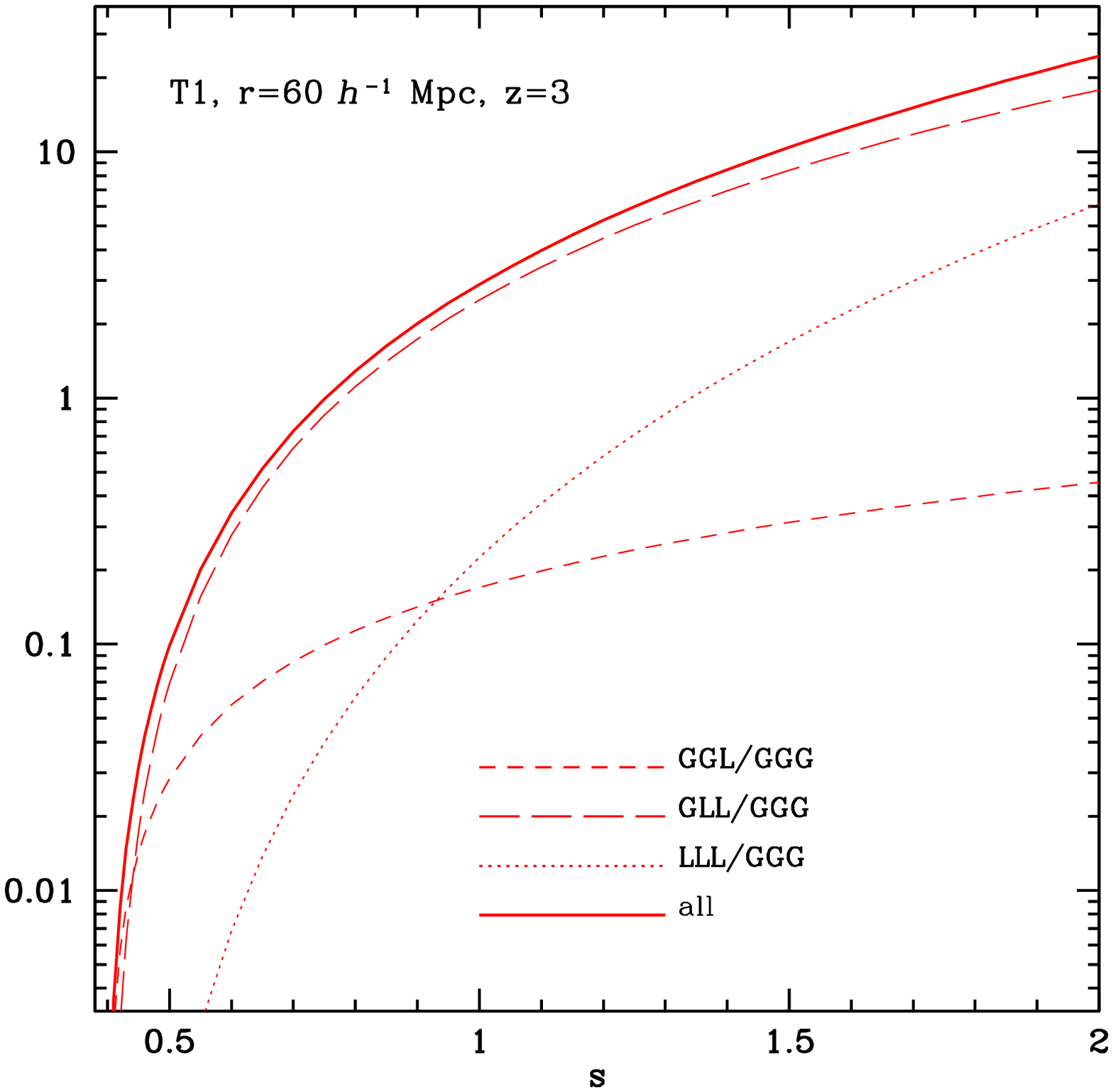}
\caption{\label{fig:thpt-vs-s} Ratio (absolute value, as in the lower panels
of Figs.~\ref{fig:t1}--\ref{fig:t4}) of the lensing 
contributions GGL, GLL, and LLL to the unlensed GGG terms, as a function of the 
number count slope $s$. The triangle, T1 with $r=60\:\Mpc$ at $z=3$, is kept 
fixed.}
\end{center}
\end{figure}
Going back to equations~(\ref{eq:GGL-A})--(\ref{eq:LLL-D}), it is clear that the 
magnification terms in different groups depend differently on the value of $s$ 
(via $c_1$ and $c_2$):
the GGL terms scale as $c_{1}$ and $c_2$, the GLL terms as $c_1^2$ and $c_1c_2$ 
while the LLL terms as $c_1^3$ and $c_1^2c_2$. Hence, depending on the value of 
$s$ of the sample, different groups of terms will dominate. This is shown in 
Fig.~\ref{fig:thpt-vs-s}, where we show the three sets of contributions as a 
function of $s$, for a fixed triangle at $z=3$ (the T1 example with 
$r=60\Mpc$). Note that the GGL and LLL terms are odd with respect to $(s-0.4)$, 
i.e., they change sign for $s < 0.4$, while the GLL terms are even. The dominant
individual contribution in this case turns out to be, as expected, the 
$\<\d_1\d_2\k_2\k_3\>$ (GLL-C) term, for all values of $s$. Note that this
term increases rapidly with $s$ around the value 
$s=0.6$ chosen for the plots of Fig.~\ref{fig:t1}--\ref{fig:t4}. Thus, for a sample with high number count slope, the
prospects of detection are significantly more optimistic than shown in 
Fig.~\ref{fig:t1}--\ref{fig:t4}. In the event of a detection, the scaling 
with $s$ should be 
observed by measuring the 3PCF with different apparent magnitude cuts in the 
survey, corresponding to different values of $s$. In principle, this would
allow to disentangle contributions with different scaling behavior, e.g.
GLL and LLL terms.

\begin{figure}[ht]
\begin{center}
\includegraphics[width=.48\textwidth]{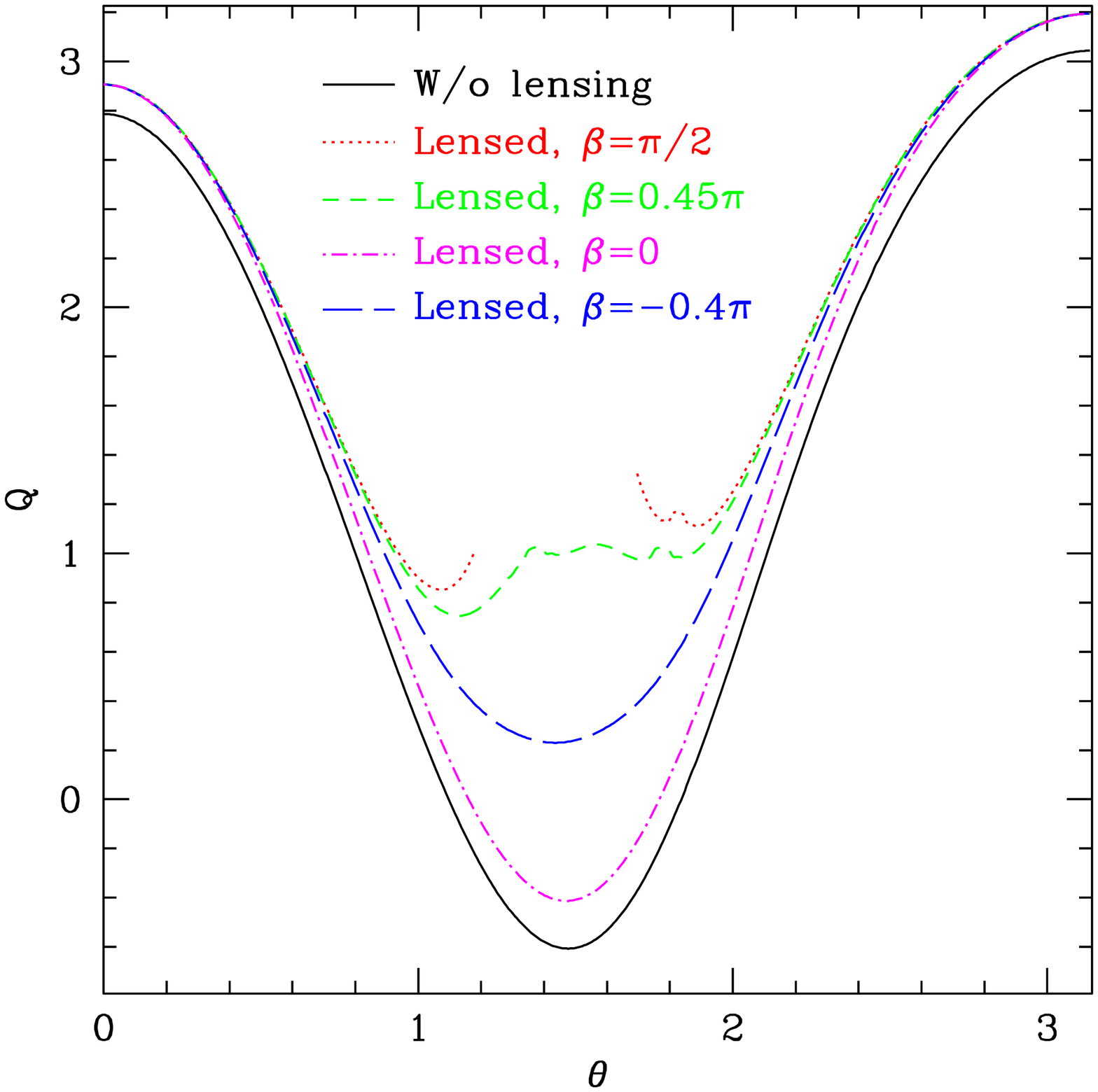}
\caption{\label{fig:Q} The reduced three-point function $Q$ as a function of
$\theta = \sphericalangle (r_{12},r_{13})$ for a triangle
with $r_{12}=20\:\Mpc$, $r_{13}=80\:\Mpc$ at $z=3$. The solid (black) curve 
shows the intrinsic $Q$, while the other curves show the total $Q$ including
magnification effects for different orientations of the triangle
(see text). We assumed $s=1.2$ for this figure.}
\end{center}
\end{figure}

\subsection{Reduced three-point function}
\label{sec:Q}

The reduced three-point function is commonly used to measure the non-Gaussianity
of the galaxy distribution, as it contains all the information of the
full 3PCF $\zeta$, but is, to first order, independent of the growth and 
amplitude of matter fluctuations. It is given by
\beq
Q(\xv_1, \xv_2, \xv_3) = \frac{\zeta(\xv_1,\xv_2,\xv_3)}
{\xi_{12}\xi_{13} + \xi_{12}\xi_{23} + \xi_{13}\xi_{23}}.
\eeq
The reduced bispectrum is defined in an analogous way. In order to exemplify
the effects of weak lensing on the reduced 3PCF, we show the intrinsic (unlensed)
as well as lensed $Q$ as a function of $\theta=\sphericalangle(r_{12},r_{13})$, 
the angle at vertex 1, for a triangle at $z=3$ (Fig.~\ref{fig:Q}). 
The lensed $Q$ is shown for different orientations of the triangle, specified
by the angle $\beta$ between the normal to the plane of 
the triangle and the line of sight. $\beta$ is chosen so that if $\beta > 0$, 
vertex 3 is further away
than 1 and 2, while it is closer for $\beta < 0$. We choose $\v{r}_{12}$ to be 
perpendicular to the line of sight with $r_{12}=20\:\Mpc$, and $r_{13}=80\:\Mpc$.
For this figure, we chose a value of $s=1.2$ for the number count slope, 
corresponding to $(5s-2)/b \approx 1.3$. This
value, while high, is not unexpected for bright magnitude cuts (e.g., Fig.~1
in \cite{HuiGaztanagaLoVerde2007}). 

Following the discussion earlier, we expect the largest lensing contributions
when $\beta=\pm \pi/2$, i.e. when the triangle is oriented along the plane
containing the line of sight, since in that case the projected separations are
the smallest. In calculating the lensing corrections,
we excluded configurations where the perpendicular separation of any two points
became less than $10\:\Mpc$, since in those cases the perturbation theory 
approach is expected to break down. This results in the gap in the curve
for $\beta=\pi/2$. In that orientation $r_{13\perp}\rightarrow 0$
for $\theta=\pi/2$. As the correlation function rises for $r\rightarrow 0$, we
expect the lensing contribution to rise sharply in this range.

Apparently, the lensing effect on $Q$ is appreciable in this configuration.
Even for $\beta=0$, i.e. a triangle lying in the sky plane, there is still
a finite lensing contribution from the LLL terms. For inclined orientations,
GGL and GLL terms contribute significantly.
There is an enhancement of $Q$ around $\theta\sim\pi/2$ where vertex~3 passes
behind ($\beta > 0$) or in front of ($\beta < 0$) vertices~1 and 2.
For $\beta$ approaching $+\pi/2$, there is noticeable structure in the
lensing contribution. This is due to the sum over GGL and GLL terms 
peaking at different values of $\theta$.

Note that the observed $Q$ will be given by the 3PCF including lensing
effects divided by the product of the \textit{observed} two-point correlation
functions which will also include contributions from magnification bias.
Since we focus on the lensing effects on the 3PCF, we did not include
them here.

\subsection{Observability of the magnification contribution}
\label{sec:obs}

In order to assess the observability of the lensing effects discussed here, 
we have to estimate the expected statistical uncertainty on the observed three-point
function. In Appendix~\ref{app:error}, we derive a simple and very approximate
expression for this uncertainty, assuming correlations are small (as is the
case on large scales). To measure the three-point function for a fixed triangle,
one defines bins in configuration space, so that for a given galaxy at vertex 1, 
triangles are counted for galaxies in the volume $dV_2$ around vertex 2 and
$dV_3$ around vertex 3. For a survey with $N_g$ galaxies with a galaxy volume 
density of $n_g$, we then obtain:
\beq
\sigma(\zeta) = N_g^{-1/2}\sqrt{\frac{1}{n_g^2 dV_2 dV_3} + 36}.
\label{eq:sigmazeta}
\eeq
For the rough estimates desired here, let us assume a typical bin size of
$dV_2 \sim dV_3 \sim 10^4\:\cMpc$, corresponding to a linear dimension of
$\sim 20\:\Mpc$. Using the values for the galaxy density and volume 
introduced above for the three ideal surveys at redshifts $z=1$, $z=3$ and 
$z=5$ we obtain uncertainties on the specific configurations given, 
respectively, by $\sigma(\zeta)\sim 4\cdot10^{-5}$, 
$\sigma(\zeta)\sim 5\cdot10^{-5}$ and $\sigma(\zeta)\sim 1.4\cdot10^{-5}$.
Note that these uncertainties correspond to triangular configuration bins with
a given orientation, so that they can be directly related to the absolute values
for the 3PCF plotted in Fig.~\ref{fig:t1}, \ref{fig:t2}, \ref{fig:t3} and 
\ref{fig:t4}. This suggests that magnification effects should be potentially 
detectable in the galaxy 3PCF at redshifts below $z \approx 3$, and clearly 
measurable at $z \gtrsim 4$. 

We expect, on the other hand, that when the correlation function is measured
by averaging over all possible orientations for a given triangular shape, the 
magnification effect will be significantly reduced. However, in this case the 
expected error will be reduced as well due to the higher statistics. A proper 
assessment of the magnitude of the correction given a specific 3PCF estimator 
is however beyond the scope of this work.

Finally we should mention that another possibility to detect the lensing effects 
in much smaller surveys is given by cross-correlating galaxies or quasars 
at different redshifts, or even across different surveys. In this case, there is
no intrinsic correlation due to the large line-of-sight separation, while the 
magnification terms are still significant at small transverse separations. 
Such a three-point magnification has indeed already been detected in 
the SDSS \cite{Gaztanaga2003} in terms of the galaxy-quasar skewness. However, 
no direct comparison between our results and those measurements can be made as 
we limit ourselves to the correlation function on much larger scales, where
perturbation theory is applicable.

\section{Conclusions}\label{sec:4}

In this work we analyzed how weak gravitational lensing affects the observed
three-point correlation function of galaxies and quasars. As a congruence of 
photons travels from
the source to the observer, the intervening distribution of matter acts 
by focusing or defocusing it, thus altering the magnitude of the object observed.
As a consequence of this effect, any magnitude limited sample -- and any 
correlation function measured from it -- will be somewhat contaminated by
magnification bias. 

In a consistent second-order treatment in perturbation theory, weak gravitational
lensing contributes 14 different terms. 
The lensing contribution to the 3PCF strongly depends on the configuration of
the triangle considered, and in particular on its projection on the sky plane. 
If the projection is held fixed while the 
triangle is stretched, magnification bias can account for an appreciable fraction
of the observed 3PCF at intermediate to high redshifts. On the other 
hand, for triangles oriented so that the projection on the sky plane is of 
similar size as the triangle itself, the magnification corrections to the 3PCF 
are smaller, reaching percent level at intermediate redshifts and 
of order 10\% at high redshifts. The lensing contributions are a rapidly varying
function of the number count slope $s$ which depends on the sample considered.
While we adopted an average value of $s=0.6$ here, the lensing effects increase
significantly for larger values of $s$ (e.g., Fig.~\ref{fig:thpt-vs-s} and \ref{fig:Q}).
A rough estimate of the signal-to-noise suggests 
that the effect of magnification bias on the galaxy and quasar 3PCF 
may be detectable by itself at redshift $z\approx3$ and above.

The results presented here suggest that such magnification corrections need to be taken into 
account when measurements of non-Gaussianity are carried out using magnitude 
limited samples from galaxy and quasar catalogs at high redshifts. If neglected, they could in 
fact bias the determination of the bias parameters $b_1$ and $b_2$, and
affect the sensitivity of galaxy surveys to primordial non-Gaussianity.
However, one can also argue that the distinctive dependence of the magnification 
effects on $s$ as well as $b_1$, $b_2$ may be used in the future to 
unambiguously identify the 
lensing contribution, and to disentangle the different terms. Ultimately, it 
might be possible to use the lensing of the 3PCF as a tool
to measure the intervening distribution of matter and the linear growth factor. 
A more detailed analysis of such aspects will be the focus of forthcoming work.

Since this work represents just a first step in the investigation and the possible exploitation of the lensing of the 3PCF, here we only considered a few specific triangular configurations. 
However, one would expect that the lensing effects on the 
\textit{projected} 3PCF, i.e.
the three-point correlation of galaxies in a given redshift bin for a 
certain projected triangle, will be much larger. This is because many triangles
with large line-of-sight separations within the redshift slice will contribute.
On the other hand, by counting galaxies in a wide redshift bin, a spectroscopic
redshift for each galaxy is no longer necessary, and much larger galaxy 
statistics become available.
A generalization of this is to cross-correlate galaxies from different redshift
slices, where the intrinsic correlation is negligible. The lensing effects
on the projected 3PCF will be studied in a future paper.

Also, it would be desirable to extend the modeling of lensing effects to smaller
separations, were they are expected and observed to grow rapidly \cite{Gaztanaga2003}. In principle, one can
use non-linear models for the matter power spectrum in the expressions
derived here [equations~(\ref{eq:ggl})--(\ref{eq:kg})] to extend the
range of validity. However, this is not
completely consistent, since the separation into terms such as $\<\d\k\k\>$,
$\<\d\d\k\k\>$ is in itself perturbative to leading order in the matter overdensity. This 
caveat is equally valid for the
calculation of magnification effects on the two-point correlation 
function \cite{Matsubara2000,VallinottoEtal2007,HuiGaztanagaLoVerde2007}.
A full non-linear calculation of lensing effects on galaxy correlation functions
would be considerably more involved and has not been attempted yet.

Finally, the treatment presented here for the magnification bias of the galaxy and 
quasar 3PCF is a generalization of the one given in \cite{Moessner1997, VallinottoEtal2007,HuiGaztanagaLoVerde2007} for the 2PCF. We point out that the perturbation
framework used in Sec.~\ref{sec:2b} to derive the magnification corrections is 
completely general: given enough patience and time, it is
straightforward to extend the treatment given here to correlation functions of 
higher order. One might expect that higher order 
correlation functions will be more and more affected by magnification bias.

\acknowledgments

We thank Felipe Mar\'in for useful discussions.
E.S. acknowledges the hospitality of the Aspen Center for Physics where part 
of this work has been completed. This research was supported by the DOE. A.V. thanks Gary Mamon and Jean-Philippe Uzan for useful comments and conversations and thanks the Particle Astrophysics Center of the Fermi National Accelerator Laboratory, where part of this work was carried out, for the hospitality. A.V. is supported by the Agence Nationale pour la Recherche. F.S. thanks the Institute of Physics at Humboldt University, Berlin, for hospitality, and Marek~Kowalski, Thomas~Lohse, and Ulli~Schwanke for inspiring discussions. F.S. was supported by the Kavli Institute for Cosmological Physics at the University of Chicago through grants NSF PHY-0114422 and NSF PHY-0551142 and an endowment from the Kavli Foundation and its founder Fred Kavli.

\bibliography{Bibliography}

\begin{widetext}
\appendix

\section{Matter and Galaxy 3PCF}
\label{app:3PCFaux}

\subsection{Conventions}\label{app:conventions}

In all derivations, we use the following convention for the Fourier transform: 
\beq
\d_\kv=\int \frac{d^3 x}{(2\pi)^3}e^{i\kv\cdot\xv}\d(\xv),\qquad
\d(\xv)=\int d^3 k e^{-i\kv\cdot\xv}\d_\kv.
\eeq
The power spectrum is defined as 
\beq
\<\d_{\kv_1}\d_{\kv_2}\>\equiv \d_D(\kv_1+\kv_2)P(k_1),
\eeq
while the bispectrum is defined as 
\beq
\<\d_{\kv_1}\d_{\kv_2}\d_{\kv_3}\>\equiv \d_D(\kv_1+\kv_2+\kv_3)B(k_1,k_2,k_3).
\eeq
The dependence of $P(k_1)$ and of $B(k_1,k_2,k_3)$ on the redshifts/comoving distances is implicitly understood and omitted throughout the text. Since we are consistently using the linear power spectrum, such dependence can simply be accounted for by multiplying by the appropriate growth factors. 

To obtain expressions in the other commonly used Fourier convention, where
$\d(\xv) = \int d^3k/(2\pi)^3 \d_{\kv}\exp(i\kv\cdot\xv)$,
just replace 
$P(k)$ with $P(k)/(2\pi)^3$. The same is true for the correlation functions 
defined below. Further, we again use the notation $\xv_{ij} = \xv_i-\xv_j$, 
$x_{ij}=|\xv_i-\xv_j|$.

\subsection{Computation of the matter 3PCF -- Auxiliary Functions}

The matter two-point correlation function is obtained from the matter power 
spectrum as 
\beq
\xi(x)=4\pi\int dk\: k^2 P(k)\frac{\sin(kx)}{kx}.
\eeq
Assuming the tree-level expression for the bispectrum given in 
equation~(\ref{Btreelevel}), Jing and B\"orner \cite{JingBoerner1997} showed that one 
can write 
\bea
\zeta(\xv_1,\xv_2,\xv_3) & = & \frac{10}{7}\xi(x_{12})\xi(x_{13})
-[\eta_2(x_{12})\eta_0(x_{13})+\eta_0(x_{12})\eta_2(x_{13})]
~\xv_{12}\cdot\xv_{13}
\nonumber\\
& & +\frac{4}{7}\left[\epsilon(x_{12})\epsilon(x_{13})(\xv_{12}\cdot\xv_{13})^2
+\eta_2(x_{13})\epsilon(x_{12})x_{12}^2+\eta_2(x_{12})\epsilon(x_{13})x_{13}^2
+3\eta_2(x_{12})\eta_2(x_{13})\right]
\nonumber\\
& &  + {\rm 2~perm.},
\eea
where 
\bea
\eta_l(x)&\equiv& 4\pi\int dk\: k^2 
\frac{P(k)}{k^l}\frac{kx\cos(kx)-\sin(kx)}{k\: x^3},\\
\epsilon(x) & \equiv & 4\pi\int dk P(k)
\frac{3[\sin(kx)-kx\cos(kx)]-k^2x^2\sin(kx)}{k\: x^5}.
\eea
This expression is quite convenient since it allows us to determine the
3PCF from the functions $\xi$, $\eta_l$ and $\epsilon$ which have to be computed 
only once. These functions are plotted in Fig.~\ref{fig:aux}.

\begin{figure*}[t]
\begin{center}
\includegraphics[width=.48\textwidth]{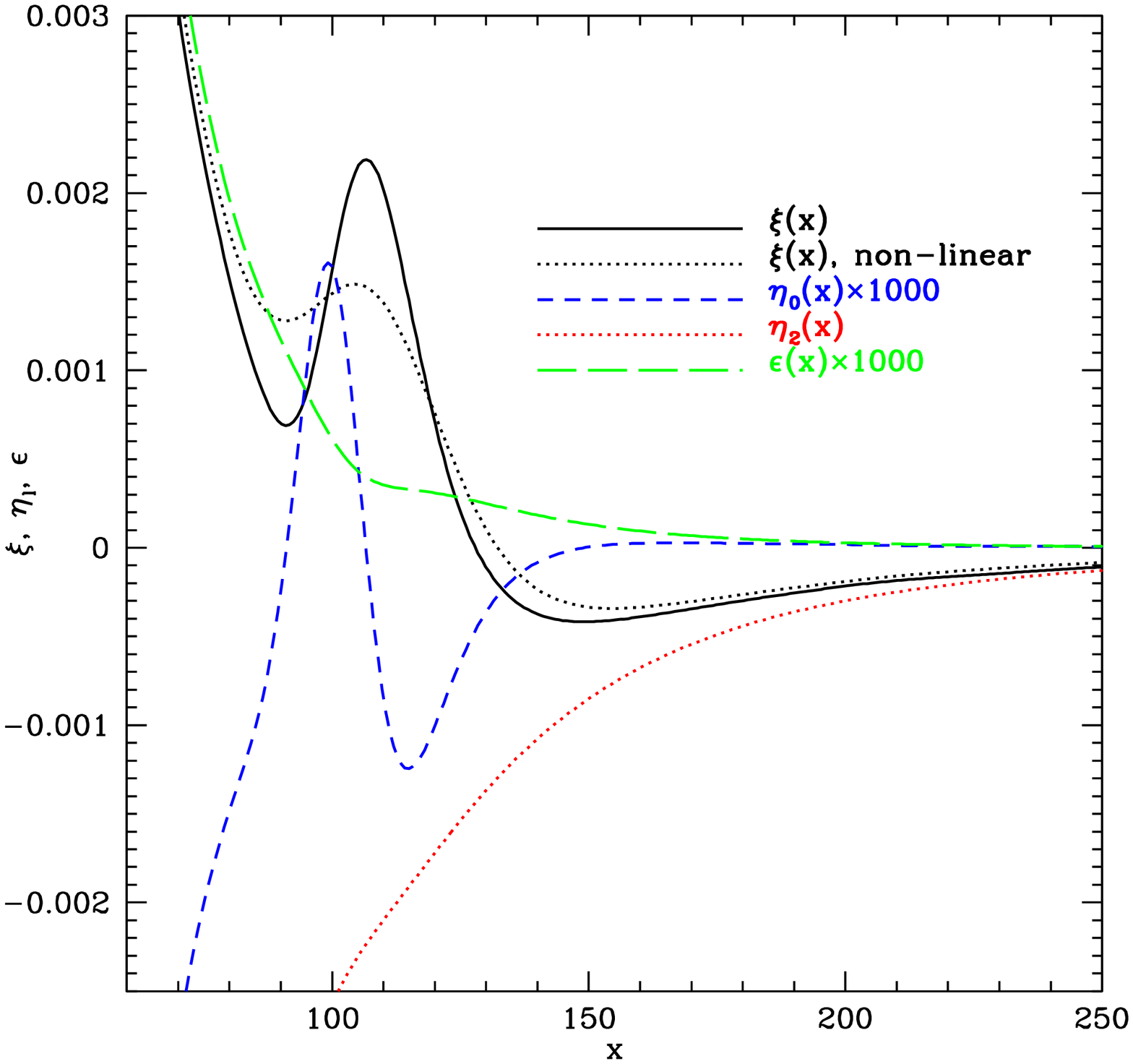}
\includegraphics[width=.48\textwidth]{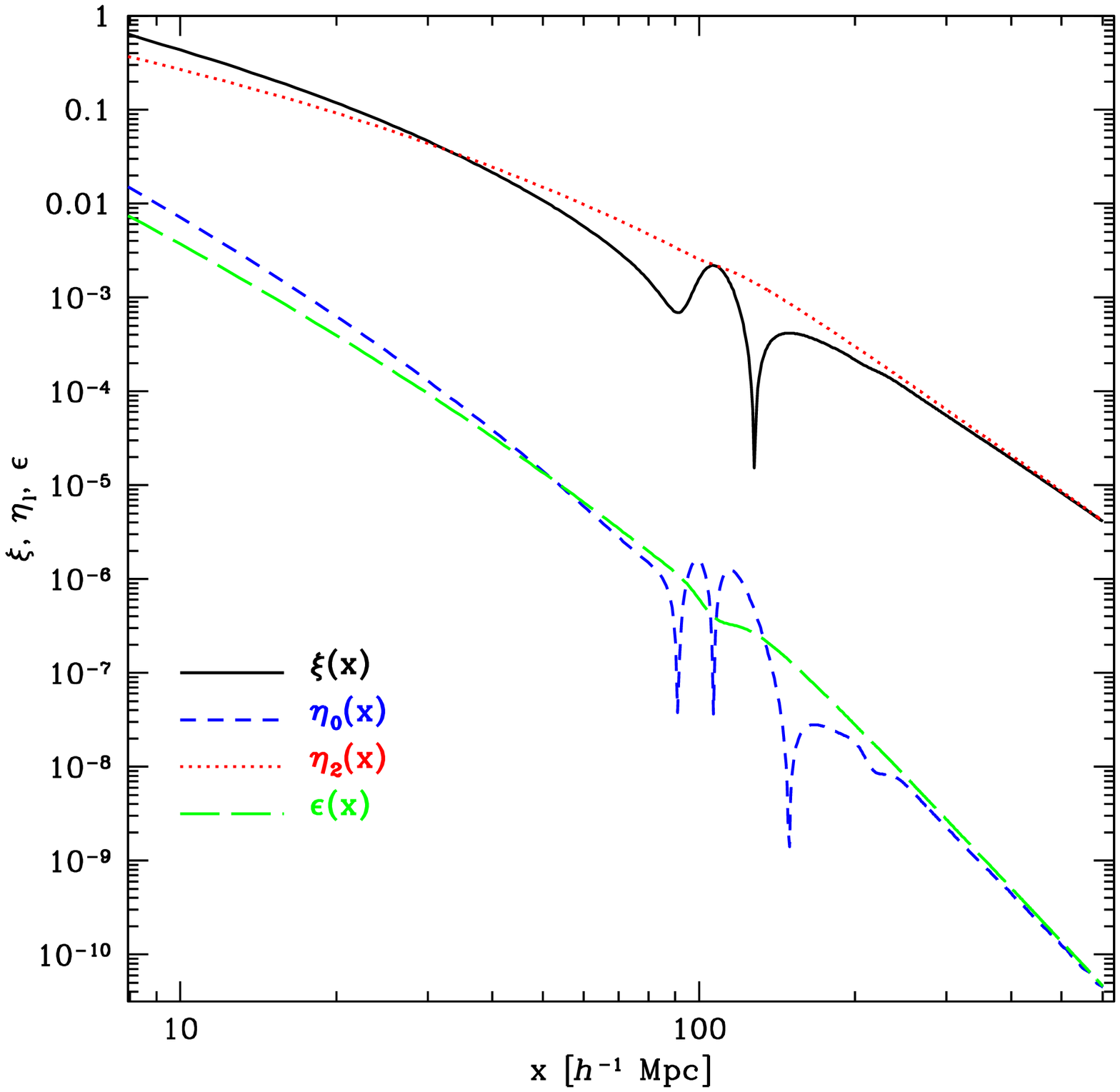}
\caption{\label{fig:aux} Linear correlation function $\xi(x)$ and auxiliary 
functions $\eta_0(x)$, $\eta_2(x)$ and $\epsilon(x)$ as a function of $x$, 
evaluated at redshift $z=0$. In the left panel, the non-linear correlation 
function is also shown. The right panel shows the absolute value of the
functions.}
\end{center}
\end{figure*}

\subsection{Transverse 2PCF, 3PCF and auxiliary functions}

When applying the Limber approximation, one can obtain very similar expressions
for the magnification contributions in terms of the \textit{projected} two-
and three-point correlation functions, $\xit$ and $\zetat$, respecively.
The projected two-point function $\xit$ is defined as the 2-dimensional
Fourier transform of the power spectrum in terms of the transverse wavenumber
$\kvp$ (equivalent to forcing $k_z=0$):
\bea
\widetilde{\xi}(\xpe)&\equiv& (2\pi) \int d^2\kpe P(\kpe) e^{-i\kvp\cdot\xvp} 
\label{eq:xitdef} \\
&=& (2\pi)^2\int dk\: k\: P(k) J_0(k \xpe),
\eea
where $J_0(x)$ is the 0th order Bessel function.
The transverse three-point function $\zetat$ is defined analogously as
\beq\label{eq:zeta_2d}
\zetat(\xv_{1\perp},\xv_{2\perp},\xv_{3\perp})
\equiv 
(2\pi)^2
\int\!\!d^2k_{1\perp}e^{-i\kv_{1\perp}\cdot\xv_{1\perp}}
\int\!\!d^2k_{2\perp}e^{-i\kv_{2\perp}\cdot\xv_{2\perp}}
B(\kv_{1\perp},\kv_{2\perp}).
\eeq
Note that $\xit$ has dimensions of length, and $\zetat$ of length squared.
In order-of-magnitude terms, $\xit(\xpe) \sim \xpe \xi(\xpe)$, 
$\zetat(\xpe,\xpe)\sim \xpe^2 \zeta(\xpe,\xpe)$.
In analogy with what is done
above for the 3-dimensional 3PCF, one can write
\bea
\widetilde{\zeta}(\xv_{1\perp},\xv_{2\perp},\xv_{3\perp})
& = & \frac{10}{7}\widetilde{\xi}(x_{12\perp})\widetilde{\xi}(x_{13\perp})
-\left[\widetilde{\eta}_2(x_{12\perp})\widetilde{\eta}_0(x_{13\perp})
+\widetilde{\eta}_0(x_{12\perp})\widetilde{\eta}_2(x_{13\perp})\right]~
\xv_{12\perp}\cdot\xv_{13\perp}
\nonumber\\
& &
+\frac{4}{7}\left[\widetilde{\epsilon}(x_{12\perp})
\widetilde{\epsilon}(x_{13\perp})
(\xv_{12\perp}\cdot\xv_{13\perp})^2
+\widetilde{\eta}_2(x_{13\perp})\widetilde{\epsilon}(x_{12\perp})x_{12\perp}^2
\right.
\nonumber\\
& & 
\left.
+\widetilde{\eta}_2(x_{12\perp})\widetilde{\epsilon}(x_{13\perp})x_{13\perp}^2
+2\widetilde{\eta}_2(x_{12\perp})\widetilde{\eta}_2(x_{13\perp})\right]
+ {\rm 2~perm.},
\eea
where
\bea
\widetilde{\eta}_l(x)&\equiv&\frac{(2\pi)^2}{x}\int dk\: k^2
\frac{P(k)}{k^l}J_1(kx),\\
\widetilde{\epsilon}(x)&\equiv&\frac{(2\pi)^2}{x^3}\int dk\: P(k)\:
[k xJ_0(kx)-2J_1(kx)].\label{eq:eps_2d}
\eea
Here, $J_l(x)$ denote Bessel functions of integer order. These functions are 
plotted in Fig.~\ref{fig:aux_2d}. Since all auxiliary functions are proportional
to the linear matter power spectrum, they scale with the square of the matter 
growth factor. For brevity, we suppress the redshift- or $\chi$-dependence of 
these functions here and in the following.

\begin{figure*}[t]
\begin{center}
\includegraphics[width=.48\textwidth]{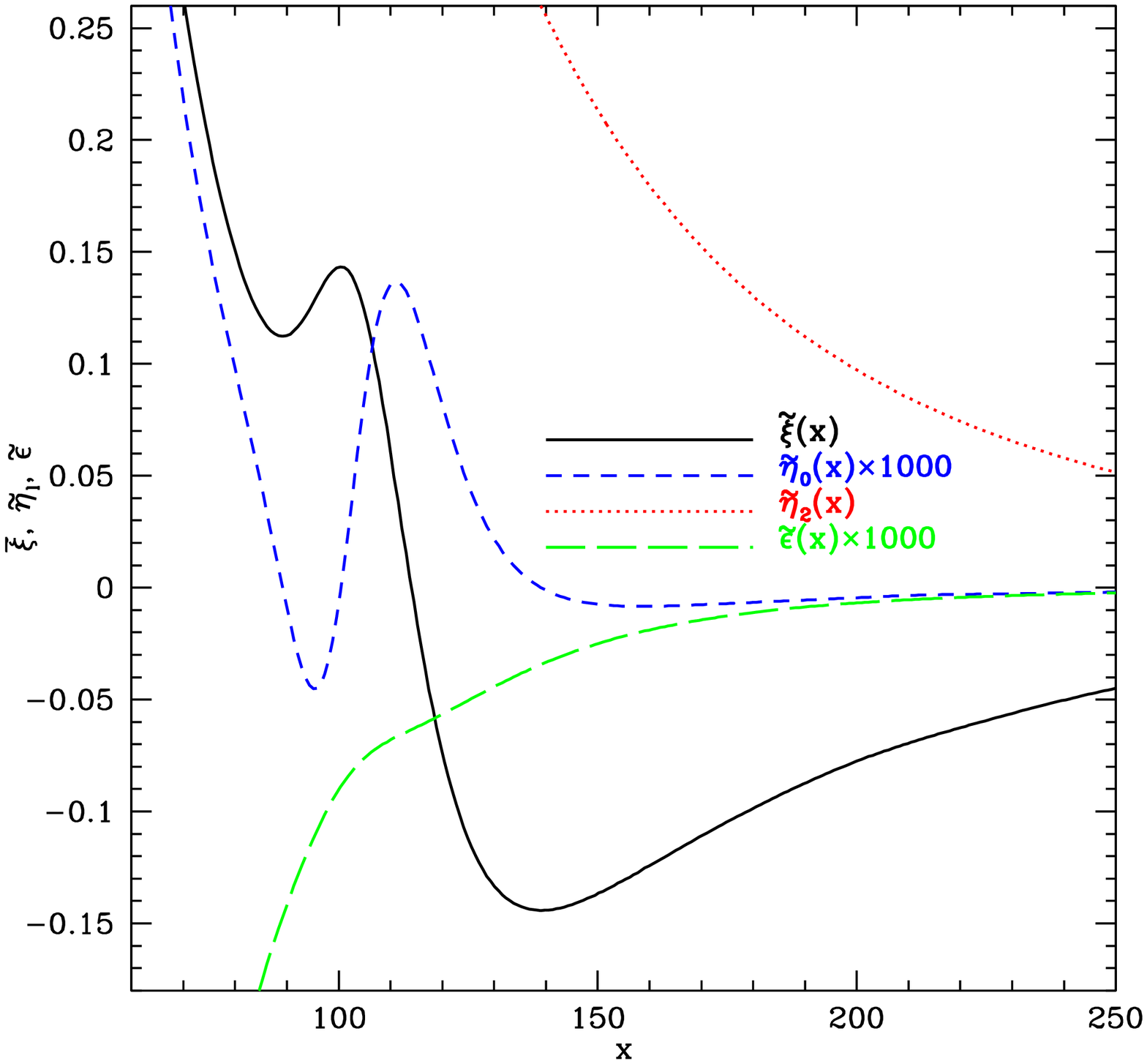}
\includegraphics[width=.48\textwidth]{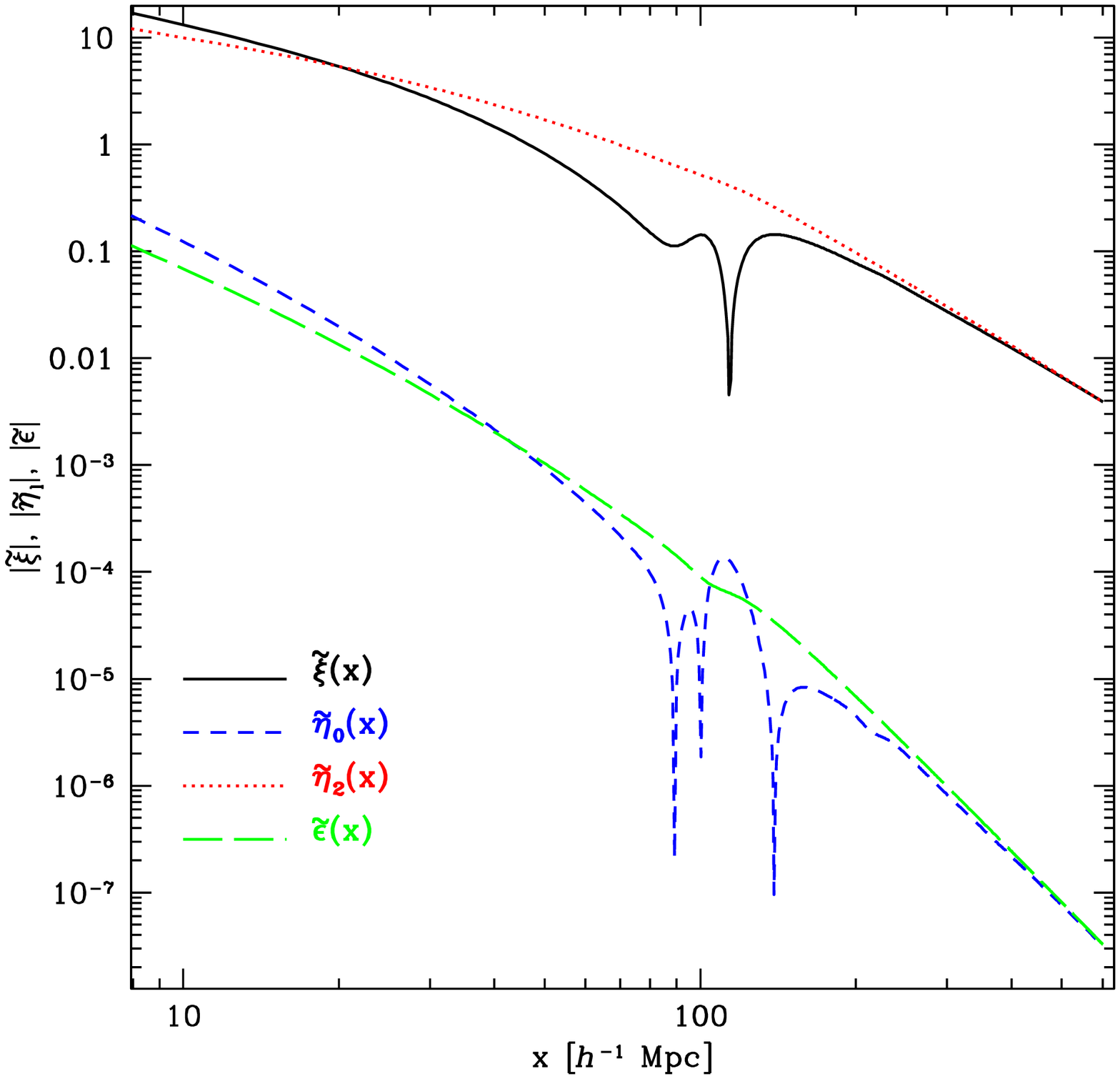}
\caption{\label{fig:aux_2d} ``Transverse'' linear correlation function 
$\widetilde{\xi}(x)$ and auxiliary functions $\widetilde{\eta}_0(x)$, 
$\widetilde{\eta}_2(x)$ and $\widetilde{\epsilon}(x)$ as a function of $x$.
The right panel shows the absolute value of the functions.}
\end{center}
\end{figure*}

\section{Derivation of convergence and shear correlators}
\label{app:shearconv}

This appendix contains the derivation of convergence $\kappa$ and shear $\gamma$
as well as representative derivations of some of the results shown in 
equations~(\ref{eq:ggl}--\ref{eq:kg}).

\subsection{Convergence and shear}

The projected lensing potential $\hat{\Psi}$ is defined as
\beq
\hat{\Psi}(\xv) = -\int_0^{\chi_s} d\chi W_L(\chi_s,\chi) \left [ 
\Phi(\yv(\chi); \chi) - \Psi(\yv(\chi); \chi) \right ],
\eeq
where $\Phi$, $\Psi$ are the gravitational potentials and $\Psi=-\Phi$ at late 
times in $\Lambda$CDM and most dark energy models. $\chi_s$ is the distance to the source, and $\yv(\chi)=(\chi/\chi_s) \xv$ 
parametrizes the line of sight. The convergence $\k(\xv)$ and the complex shear 
$\g(\xv)$ are then given by
\bea
2\k &=& \hat{\Psi}_{,11} + \hat{\Psi}_{,22}, \\
2\g &=& \hat{\Psi}_{,11} - \hat{\Psi}_{,22} + 2 i \hat{\Psi}_{,12}.
\eea
We now use the Poisson equation,
\beq
k^2 \Phi = -\nabla^2 \Phi = \frac{3}{2}\Omega_m H_0^2\frac{1}{a} \d 
\equiv\frac{C}{a}\:\d,
\eeq
to write $\k$ and $\g$ in terms of the matter overdensity $\d$. For the 
convergence, this yields:
\bea
\k(\xv) &=& C \int_0^{\chi_s}d\chi \frac{W_L(\chi_s,\chi)}{a(\chi)}
\d(\yv(\chi); \chi) \\
 &=& C \int_0^{\chi_s}d\chi \frac{W_L(\chi_s,\chi)}{a(\chi)}
\int d^3 k\: e^{-i \kv\cdot\yv(\chi)} \d_\kv(\chi).
\eea
For the shear, we obtain:
\beq
\g(\xv) = C
\int_0^{\chi_s}d\chi\frac{W_L(\chi_s,\chi)}{a(\chi)}
\int d^3 k\frac{k_1^2-k_2^2+2ik_1k_2}{k^2}\d_\kv(\chi) e^{-i\kv\cdot\yv(\chi)}.
\eeq

\subsection{Derivation of correlators}

As shown in section~\ref{sec:2}, the different correlators involving $\k$ and 
$\g$ simplify considerably when using the Limber approximation. We now show this 
in more detail in a few sample derivations. First, consider 
$\<\d(\xv_1)\k(\xv_2)\>$, where we assume as usual $\chi_1 < \chi_2$.
\bea
\<\d(\xv_1)\k(\xv_2)\> &=& C \int_0^{\chi_2} d\chi \frac{W_L(\chi_2,\chi)}{a(\chi)}
\xi(|\xv_1 - \yv_2|) \nonumber\\
 &=& C \int_0^{\chi_2}d\chi \frac{W_L(\chi_2,\chi)}{a(\chi)}
 \int d^3 k\: e^{-i \kv\cdot(\xv_1-\yv_2)} P(k).
\eea
The first expression can be used to evaluate this correlator directly. 
However, in the $k$-space formulation, we can apply the Limber approximation:
the typical longitudinal separations $\xv_{1z}-\yv_{2z}$
are much larger than the perpendicular separation, hence the exponential
will oscillate rapidly unless $k_z \lesssim 1/(\xv_{1z}-\yv_{2z})$. Thus, we set
$k_z=0$, which yields a Dirac $\d$-function in $\chi-\chi_1$, fixing the
lens to be at the same distance as point 1:
\bea
\<\d(\xv_1)\k(\xv_2)\> 
&\stackrel{\mbox{\tiny L.A.}}{\simeq}& 
C \frac{W_L(\chi_2,\chi_1)}{a(\chi_1)} 
\int d^2 k_\perp\: e^{-i \kvp\cdot(\xv_{1\perp}-\xv_{2\perp})} P(k_\perp) \\
&=& C \frac{W_L(\chi_2,\chi_1)}{a(\chi_1)} \widetilde{\xi}(x_{12\perp}),
\eea
where we have used the definition of the projected two-point function, 
equation~(\ref{eq:xitdef}).

In a similar way, we obtain expressions for the correlators $\<\d\g\>$ (again 
assuming $\chi_1 < \chi_2$):
\beq
\<\d(\xv_1)\g(\xv_2)\>  =  C
\int_0^{\chi_2}d\chi\frac{W_L(\chi_2,\chi)}{a(\chi)}
\int d^3 k P(k)\frac{k_1^2-k_2^2+2ik_1k_2}{k^2}
e^{-i\kv\cdot(\yv_2-\xv_1)}. \label{eq:dg_exact}
\eeq
Using the Limber approximation and setting $k^2\rightarrow k_\perp^2$ in 
the denominator as well, we find
\beq
\<\d(\xv_1)\g(\xv_2)\> \stackrel{\mbox{\tiny L.A.}}{\simeq} 2\pi C
\frac{W_L(\chi_2,\chi_1)}{a(\chi_1)}
\int d^2 k_\perp P(k_\perp)
\frac{k_1^2-k_2^2+2ik_1k_2}{k^2_\perp}
 e^{-i\kvp\cdot(\xv_{2\perp}-\xv_{1\perp})}.
\eeq
The integral can be expressed as 
\bea
2\pi\int d^2 k_\perp P(k_\perp)
\frac{k_1^2-k_2^2+2ik_1k_2}{k^2_\perp}
e^{-i\kvp\cdot\xvp}
& = & \nonumber \\ 
2\pi\int d k_\perp k_\perp P(k_\perp)
\int d \phi [2\cos^2\phi-1]
e^{-ik_\perp x_\perp \cos\phi} 
& = & 
\nonumber\\
\frac{2\pi}{x_\perp}\int d k_\perp P(k_\perp)
[k_\perp x_\perp J_0(k_\perp x_\perp)-2J_1(k_\perp x_\perp)]
& = & 
 x_\perp^2 \widetilde{\epsilon}(x_\perp),
\eea
where the function $\widetilde{\epsilon}(x_\perp)$ is defined in 
equation~(\ref{eq:eps_2d}). Hence,
\beq
\<\d(\xv_1)\g(\xv_2)\> \stackrel{\mbox{\tiny L.A.}}{\simeq} C
\frac{W_L(\chi_2,\chi_1)}{a(\chi_1)}
x_{12,\perp}^2 \widetilde{\epsilon}(x_{12,\perp}).
\label{eq:dg_limber}
\eeq

Similarly, for the $\<\k\g\>$ term we get: 
\beq
\<\k(\xv_1)\g(\xv_2)\> = C^2\!\!
\int_0^{\chi_1}d\chi\frac{W_L(\chi_1,\chi)}{a(\chi)}
\int_0^{\chi_2}d\chi'\frac{W_L(\chi_2,\chi')}{a(\chi')}
\int d^3 k P(k)\frac{k_1^2-k_2^2+2ik_1k_2}{k^2}
e^{-i\kv\cdot(\yv_2-\yv_1)},
\label{eq:kg_exact}
\eeq
which, applying the Limber approximation, yields
\beq
\<\k(\xv_1)\g(\xv_2)\> \stackrel{\mbox{\tiny L.A.}}{\simeq}  C^2
\int_0^{\chi_1}d\chi
\frac{W_L(\chi_1,\chi)W_L(\chi_2,\chi)}{a^2(\chi)}
y_{12,\perp}^2 \widetilde{\epsilon}(y_{12,\perp}).
\eeq

In all cases, the auxiliary functions are to be evaluated at $\chi_1$, or
at $\chi$ if inside an integral over $\chi$.
Fig.~\ref{fig:dketc} shows the four two-point correlators as function of the
transverse separation $x_{12\perp}$, for $\xv_1$ at $z=3$, and
keeping $\Delta\chi_{12}=\chi_2-\chi_1$ fixed. Both the exact expressions and
the Limber approximation results are shown (see the next section for a discussion).
In the case of $\<\d\k\>$, $\<\d\g\>$, we also show the functions
for different values of the longitudinal separation $\Delta\chi_{12}=\chi_2-\chi_1$.
Clearly, these correlators increase strongly for increasing longitudinal
separation, as they are proportional to $W_L(\chi_2,\chi_1)\sim \Delta\chi_{12}$.
For $\<\k\k\>$ and $\<\k\g\>$, the dependence on the longitudinal separation
of 1 and 2 is very small (as long as $\Delta\chi_{12}\ll \chi_1,\:\chi_2$),
since it only enters the arguments of the lensing window functions in the
integral.

\begin{figure*}[t]
\begin{center}
\includegraphics[width=.6\textwidth]{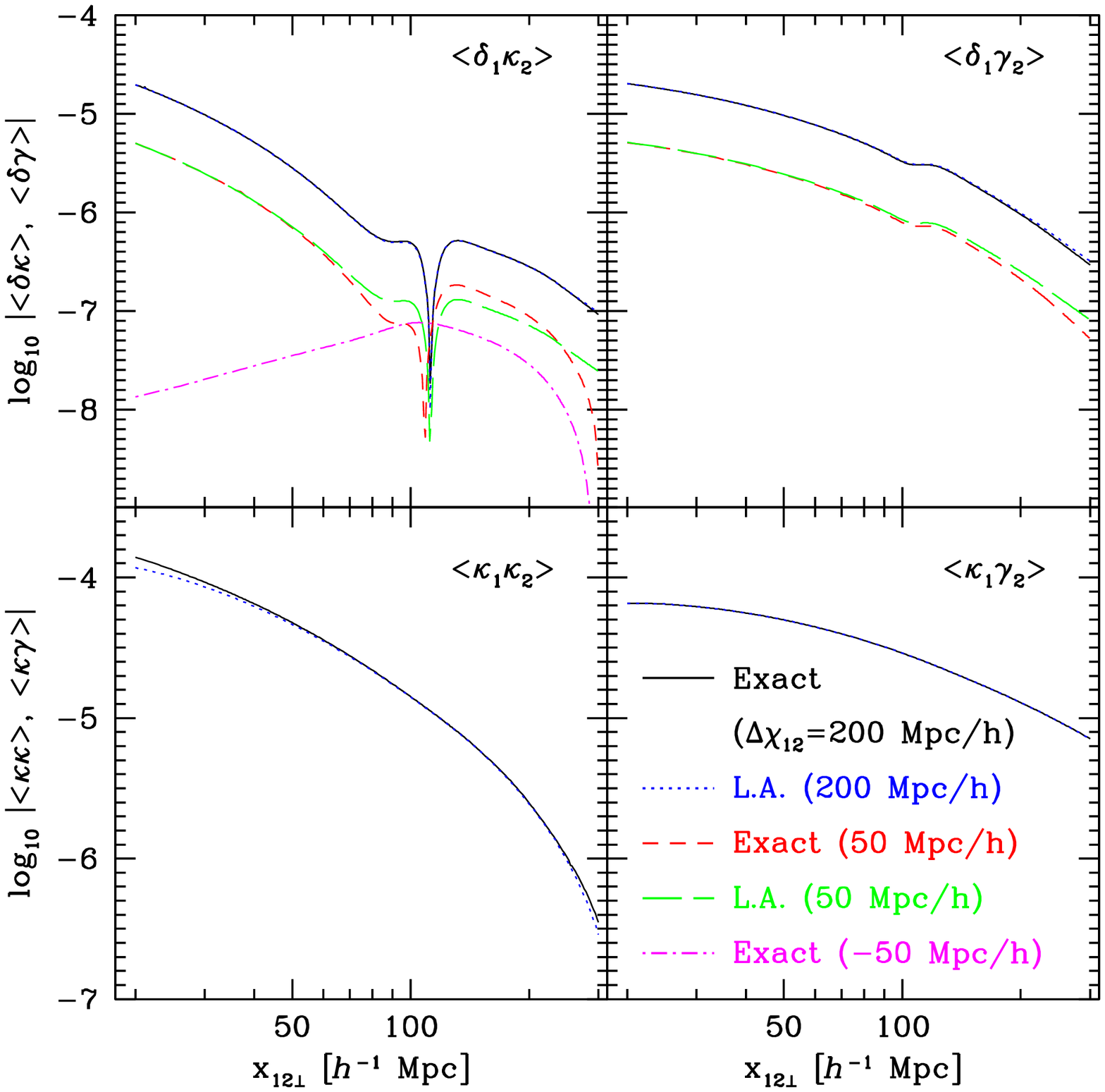}
\caption{\label{fig:dketc} $\<\d(\xv_1)\k(\xv_2)\>$,
$\<\d(\xv_1)\g(\xv_2)\>$, $\<\k(\xv_1)\k(\xv_2)\>$ and $\<\k(\xv_1)\g(\xv_2)\>$
as function of the perpendicular separation $x_{12\perp}$. Here, $\xv_1$ is
kept fixed at $z=3$, while $\xv_2$ is separated by 
$\Delta\chi_{12}=\chi_2-\chi_1=200,\:50,\:-50\:\Mpc$. In the last case, $\xv_2$
is \textit{closer} than $\xv_1$.
Shown are the results for the exact integral (in case of $\<\d\g\>$ and $\<\k\g\>$,
the first-order calculation) and when using the Limber approximation (L.A.).}
\end{center}
\end{figure*}

\section{Limber approximation}
\label{app:Limber}

\subsection{Discussion}

The Limber approximation is very useful to gain physical insight into the
magnification effects, and for order-of-magnitude estimates. However, before
using the simplified expressions for exact calculations, one should
assess the accuracy of this approximation. We calculated the exact
integrals for all of the terms in equations~(\ref{eq:ggl})--(\ref{eq:kg}), except
the $\<\k\k\k\>$, $\<\d\g\>$, and $\<\k\g\>$ terms. For the latter two terms
involving the shear, the exact calculation is very difficult, and we 
include the first order correction which is good to $O(k_z^4/\kpe^4)$ instead. 
The calculation of this correction
is outlined in the next section. Note that the first order correction
to the shear terms are quite small, so that there is no need to go to higher 
orders.

Fig.~\ref{fig:dketc} shows a comparison of the exact expressions and the
Limber approximation. For $\<\d\k\>$ and $\<\d\g\>$, the Limber approximation
is not accurate anymore at large transverse separations. Since we keep the
longitudinal separation $\Delta\chi_{12}$ fixed, large $x_{12\perp}$ 
correspond to more and more transverse $\xv_{12}$. Correspondingly, the
deviations become smaller for larger $\Delta\chi_{12}$, as is apparent from
Fig.~\ref{fig:dketc}. The reason for this is that the Limber approximation
assumes that separation vectors with $x_{\parallel} \lesssim x_\perp$ contribute
to the correlation. However, if $x_{12\perp} > x_{12\parallel}$, then this
overcounts the contributing modes and overestimates $\<\d\k\>$, $\<\d\g\>$.
As we lower $x_{12\parallel}=\Delta\chi_{12}$ to 0 and negative values,
the result of the Limber approximation goes linearly to 0, while there is actually still
a finite correlation at negative $\Delta\chi_{12}$, i.e. when we correlate
an overdensity at $\xv_1$ \textit{behind} the lens magnifying $\xv_2$. Assuming
the lens is a distance $\De\chi_L$ in front of $\xv_2$, an order of magnitude estimate 
gives $\<\d_1\k_2\> \sim (H_0 \De\chi_L)(H_0 x_{12\perp}) \xi(r_L)$,
where $r_L = \sqrt{x_{12\perp}^2 + (\De\chi_L + |\De\chi_{12}|)^2}$. For 
$x_{12\perp}= 100\:\Mpc$, $\De\chi_{12}=-50\:\Mpc$, and setting 
$\De\chi_L\sim 50\:\Mpc$, this yields $|\<\d_1\k_2\>| \sim 10^{-8}$.
This is in rough agreement with the exact calculation, shown by the dash-dotted
curve of $\<\d_1\k_2\>$ for $\Delta\chi_{12}=-50$ in Fig.~\ref{fig:dketc}.
Clearly, $\<\d_1\k_2\>$ becomes very small for negative $\Delta\chi_{12}$ and 
in fact decreases rapidly for increasingly negative values. Terms like this 
would only contribute for almost transverse separations, where the $\<\d\k\>$ term 
is in any case very small. This justifies the neglection of terms involving
$\<\d_2\k_1\>$ etc. in the derivation of equations~(\ref{eq:GGL-A})--(\ref{eq:LLL-D}).

For the $\<\k\k\>$ and $\<\k\g\>$ terms, the Limber approximation is 
generally very good (we do not show them for other values of $\Delta\chi_{12}$
as they are essentially the same). For $\<\k\k\>$, there are very small
deviations at small $x_{12\perp}$, corresponding to almost longitudinal
$\xv$. In this case, the Limber approximation underestimates the correlation,
counting only purely transverse modes ($y_{12\parallel} < y_{12\perp}$), while 
in the exact calculation non-transverse modes contribute as well.

To summarize, we found that the Limber approximation holds well for quantities 
that, after the approximation, still involve an integral over the line of sight,
like $\<\k\k\>$, $\<\k\g\>$, and $\<\k\k\k\>$. The 
integral serves to average out the effect of the approximation. For terms that 
only involve a single evaluation of the correlation functions, e.g., $\<\d\k\>$, 
$\<\d\g\>$, $\<\d\k\k\>$, the Limber approximation shows deviations, especially 
when the Limber-approximated expression goes to zero, while the actual 
correlation is finite. For these quantities, the exact integral should be used.

\subsection{Shear terms beyond the Limber approximation}

While the evaluation of the exact integrals is straightforward for the terms
involving the convergence $\k$, going beyond the Limber approximation requires
more work for the shear terms -- however, it provides some insight as well.
Going back to equation~(\ref{eq:dg_exact}) and using cylindrical coordinates 
$(\kpe,k_z,\phi)$, we can write $\<\d_1\g_2\>$ as:
\bea
\<\d(\xv_1)\g(\xv_2)\> & = & C \int_0^{\chi_2}d\chi\frac{W_L(\chi_2,\chi)}{a(\chi)} I(\rv = \yv_2-\xv_1,\chi) \label{eq:dg_exact2} \\
I(\rv,\chi) &=& \int \kpe d\kpe \int dk_z \: P(k=\sqrt{\kpe^2 + k_z^2};\chi) \int d\phi 
\frac{k_{\perp1}^2-k_{\perp2}^2+2ik_{\perp1}k_{\perp2}}{k^2} \exp \left ( -i(\kvp\cdot\rv_\perp + k_z\,r_\parallel) \right ) \\
&=& \int \kpe d\kpe \int dk_z \: \exp ( -i k_z\,r_\parallel )\: P(k;\chi) \frac{k_{\perp}^2}{k^2} 
\int d\phi \: (2\cos^2\phi - 1 + 2i \cos\phi\sin\phi) \exp(-i(\kvp\cdot\rv_\perp ))\quad\quad \\
&=& \int \kpe d\kpe \int dk_z \: \exp ( -i k_z\,r_\parallel )\: P(k;\chi) 
\frac{k_{\perp}^2}{k_{\perp}^2 + k_z^2} 
(2\pi)\left (J_0(\kpe\rpe) - \frac{2}{\kpe\rpe}J_1(\kpe\rpe) \right ),
\label{eq:Iexact}
\eea
where we have aligned $\kv_\perp$ so that $\phi$ is the angle between $\kv_\perp$ and 
$\rv_\perp$. The last integral is difficult to evaluate. In order to make progress,
we use the fact that typically $r_\parallel \gg \rpe$, so that $k_z$ should be much
less than $\kpe$ to get a significant contribution. Thus, we expand the integrand
in powers of $k_z^2/\kpe^2$. For the power spectrum, this expansion yields:
\beq
P(k) = P(\kpe) + \frac{1}{2\kpe}\frac{\partial P}{\partial k}|_{k_z=0} k_z^2
+ O\left ( \frac{k_z}{\kpe}\right )^4.
\eeq
Here and in the following, we have suppressed the $\chi$-evolution of the power
spectrum. Note that in this expansion there is no linear (and in general no odd) term in $k_z$. We now obtain
\bea
I(\rv) &\simeq& \int \kpe d\kpe \int dk_z \: \exp ( -i k_z\,r_\parallel )\: 
\left (P(\kpe) + \frac{1}{2\kpe}\frac{\partial P}{\partial k}|_{k_z=0} k_z^2 \right )
\left (1 - \frac{k_z^2}{\kpe^2} \right )\nonumber\\
& & \times (2\pi) \left ( J_0(\kpe\rpe) - \frac{2}{\kpe\rpe}J_1(\kpe\rpe) \right ).
\eea
We then obtain zeroth and first order expressions for the integral, the first order
expression being good to $(k_z/\kpe)^4$:
\bea
I^{(0)}(\rv) &=& \int \kpe d\kpe 
P(\kpe) (2\pi) \left ( J_0(\kpe\rpe) - \frac{2}{\kpe\rpe}J_1(\kpe\rpe) \right )
\:\int dk_z \: \exp ( -i k_z\,r_\parallel ) \\
I^{(1)}(\rv) &=& \int \kpe d\kpe \:
\left (- \frac{P(\kpe)}{\kpe^2} + \frac{1}{2\kpe}\frac{\partial P}{\partial k}|_{k_z=0}  \right ) (2\pi) \left ( J_0(\kpe\rpe) - \frac{2}{\kpe\rpe}J_1(\kpe\rpe) \right )\nonumber\\
& & \times \int dk_z \: \exp ( -i k_z\,r_\parallel )\: k_z^2.
\eea
The integrals over $k_z$ result in Dirac $\d_D$ functions and derivatives.
We see that the zeroth order term is exactly what we obtained earlier in 
the Limber approximation, equation~(\ref{eq:dg_limber}):
\beq
I^{(0)}(\rv) = \frac{(2\pi)^2}{\rpe} \int dk 
P(k) \left ( k\rpe\: J_0(k\rpe) - 2 J_1(k\rpe) \right ) = \rpe^2\:\widetilde{\epsilon}(\rpe)\:\d_D(r_\parallel),
\eeq
where $r_\parallel=\chi-\chi_1$ in equation~(\ref{eq:dg_exact2}), so that the $\d_D$
function fixes $\chi=\chi_1$. For the first correction to the Limber approximation,
we get
\beq
I^{(1)}(\rv) = \frac{(2\pi)^2}{\rpe} \int \frac{dk}{k^2} \: P(k)
\left (\frac{1}{2}\frac{\partial \ln\:P}{\partial \ln\:k} -1 \right )  
\left ( k\rpe\:J_0(k\rpe) - 2 J_1(k\rpe) \right ) \,\cdot\,
\frac{\partial^2}{\partial r_\parallel^2} \d_D(r_\parallel)
\equiv \widetilde{\nu}(\rpe)\,\cdot\,\frac{\partial^2}{\partial r_\parallel^2} \d_D(r_\parallel),
\eeq
defining another auxiliary function $\widetilde{\nu}(r)$. By partial integration
in equation~(\ref{eq:dg_exact2}) for $\<\d\g\>$ and equation~(\ref{eq:kg_exact})
for $\<\k\g\>$,
we can calculate the first correction to the Limber approximation for the
shear terms, $\<\d\g\>^{(1)}$, $\<\k\g\>^{(1)}$:
\bea
\<\d\g\>^{(1)} &=& -C \frac{\partial^2}{\partial \chi_1^2}\left (
\frac{W_L(\chi_2,\chi_1)}{a(\chi_1)} \right ) \: \widetilde{\nu}(x_{12\perp}; \chi_1) \\
\<\k\g\>^{(1)} &=& -C^2 \int_0^{\chi_1} d\chi \frac{\partial^2}{\partial \chi^2}\left (
\frac{W_L(\chi_2,\chi) W_L(\chi_1,\chi)}{a^2(\chi)} \right ) \: \widetilde{\nu}(y_{12\perp}; \chi), 
\eea
where $y_{12\perp} = (\chi/\chi_1) x_{12\perp}$ and we have made the dependence
on the evolution $\chi$ explicit again. The derivatives of the lensing weight
factors can be done easily, yielding polynomials in $\chi_1$, $\chi$ and factors
of $H(z)$ and $dH/dz$.

The effect of the first order correction on $\<\d_1\g_2\>$ and $\<\k_1\g_2\>$
is shown in Fig.~\ref{fig:dketc} (right two plots). While for $\<\d_1\g_2\>$
there is some effect at large $x_{12\perp}$ (similarly to $\<\d_1\k_2\>$),
the correction to $\<\k_1\g_2\>$ is negligible. In any case, we conclude
that the first-order calculation is sufficient, and there is no need to go
to higher orders in equation~(\ref{eq:Iexact}).

\section{A rough error estimate for the three-point correlation function}
\label{app:error}

In order to estimate the error on the three-point correlation function (3PCF),
we start from the number of triples $DDD$ observed, i.e. the number of 
triangles in a certain geometry bin formed by galaxies in a survey. Say the
survey observes a total volume $V$
containing $N_g$ galaxies. We specify a triangle bin in such a way that, after
fixing point 1, we count all triangles where a galaxy is found in the volume
element $dV_2$ around point 2 and a galaxy found in $dV_3$ around point 3 
(points 2 and 3 being determined by the triangle geometry considered).
Then, the expectation value of triples is given by:
\bea
\langle DDD\rangle &=& N_3\: [1 + \xi(r_{12}) + \xi(r_{13}) + \xi(r_{23})
+ \zeta(r_{12},r_{13},r_{23})],\label{eq:triplecount} \\
N_3 &\equiv& N_g\: (n_g dV_2)\:(n_g dV_3) = N_g^3 \frac{dV_2}{V}\frac{dV_3}{V},
\eea
where $n_g$ is the average volume density of galaxies, estimated by $n_g=N_g/V$. 
An estimator for the 3PCF $\zeta$ is given in \cite{Peebles}, \S~55:
\beq
\hat{\zeta} = \frac{DDD - DDR}{RRR} + 2,
\eeq
where $D$ stands for points drawn from the data, while $R$ stands for points 
drawn from a random distribution (with the same boundaries). Calculalting the 
variance of this estimator, we obtain:
\bea
\sigma^2(\hat{\zeta}) &\equiv& \langle\hat{\zeta}^2\rangle - \langle\hat{\zeta}
\rangle^2\\
&=& \evfrac{(DDD)^2}{(RRR)^2} - \evfrac{DDD}{RRR}^2 
 - 2 \evfrac{DDDDDR}{(RRR)^2}
\nonumber\\ & &  
+ 2\evfrac{DDD}{RRR}\evfrac{DDR}{RRR} 
 +\evfrac{(DDR)^2}{(RRR)^2} - \evfrac{DDR}{RRR}^2.
\label{eq:sigmazfull}
\eea
To proceed further, we make the following approximations. The $RRR$ in the 
denominator is assumed to be uncorrelated with the numerators. Further, we set:
\beq
\langle (RRR)^2 \rangle = \langle RRR\rangle^2 = N_3^2.
\eeq
In addition, if we assume that the galaxy correlations are weak, i.e. that
$\langle DDD\rangle \approx N_3$, then the second and third set of terms in 
\refeq{sigmazfull} vanish. We then obtain:
\bea
\sigma^2(\hat{\zeta}) & = & \frac{1}{N_3^2} \left ( \langle (DDD)^2 \rangle - 
\langle DDD \rangle^2 \right )
\nonumber\\
& = & \frac{\sigma^2(\langle DDD\rangle)}{N_3^2}.
\eea

In this approximation, the error on the 3PCF $\zeta$ is simply the 
relative error on the number of triples $DDD$. The latter was
calculated in \cite{Mo-etal} and is given to good accuracy by:
\beq
\sigma^2(\langle DDD\rangle) = \langle DDD\rangle + \frac{36}{N_g}\langle DDD\rangle^2,
\label{eq:DDD}
\eeq
The first term is the standard Poisson term expected for any number count. The 
second one was found for the first time in the derivation of \cite{Mo-etal}. 
Now we again neglect the galaxy correlations in \refeq{triplecount}
and set $\langle DDD\rangle \approx N_3 = N_g\:n_g^2 dV_2 dV_3$, to obtain:
\beq
\sigma(\zeta) = \sqrt{\frac{1}{\langle DDD\rangle} + \frac{36}{N_g}}
= N_g^{-1/2}\sqrt{\frac{1}{n_g^2 dV_2 dV_3} + 36}.
\label{eq:sigmazeta2}
\eeq
The first term again is the Poisson term: 
$\sigma(\zeta)_{\rm Poisson} = N_{\rm triples}^{-1/2}$, analogous to 
$\sigma(\xi) \approx N_{\rm pairs}^{-1/2}$ as found in \cite{Peebles}, \S~48,
for the two-point correlation function. 
The second term in equation~(\ref{eq:DDD}) was added in \cite{Mo-etal},
along with a similar term for the variance of pair counts.

\vspace{0.5 cm}
\end{widetext}

\end{document}